\documentclass[a4paper,fleqn,usenatbib]{mnras}

\usepackage{newtxtext,newtxmath}

\usepackage{graphicx}	
\usepackage{amsmath}	
\usepackage{amssymb}	
\usepackage{multirow}
\usepackage{mathtools}
\usepackage{IEEEtrantools}
\usepackage{float}
\usepackage{rotating}
\usepackage{bm}
\usepackage{footmisc}
\usepackage{booktabs}
\usepackage{threeparttable}
\usepackage{hyperref}
\usepackage{scalefnt}
\LetLtxMacro{\oldtextsc}{\textsc}
\renewcommand{\textsc}[1]{\oldtextsc{\scalefont{1.2}#1}}

\usepackage{relsize}

\usepackage{xcolor}
\usepackage{pdflscape}
\usepackage{array}
\usepackage{threeparttable}
\usepackage{caption}
\usepackage{subcaption}
\usepackage{afterpage}

\newcommand{\cloudy}{\textsc{cloudy}}
\newcommand{\beagle}{\textsc{beagle}}
\newcommand{\cloudspec}{\textsc{cloudspec}}
\newcommand{\synspec}{\textsc{synspec}}
\newcommand{\popstar}{\textsc{popstar}}

\newcommand{\starburst}{\textsc{starburst}\small{99}}
\newcommand{\mappings}{\textsc{mappings\,v}}

\newcommand{\bpass}{\textsc{bpass}}
\newcommand{\fast}{\textsc{fast}}

\newcommand{\CB}{{\small C\&B}}
\newcommand{\HST}{\textit{HST}}
\newcommand{\JWST}{\textit{JWST}}

\newcommand{\Msun}{\hbox{M$_{\rm{\odot}}$}}
\newcommand{\Rsun}{\hbox{R$_{\rm{\odot}}$}}
\newcommand{\mup}{\hbox{$m_{\rm{up}}$}}

\newcommand{\Rt}{\hbox{$R_\mathrm{t}$}}
\newcommand{\Tstar}{\hbox{$T_{\star}$}}
\newcommand{\vinf}{\hbox{$\varv_\infty$}}
\newcommand{\tprime}{\hbox{$t^\prime$}}
\newcommand{\hii}{\hbox{H{\sc ii}}}
\newcommand{\hi}{\hbox{H{\sc i}}}
\newcommand{\zsun}{\hbox{${Z}_{\odot}$}}
\newcommand{\zpsun}{\hbox{${Z}_{\odot}^0$}}

\newcommand{\nh}{\hbox{$n_{\mathrm{H}}$}}
\newcommand{\Nh}{\hbox{$N_{\mathrm{H}}$}}
\newcommand{\Nhzero}{\hbox{$N(\mathrm{H}^0)$}}
\newcommand{\NhIB}{\hbox{$N_{\mathrm{H}}^{\mathrm{IB}}$}}
\newcommand{\rIB}{\hbox{$r^{\mathrm{IB}}$}}

\newcommand{\xid}{\hbox{$\xi_{\rm{d}}$}}
\newcommand{\fesc}{\hbox{$f_{\rm{esc}}$}}
\newcommand{\xidsol}{\hbox{$\xi_{\rm{d\odot}}$}}
\newcommand{\CO}{\hbox{C/O}}
\newcommand{\OH}{\hbox{O/H}}

\newcommand{\COsol}{\hbox{(C/O)$_\odot$}}

\newcommand{\logoh}{\hbox{12 + log(\OH)}}
\newcommand{\logohgas}{\hbox{12 + log(\OH)$_\mathrm{gas}$}}
\newcommand{\Uav}{\hbox{$\langle U\rangle$}}

\newcommand{\tauref}{\hbox{$\tau_{570}$}}

\newcommand{\Te}{\hbox{$T_{\rm{e}}$}}

\newcommand{\ciia}{\hbox{C\,\textsc{ii}\,$\lambda\lambda1036,1037$}}
\newcommand{\siliiia}{\hbox{Si\,\textsc{iii}\,$\lambda1206$}}
\newcommand{\nv}{\hbox{N\,\textsc{v}\,$\lambda1240$}}

\newcommand{\lya}{\hbox{Ly$\alpha$}}
\newcommand{\siliia}{\hbox{Si\,\textsc{ii}\,$\lambda1260$}}

\newcommand{\loi}{\hbox{O\,\textsc{i}}}

\newcommand{\ciib}{\hbox{C\,\textsc{ii}\,$\lambda1335$}}

\newcommand{\civd}{\hbox{C\,\textsc{iv}\,$\lambda\lambda1548,1551$}}
\newcommand{\civ}{\hbox{C\,\textsc{iv}\,$\lambda1549$}}

\newcommand{\heii}{\hbox{He\,\textsc{ii}\,$\lambda1640$}}
\newcommand{\oiiid}{\hbox{O\,\textsc{iii}]$\lambda\lambda1661,1666$}}
\newcommand{\oiii}{\hbox{O\,\textsc{iii}]\,$\lambda1664$}}

\newcommand{\niii}{\hbox{N\,\textsc{iii}]\,$\lambda1750$}}

\newcommand{\ciiid}{\hbox{[C\,\textsc{iii}]$\lambda1907$+C\,\textsc{iii}]$\lambda1909$}}
\newcommand{\ciii}{\hbox{C\,\textsc{iii}]\,$\lambda1908$}}

\newcommand{\nevopt}{\hbox{[Ne\,\textsc{v}]\,$\lambda3426$}}
\newcommand{\lnev}{\hbox{[Ne\,\textsc{v}]}}

\newcommand{\oiid}{\hbox{[O\,{\sc ii}]$\lambda\lambda3726,3729$}}
\newcommand{\oiiopt}{\hbox{[O\,\textsc{ii}]$\lambda3727$}}

\newcommand{\heiopta}{\hbox{He\,\textsc{i}\,$\lambda3889$}}
\newcommand{\heiiopt}{\hbox{He\,\textsc{ii}$\lambda4686$}}
\newcommand{\hb}{\hbox{H$\beta$}}
\newcommand{\oiiiopt}{\hbox{[O\,\textsc{iii}]$\lambda5007$}}

\newcommand{\oiiiopttot}{\hbox{O\,\textsc{iii}\,$\lambda\lambda4959,5007$}}
\newcommand{\oiopt}{\hbox{[O\,\textsc{i}]\,$\lambda6300$}}
\newcommand{\ha}{\hbox{H$\alpha$}}
\newcommand{\niiopt}{\hbox{[N\,\textsc{ii}]$\lambda6584$}}
\newcommand{\heioptb}{\hbox{He\,\textsc{i}\,$\lambda6678$}}
\newcommand{\heioptc}{\hbox{He\,\textsc{i}\,$\lambda7065$}}

\newcommand{\ariiiopt}{\hbox{[Ar\,\textsc{iii}]\,$\lambda7135$}}
\newcommand{\arivopt}{\hbox{[Ar\,\textsc{iv}]\,$\lambda4740$}}
\newcommand{\lariiiopt}{\hbox{[Ar\,\textsc{iii}]}}
\newcommand{\larivopt}{\hbox{[Ar\,\textsc{iv}]}}
\newcommand{\oivfir}{\hbox{[O\,\textsc{iv}]$\,25.9\,\mu$m}}
\newcommand{\oiiifir}{\hbox{[O\,\textsc{iii}]$\,51.8\,\mu$m}}

\newcommand{\lhei}{\hbox{He\,\textsc{i}}}

\newcommand{\loiopt}{\hbox{[O\,\textsc{i}]}}
\newcommand{\loiii}{\hbox{O\,\textsc{iii}]}}
\newcommand{\oratio}{\hbox{O$_\textrm{32}$}}

\newcommand{\lciii}{\hbox{C\,\textsc{iii}]}}
\newcommand{\lciv}{\hbox{C\,\textsc{iv}}}

\newcommand{\loiiopt}{\hbox{[O\,\textsc{ii}]}}
\newcommand{\loiiiopt}{\hbox{[O\,\textsc{iii}]}}

\newcommand{\lheii}{\hbox{He\,\textsc{ii}}}
\newcommand{\lniiopt}{\hbox{[N\,\textsc{ii}]}}
\newcommand{\lnv}{\hbox{N\,\textsc{v}}}

\definecolor{col1}{HTML}{F572B6}
\definecolor{col2}{HTML}{DBAC45}
\definecolor{col3}{HTML}{700002}

\title[Production and escape of ionizing radiation]{Constraints on the production and escape of ionizing radiation from the emission-line spectra of metal-poor star-forming galaxies}

\author[A.Plat et al.]
{A.~Plat,$^{1}$\thanks{E-mail: plat@iap.fr} S.~Charlot,$^1$ G.~Bruzual,$^2$ A.~Feltre,$^{1,3,4}$ A.~Vidal-Garc\'{\i}a,$^{1,5}$ C.~Morisset,$^6$
\newauthor J.~Chevallard$^1$ and H.~Todt$^7$
\\ 
\\
$^{1}$Sorbonne Universit\'e, CNRS, UMR7095, Institut d'Astrophysique de Paris, F-75014, Paris, France\\
$^{2}$Instituto de Radioastronom{\'i}a y Astrof{\'i}sica, UNAM, Campus Morelia, Michoacan, M{\'e}xico, C.P. 58089, M{\'e}xico\\
$^{3}$SISSA, via Bonomea 265, I-34136 Trieste, Italy\\
$^{4}$Univ.\,Lyon, Univ.\,Lyon1, ENS de Lyon, CNRS, Centre de Recherche Astrophysique de Lyon, UMR5574, 69230 Saint-Genis-Laval, France\\
$^{5}$LERMA, Observatoire de Paris, Ecole Normale Sup\'erieure, PSL Research University, CNRS, UMR8112, F-75014 Paris, France\\
$^{6}$Instituto de Astronom{\'i}a, UNAM, Apdo. postal 106, C.P. 22800 Ensenada, Baja California, M\'exico\\
$^{7}$Institute of Physics and Astronomy, University of Potsdam, Karl-Liebknecht-Str. 24/25, 14476 Potsdam, Germany
}

\date{Accepted 2019 September 13. Received 2019 September 9; in original form 2019 July 12}

\pubyear{2018}

\begin{document}
\label{firstpage}
\pagerange{\pageref{firstpage}--\pageref{lastpage}}
\maketitle

\begin{abstract}

We explore the production and escape of ionizing photons in young galaxies by investigating the ultraviolet and optical emission-line properties of models of ionization-bounded and density-bounded \hii\ regions, active-galactic-nucleus (AGN) narrow-line regions and radiative shocks computed all using the same physically-consistent description of element abundances and depletion on to dust grains down to very low metallicities. We compare these models with a reference sample of metal-poor star-forming galaxies and Lyman-continuum (LyC) leakers at various redshifts, which allows the simultaneous exploration of more spectral diagnostics than typically available at once for individual subsamples. We confirm that current single- and binary-star population synthesis models do not produce hard-enough radiation to account for the high-ionization emission of the most metal-poor galaxies. Introducing either an AGN or radiative-shock component brings models into agreement with observations. A published model including X-ray binaries is an attractive alternative to reproduce the observed rise in \heiiopt/\hb\ ratio with decreasing oxygen abundance in metal-poor star-forming galaxies, but not the high observed \heiiopt/\hb\ ratios of galaxies with large EW(\hb). A source of harder ionizing radiation appears to be required in these extreme objects, such as an AGN or radiative-shock component, perhaps linked to an initial-mass-function bias toward massive stars at low metallicity. This would also account for the surprisingly high \loiopt/\loiiiopt\ ratios of confirmed LyC leakers relative to ionization-bounded models. We find no simple by-eye diagnostic of the nature of ionizing sources and the escape of LyC photon, which require proper simultaneous fits of several lines to be discriminated against.

\end{abstract}



\section{Introduction}\label{sec:intro}

The nebular emission from primeval galaxies represents one of our best hopes to constrain the physical processes that dominated reionization of our Universe. Beyond the physical conditions of pristine gas, emission lines are sensitive to different components expected to characterise primeval galaxies: hot massive stars, often considered as the main source of ionizing radiation; active galactic nuclei (AGN), arising from gas accretion onto primordial black holes; radiative shocks, induced by large-scale gas flows; and the leakage of Lyman-continuum (LyC) photons through a porous interstellar medium (ISM), contributing to reionization \citep[see, e.g., the review by][]{stark16}. In waiting for advent of the {\it James Webb Space Telescope} (\JWST), which will enable deep rest-frame ultraviolet and optical emission-line spectroscopy of galaxies into the reionization era at redshifts $z\sim10$--15, more nearby metal-poor galaxies approaching the properties of primeval galaxies offer a useful laboratory in which to test our ability to interpret emission-line spectra.

Observationally, a fast-growing number of studies have progressively uncovered the spectroscopic properties of distant, metal-poor star-forming galaxies at redshifts out to $z\gtrsim7$ \citep[e.g.,][]{erb10, stark14, stark15, rigby15, amorin17, laporte17, schmidt17, mainali17, vanzella17, berg18, nakajima18, nanayakkara19, tang19} and in parallel those of nearby analogues of these pristine galaxies \citep[e.g.,][]{berg16, berg19, senchyna17, senchyna19}. Most of these studies focused on identifying promising tracers and diagnostics of the early chemical enrichment and gas conditions in primeval galaxies, such as for example the \heii, \ciii\ and \civ\ lines and \CO\ abundance ratio. Other studies were more specifically aimed at probing LyC-photon leakage from young star-forming galaxies, using clues such as a small velocity spread of the \lya\ double-peaked emission or large ratios of high- to low-ionization lines \citep[e.g.,][]{jaskot13, izotov16oct, izotov17oct, izotov18aug, debarros16, vanzella16}. Meanwhile, on the theoretical front, much effort was invested into characterising the ionizing spectra and ultraviolet emission-line signatures of young stellar populations \citep[e.g.,][]{gutkin2016, vidal17, byler18}, along with the dependence of these on stellar rotation and binary interactions \citep[e.g.,][]{levesque13,stanway19}, as well as the signatures of active galactic nuclei \citep[AGN; e.g.,][]{Feltre2016,Hirschmann2017,Hirschmann2019,nakajima18} and shock-ionized gas \citep[e.g.,][]{allen08, izotov12, 3MdBs19}.

In practice, the observational studies mentioned above provide a valuable set of, in some respects, independent investigations focusing individually on the analysis of a specific set of emission lines with specific models (for the production of radiation and the photoionization calculations), parametrized in a specific way (e.g., element abundances and depletions; inclusion or not of dust physics in the photoionization calculations), depending on the nature and redshift of the sample and the spectrograph employed. These analyses have brought important lessons, such as the usefulness of the \ciii\ and even \civ\ lines as signposts of galaxies in the reionization era given the expected strong attenuation of \lya\ \citep[e.g.,][]{stark14,senchyna19} and the potentiality of the \civ/\heii\ luminosity ratio for identifying AGN  \citep[e.g.,][]{Feltre2016,nakajima18}, the \ciii/\heii\ ratio for identifying shock-ionized gas \citep{jaskot16} and the \oiiiopt/\oiiopt\ ratio for identifying LyC-photon leakage \citep[e.g.,][]{nakajimaetouchi14, izotov16oct}. 

A current difficulty in reaching a robust picture from this progress on several fronts in parallel is that the conclusions drawn from fitting a restricted set of emission lines using specific models may not be consistent with findings based on other lines and different models. This may be particularly important, for example, in the context of interpreting the exceedingly large strengths of He\,\textsc{ii} recombination lines (requiring photon energies $E_\mathrm{ion}>54.4$\,eV) found in very metal-poor, actively star-forming galaxies, which seem to elude standard model predictions \citep[e.g.,][]{shirazi12,senchyna17,berg18,nanayakkara19,stanway19}. Based on various arguments, contributions from very massive stars \citep[e.g.,][]{Graefener15}, stripped stars produced by close-binary evolution or X-ray binaries \citep[e.g.,][]{SenchynaStark19, schaerer19}, AGN \citep[e.g.,][]{nakajima18} and radiative shocks \citep[e.g.,][]{izotov12} have been proposed to account for the required hard radiation. Also, while the \oiiiopt/\oiiopt\ diagnostic to characterise galaxies leaking LyC photons may not be as reliable as expected and alternative diagnostics based, e.g., on He\,\textsc{i} lines have been suggested, different types of investigations of the ultraviolet and optical signatures of LyC leakage have so far focused on rather limited sets of emission lines \citep[e.g.,][]{jaskot13, zackrisson13, zackrisson17, stasinska15, jaskot16, izotov17oct}. These examples illustrate the need for a homogeneous investigation of emission-line diagnostics of metal-poor star-forming galaxies with a wide collection of intercomparable models (of the type of that proposed by \citealt{stasinska15} for a few optical lines).

In this paper, we examine a full set of ultraviolet/optical observables of metal-poor star-forming galaxies with a library of nebular-emission models enabling the exploration of a wide range of physical parameters. To conduct this analysis, we build a reference sample of ultraviolet and optical observations of  metal-poor star-forming galaxies and confirmed and candidate LyC leakers (and other star-forming galaxies and AGN) in a wide redshift range. This sample allows us to simultaneously explore diagnostic diagrams involving more emission lines than typically available at once for individual subsamples. We use this sample to investigate potentially discriminating signatures of single- and binary-star populations (using the most recent versions of the \citealt{Bruzual2003} and \citealt{BPASSv21} models), narrow-line regions of AGN \citep{Feltre2016} and radiative shocks \citep{3MdBs19} on the emission-line properties of metal-poor star-forming galaxies, adopting throughout the same parametrization of nebular-gas abundances \citep{gutkin2016}. 

Our analysis confirms that current single- and binary-star population synthesis models do not produce hard-enough ionizing radiation to account for the strong \lheii\ emission seen in some of the most metal-poor galaxies, although with slightly better agreement than concluded recently by \citet{stanway19}. We show that an AGN or radiative-shock component allows models to reproduce observations in nearly all the ultraviolet and optical line-ratio diagrams we investigate. We also consider X-ray binaries as a potential source of ionizing radiation, using the model recently proposed by \citet{schaerer19}. This can reproduce the observed rise in \heiiopt/\hb\ ratio toward low metallicities in star-forming galaxies, but not the high observed \heiiopt/\hb\ ratios of galaxies with large EW(\hb). A source of harder ionizing radiation appears to be required in these extreme objects. In the end, we find that while none of the ultraviolet and optical emission-line diagrams we consider allows simple by-eye diagnostics of the nature of ionizing sources and the escape of LyC photons in metal-poor star-forming galaxies, differences exist in the spectral signatures of these physical quantities, which should enable more stringent constraints from simultaneous fits of several lines using tools such as \beagle\ \citep{Chevallard2016}.

We present our models of ionization-bounded and density-bounded galaxies, AGN narrow-line regions and radiative shocks in Section~\ref{sec:models}. In Section~\ref{sec:obs}, we assemble the reference sample of metal-poor star-forming galaxies, LyC leakers and other star-forming galaxies and AGN, which we use in Section~\ref{sec:params} to explore the influence of the different adjustable parameters of the models on emission-line spectra. In Section~\ref{sec:constraints}, we investigate potentially discriminating emission-line diagnostics of the production and escape of ionizing radiation in metal-poor star-forming galaxies. Our conclusions are summarized in Section~\ref{sec:conclu}.


\section{Modelling approach}\label{sec:models}

In this section, we present the set of versatile models that will be used in Section~\ref{sec:params} to explore, in a physically consistent way, the influence of a wide range of parameters on the observed ultraviolet and optical nebular emission from young star-forming galaxies. We start by describing the models we adopt to compute properties of ionization-bounded galaxies. Then, we describe our approach to model density-bounded galaxies. We also appeal to existing prescriptions to include the contributions by AGN and shock-ionized gas to the nebular emission from galaxies.

\subsection{Ionisation-bounded models}\label{sec:ionb}

We adopt the approach introduced by \citet[][see also \citealt{gutkin2016}]{ChaLon01} to compute the nebular emission from ionization-bounded galaxies. This is based on the combination of a stellar population synthesis model with a photoionization code to compute the luminosity per unit wavelength $\lambda$ emitted at time $t$ by a star-forming galaxy as
\begin{equation}
L_{\lambda}(t)=\int_0^t \mathrm{d}\tprime\, \psi(t-\tprime) \, S_{\lambda}[\tprime,Z(t-\tprime)] \, T_{\lambda}(t,\tprime)\,,
\label{eq:flux_gal}
\end{equation}
where $\psi(t-\tprime)$ is the star formation rate at time $t-\tprime$, $S_\lambda[\tprime,Z(t-\tprime)]$ the luminosity produced per unit wavelength per unit mass by a single stellar generation of age $\tprime$ and metallicity $Z(t-\tprime)$ and $T_\lambda(t,\tprime)$ the transmission function of the ISM. Following \citet[][see also \citealt{vidal17}]{CharlotFall2000}, we write
\begin{equation}
\label{eq:transmission}
T_{\lambda}(t,\tprime)= T_{\lambda}^{\rm BC}(\tprime) \, T_{\lambda}^{\rm ICM}(t).
\end{equation}
where $T_{\lambda}^{\rm BC}(\tprime)$ is the transmission function of stellar birth clouds (i.e. giant molecular clouds) and $T_{\lambda}^{\rm ICM}(t)$ that of the intercloud medium  (i.e. diffuse ambient ISM). In the present study, we focus on young galaxies with ages close to the typical timescale for the dissipation of giant molecular clouds in star-forming galaxies \citep[$\sim10$\,Myr, ; e.g.,][]{Murray2010,Murray2011} and do not include any intercloud medium. The birth clouds, assumed all identical, are described as an inner \hii\ region ionized by young stars and bounded by an outer \hi\ region \citep{CharlotFall2000}. We thus write
\begin{equation}\label{eq:transBC}
T_{\lambda}(t,\tprime)= T_{\lambda}^{\rm BC}(\tprime)= T_{\lambda}^{\rm HII}(\tprime) \, T_{\lambda}^{\rm HI}(\tprime)\,.
\end{equation}

By analogy with \citet[][see also \citealt{gutkin2016}]{ChaLon01}, we compute the transmission function $T_{\lambda}^{\rm HII}(\tprime)$ of the ionized gas [$T_{\lambda}^{+}(\tprime)$ in their notation] using the photoionization code \cloudy\ (we adopt here version c17.00; \citealt{cloudyc17}). In this approach, the galaxy-wide transfer of stellar radiation through ionized gas is described via a set of `effective' parameters. The main adjustable parameters are (see \citealt{gutkin2016} for details):
\begin{enumerate}
\item The (hydrogen) gas density, \nh.
\item The total gas metallicity, assumed to be equal to that of the ionizing stars, $Z$. We adopt the chemical-element abundances listed in table~1 of \citet{gutkin2016},\footnote{These are based on the solar chemical abundances compiled by \citet{Bressan2012} from the work of \cite{Grevesse1998}, with updates from \citet[][see table~1 of \citealt{Bressan2012}]{Caffau2011}, and small adjustments of the solar nitrogen ($-0.15$\,dex) and oxygen ($+0.10$\,dex) abundances relative to the mean values quoted in table~5 of \citet[][see \citealt{gutkin2016} for details]{Caffau2011}.} corresponding to a present-day solar (photospheric) metallicity $\zsun=0.01524$ and a protosolar metallicity (i.e. before the effects of diffusion) $\zpsun=0.01774$. Nitrogen and carbon are assumed to both have primary and secondary nucleosynthetic components. The total (primary+secondary) nitrogen abundance is related to that of oxygen via equation~(11) of \citet{gutkin2016}. 
\item The carbon-to-oxygen abundance ratio, \CO. This adjustable parameter allows secondary C production to be kept flexible [for reference, $\COsol=0.44$].
\item The dust-to-metal mass ratio, \xid, which reflects the depletion of heavy elements on to dust grains ($\xidsol=0.36$; see table~1 of \citealt{gutkin2016}).
\item The volume-averaged ionisation parameter at age $\tprime=0$, noted simply $\Uav\equiv\Uav(\tprime=0)$. The volume-averaged ionisation parameter of a spherical \hii\ region can be expressed as \citep[e.g., equation~3 of][]{Panuzzo03}
\begin{equation}\label{eq:Udef}
\Uav (\tprime)= \frac{3\alpha_B^{2/3}}{4c} \left[
    \frac{3Q(\tprime)\epsilon^2n_\mathrm{H}}{4 \pi} \right]^{1/3}\,,
\end{equation}
where $Q(\tprime)$ is the time-dependent rate of ionizing photons produced by a single stellar generation of age $\tprime$, $\epsilon$ the volume-filling factor of the gas (i.e., the ratio of the volume-averaged hydrogen density to \nh) and $\alpha_{\rm B}$ the case-B hydrogen recombination coefficient. We note that the volume-averaged ionization parameter in expression~\eqref{eq:Udef} is a factor of 9/4 larger than the zero-age ionization parameter at the Str\"omgren radius used by \citet{gutkin2016} and a factor of 3/4 smaller than the quantity defined by equation~(7) of \citet{ChaLon01}. These different model-labelling choices are transparent to the \cloudy\ calculations. Also, since \Uav\ is proportional to $[Q(0)\epsilon^2\nh]^{1/3}$, at fixed \Uav\ and \nh, there is a degeneracy in the calculations between the adopted normalisation of $Q(0)$ (via an effective mass of ionizing star cluster) and $\epsilon$.
\end{enumerate}
The \cloudy\ calculations to compute $T_{\lambda}^{\rm HII}(\tprime)$ are performed in closed geometry, adopting a small inner radius of the gaseous nebula, $r_\mathrm{in}=0.01$\,pc, to ensure spherical geometry. The photoionization calculations are stopped at the edge of the \hii\ region, when the electron density falls below 1 per cent of \nh.

As noted by \citet{vidal17}, the above standard \cloudy\ calculations do not account for  interstellar-{\em line} absorption in the ionized gas. In the following, we also wish to investigate the effects on nebular emission of interstellar-line absorption in the \hii\ interiors and \hi\ envelopes of stellar birth clouds. To compute $ T_{\lambda}^{\rm HII}(\tprime) \, T_{\lambda}^{\rm HI}(\tprime)$ in equation~\eqref{eq:transBC} in this case, we  appeal to the prescription of \citet[][see their section~4]{vidal17}, which extends the computations of \citet{gutkin2016} to account for interstellar-line absorption in stellar birth clouds. This is achieved through the combination of \cloudy\ with the general spectrum synthesis program \synspec\ \citep[e.g.,][]{Hubeny2011}\footnote{See \url{http://nova.astro.umd.edu/Synspec49/synspec.html}}  via an interactive program called \cloudspec\ \citep[][see also \citealt{Heap2001}]{Hubeny2000}. For this purpose, the \cloudy\ calculations are stopped when the kinetic temperature of the gas falls below 50\,K, assumed to define the \hi\ envelope of a typical stellar birth cloud (see \citealt{vidal17} for more details).

We require a stellar population synthesis model to compute the spectral evolution of a single stellar generation, $S_\lambda[\tprime,Z(t-\tprime)]$, in equation~\eqref{eq:flux_gal}. In most applications in this paper, we use the latest version of the \citet{Bruzual2003} stellar population synthesis model (Charlot \& Bruzual, in preparation, hereafter \CB). This differs from the version used by \citet{gutkin2016} in the inclusion of updated spectra of Wolf-Rayet (hereafter WR) stars from the Potsdam Wolf-Rayet (PoWR) model library  (see Appendix~\ref{app:wrmodels}) and of main-sequence massive stars from \citet{Chen2015}. When indicated, we also use the Binary Population and Spectral Synthesis (\bpass\,v2.2.1) models of \citet{BPASSv22} to explore the effects of binary interactions on the spectral evolution of young stellar populations, in particular the enhancement of extreme ultraviolet radiation by envelope stripping (of primary stars) and chemical homogeneisation (of rapidly rotating secondaries). We adopt throughout a \citet{Chabrier2003} initial mass function (IMF), with lower mass cutoff 0.01\,\Msun\ and upper mass cutoff in the range $100\leq\mup\leq600\,$\Msun. IMF upper mass cutoffs well in excess of 100\,\Msun\ have been suggested by models and observations of massive, low-metallicity star clusters \citep[e.g.,][see also \citealt{vink11}]{crowther16, smith16}. For the star formation history, $\psi(t-\tprime)$ in equation~\eqref{eq:flux_gal}, we adopt either a delta function (Simple Stellar Population, hereafter SSP) or constant star formation rate. We consider ages of up to 50\,Myr, as, even though 99.9 per cent of H-ionizing photons are produced at ages less than 10\,Myr in single-star models \citep[e.g.,][]{CharlotFall1993,Binette1994}, binary interactions can extend the production over longer timescales \citep[e.g.][]{stanway16}.


\subsection{Density-bounded models}\label{sec:fesc}

\begin{figure*}
\begin{center}
\resizebox{\hsize}{!}{\includegraphics{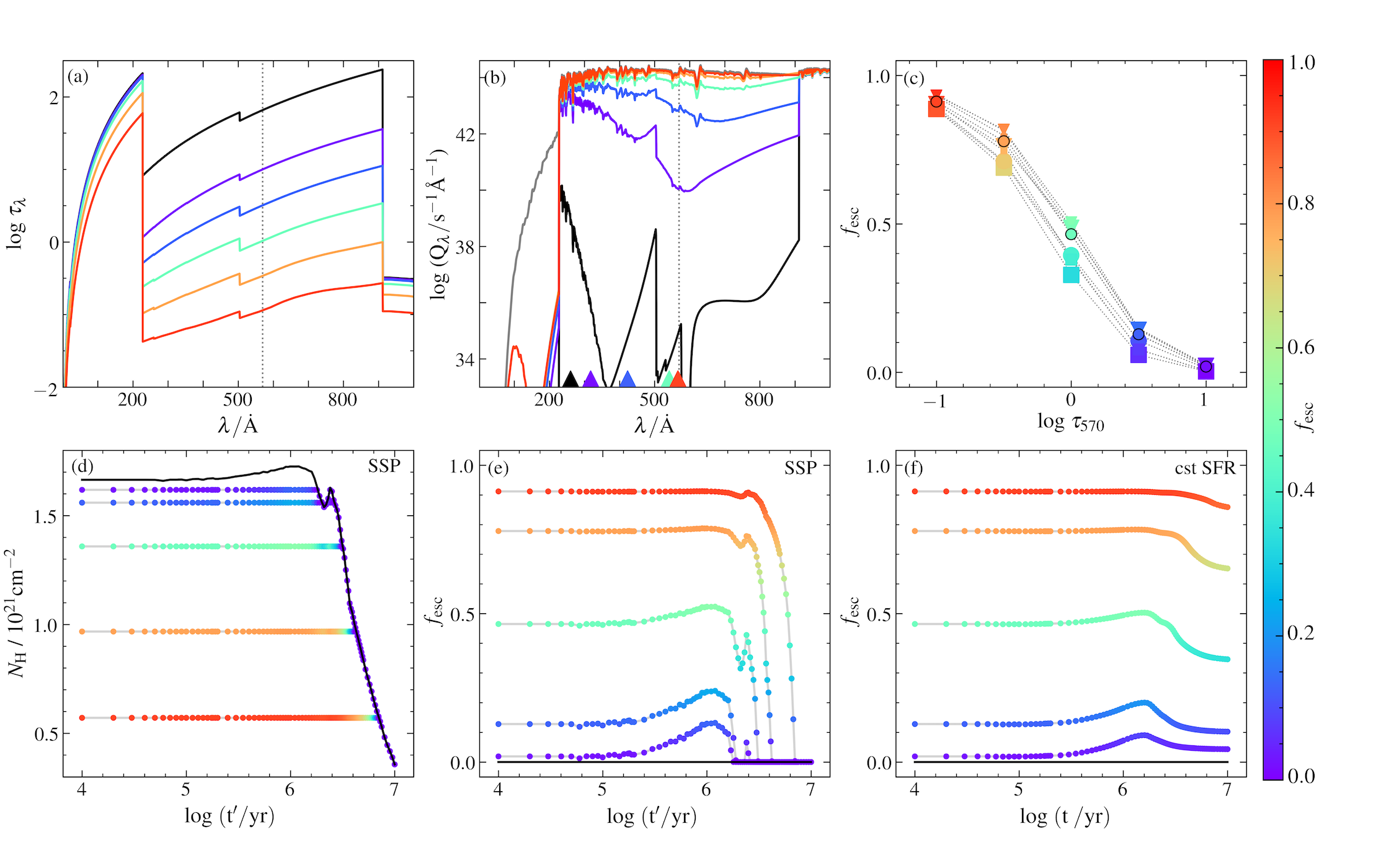}}
\end{center}
\caption{Relationship between zero-age optical depth to LyC photons, $\tau_\lambda$, fraction of escaping LyC photons, \fesc, and H-column density, \Nh, in the density-bounded models of Section~\ref{sec:fesc}. (a) $\tau_\lambda$ plotted against $\lambda$ for models with $Z=0.002$, $\log\Uav=-2.0$, $\xid=0.3$, $\mup=300\,$\Msun\ and for 5 choices of \tauref\ (from $-1.0$ to +1.0 in steps of 0.5, as identifiable from the dotted vertical line). Different colours reflect the \fesc\ values of these models (see scale on the right), while the black line corresponds to the ionization-bounded model. (b) Monochromatic  photon rate $Q_\lambda$ emerging at age zero from the models of panel (a), plotted against $\lambda$. Triangles indicate the photon-weighted mean wavelength of each spectrum. The grey curve shows the input stellar population spectrum, with photon-weighted mean wavelength marked by the dotted vertical line. (c) \fesc\ plotted against \tauref\ for the same models as in (a) (black-contoured circles) and for models with different metallicities (upside-down triangles: $Z=0.0005$; squares: $Z=0.008$) and $\log\Uav=-3.0$, $-2.0$ and $-1.0$ (in order of increasing symbol size). (d) H-column density plotted against \hii-region age for the same models as in (a). (e) \fesc\ plotted against \hii-region age for the same models as in (a). (f) Same as (e), but for a galaxy with constant star formation rate (i.e., adopting $\psi=\,$cst in equation~\ref{eq:flux_gal}). In panels (d)--(f), different colours reflect the \fesc\ values of the density-bounded models computed at discrete ages (see scale on the right), }
\label{fig:fesc}
\end{figure*}

We are also interested in the influence of LyC-photon leakage on the nebular emission from young star-forming galaxies. This leakage is generally thought to occur in two main possible ways: through holes carved into the neutral ISM by extreme galactic outflows (`picket-fenced' model), which can be traced by the presence of residual flux in the cores of saturated interstellar low-ionization absorption lines \citep[e.g., \ciia;][]{Heckman11,Alexandroff15} and reduced nebular emission-line equivalent widths \citep{zackrisson13}, but with no effect on line ratios \citep{zackrisson17}; or through density-bounded (i.e. optically thin to LyC photons) \hii\ regions, which are expected to lead to weak low-ionization emission lines, a small velocity spread of the Ly$\alpha$ double-peaked emission and large ratios of high- to low-ionization lines \citep{giammanco05, pellegrini12, jaskot13, zackrisson13, nakajimaetouchi14, Nicholls14, stasinska15, jaskot16, Alexandroff15, izotov18aug, dAgostino19b}. We note that, in addition to these two commonly cited scenarios, direct ionizing radiation from runaway massive stars could also contribute significantly to LyC leakage \citep{ConroyKratter12}. We focus here on the modelling of density-bounded \hii\ regions, which is the only LyC-leakage scenario affecting ratios of nebular emission lines and also seems to be favoured by current observations (see Section~\ref{obs:leakers} below).

\subsubsection{Modelling approach}

In the framework of photoionization modelling described in Section~\ref{sec:ionb}, we can write the (effective) time-dependent fraction of LyC photons escaping from a density-bounded \hii\ region ionized by a single stellar generation as
\begin{equation}\label{eq:fesc}
\fesc(\tprime)=\dfrac{Q^\mathrm{out}(\tprime)}{Q(\tprime)},
\end{equation}
where $Q(\tprime)$ is the rate of LyC photons produced by the stellar population at age \tprime, and $Q^\mathrm{out}(\tprime)$ the rate emerging from the nebula at that age. The quantity $Q^\mathrm{out}(\tprime)$ encompasses both the fraction of LyC photon initially produced by stars that escape from the nebula, and the LyC photons created within the nebula (via free-bound emission) that also escape from it. This latter contribution is negligible, as the ionizing recombination continuum amounts to less than 0.001 per cent of $Q(\tprime)$ for an ionization-bounded nebula.

It is convenient to parametrize density-bounded models in terms of the zero-age optical depth to LyC photons, rather than the H-column density of the \hii\ region. This is because at fixed H-column density, the optical depth, which controls the quantity \fesc\ we are interested in, can vary greatly depending on gas composition and ionization state. While \cloudy\ computes the optical depth to LyC photons in a self-consistent way, it is useful, for the purpose of describing the sensitivity of observed line ratios on model parameters (Section~\ref{sec:params}), to express the optical depth at wavelength $\lambda$ and radius $r$ at age $\tprime=0$ as the sum of the optical depths arising from the gas and dust phases,
\begin{equation}\label{eq:tautot}
\tau_\lambda(r)=\tau_{\lambda,\mathrm{gas}}(r)+\tau_{\lambda,\mathrm{dust}}(r)\,.
\end{equation}
The optical depth from neutral hydrogen and other gaseous species is \citep[e.g.][]{Osterbrock2006}
\begin{equation}\label{eq:taugas}
\tau_{\lambda,\mathrm{gas}}(r)=\sigma_\lambda(\mathrm{H}^0) N(\mathrm{H}^0,r)  + \sum_{\mathrm{X},i} \sigma_\lambda({\mathrm{X}^{+i}}) N(\mathrm{X}^{+i},r)\,,
\end{equation}
where $\sigma_\lambda(\mathrm{H}^0)$ and $\sigma_\lambda({\mathrm{X}^{+i}})$ are the monochromatic absorption cross-sections of neutral hydrogen and element X (with atomic number $\geq2$) in ionization state $+i$, and $N(\mathrm{H}^0,r)$ and $N(\mathrm{X}^{+i},r)$ the column densities of H$^0$ and X$^{+i}$ out to radius $r$. Both $N(\mathrm{H}^0,r)$ and $N(\mathrm{X}^{+i},r)$ are proportional to the gas filling factor $\epsilon$ (Section~\ref{sec:ionb}). The optical depth arising from dust can be expressed as
\begin{equation}\label{eq:taudust}
\tau_{\lambda,\mathrm{dust}}(r)=  \sigma_{\lambda,\mathrm{dust}}\,\xid Z \Nh(r)\,,
\end{equation}
where $ \sigma_{\lambda,\mathrm{d}}$ is the dust absorption cross-section at wavelength $\lambda$, and $\Nh(r)=\epsilon \nh r$ the H-column density at radius $r$.

In practice, we parametrise density-bounded models in terms of the zero-age optical depth of the \hii\ region to LyC photons with wavelength $\lambda=570\,$\AA, noted \tauref. This corresponds to the photon-rate-weighted mean wavelength of H-ionizing radiation produced by a zero-age stellar population with metallicity $Z=0.002$ and IMF upper-mass cutoff $\mup=300\,$\Msun\ in the \CB\ models. For chosen input parameters, including \tauref, we run \cloudy\ at age $\tprime=0$ in the same way as described in the previous section for ionization-bounded \hii\ regions, but stopping this time the calculation when the optical depth at $\lambda=570\,$\AA\ reaches \tauref. At the end of the calculation, we record the H-column density corresponding to this model of density-bounded nebula. Then, for all ages $\tprime>0$, we compute the nebular emission with \cloudy, stopping the calculation when the H-column density reaches that determined at $\tprime=0$, or when the electron density falls below 1 per cent of \nh. 

\subsubsection{Properties of density-bounded models}

Fig.~\ref{fig:fesc} illustrates the relationship between \tauref, \fesc\ and H-column density, \Nh, in these density-bounded models. Fig.~\ref{fig:fesc}a shows the wavelength dependence of the zero-age optical depth, $\tau_\lambda$ (equation~\ref{eq:tautot}), for models with fixed metallicity $Z=0.002$, ionization parameter $\log\Uav=-2.0$, dust-to-metal mass ratio $\xid=0.3$, IMF upper mass cutoff $\mup=300\,$\Msun, and for 5 choices of \tauref, from $-1.0$ to +1.0 in steps of 0.5. The curves are colour-coded to reflect the \fesc\ values of these models. Also shown for comparison is the ionization-bounded model with same parameters (in black). The breaks in the curves correspond to the ionization potentials of helium (at 228 and 504\,\AA) and hydrogen (at 912\,\AA), which give rise to sharp features in the ionizing spectra emerging from these \hii\ regions (Fig.~\ref{fig:fesc}b). Also, the increase in $\tau_\lambda$ at wavelengths from 228 to 912\,\AA\ implies that the photon-weighted mean wavelength of ionizing photons emerging from the \hii\ region increases from high to low \tauref\ (as indicated by the triangles at the bottom of Fig.~\ref{fig:fesc}b). This also implies that ionizing photons with wavelengths less than 912\,\AA\ can escape the nebula when the optical depth at the Lyman edge is unity.

In Fig.~\ref{fig:fesc}c, we show \fesc\ as a function of \tauref\ for models with different metallicities, $Z=0.0005$ (upside-down triangles), 0.002 (circles) and 0.008 (squares), and different ionization parameters, $\log\Uav=-3.0$, $-2.0$ and $-1.0$ (in order of increasing symbol size).  At fixed \tauref, differences in \fesc\ between these models arise from differences in the wavelength dependence of $\tau_\lambda$. Increasing $Z$ at fixed $\log\Uav$ implies a larger contribution to the optical depth by metals and dust, and hence, at fixed \tauref, a smaller one by H$^0$ (equations~\ref{eq:tautot}--\ref{eq:taudust}). This turns out to produce a flatter dependence of $\tau_\lambda$ on wavelength relative to that shown in Fig.~\ref{fig:fesc}a, which makes \fesc\ drop at fixed \tauref\ in Fig.~\ref{fig:fesc}c. Also, increasing $\log\Uav$ at fixed $Z$ (which can be achieved by raising $\epsilon$ at fixed $Q$ in equation~\ref{eq:Udef}) makes \Nh, and hence, the optical depths from metals and dust, increase, implying a smaller H$^0$ optical depth at fixed \tauref. This (and the higher ionization state of metals) again contributes to making \fesc\ drop when $\log\Uav$ increases in Fig.~\ref{fig:fesc}c. The effect is largest around the critical regime $\tauref\sim1$, where \fesc\ can change from 0.35 to 0.55 depending on the adopted metallicity and ionization parameter. 

In Fig.~\ref{fig:fesc}d, we plot \Nh\ as a function of \hii-region age for the same models with $Z=0.002$ and (zero-age) $\log\Uav=-2.0$ as in Fig.~\ref{fig:fesc}a. For the reference ionization-bounded model (black curve), \Nh\ rises until ages around 1\,Myr, as massive stars evolve on the main sequence, and then drops and exhibits a secondary peak around 2.5\,Myr, when the hard ionizing radiation from hot WR stars induces a peak in $Q(\tprime)$ (and , $\log\Uav$; equation~\ref{eq:Udef}). Then, at later ages, \Nh\ drops as the supply of ionizing photons dries up. For density-bounded models, by design, \Nh\ remains constant at all ages until $Q(\tprime)$ drops enough for the region to become ionization-bounded, reducing \fesc\ to zero (Fig.~\ref{fig:fesc}e). In the case of a galaxy with constant star formation rate (i.e., adopting $\psi=\,$cst in equation~\ref{eq:flux_gal}), \fesc\ does not reach zero at ages $\tprime\gtrsim10\,$Myr, as newly formed \hii\ regions continue to maintain leakage of LyC photons (Fig.~\ref{fig:fesc}f).

\begin{figure*}
\begin{center}
\resizebox{\hsize}{!}{\includegraphics{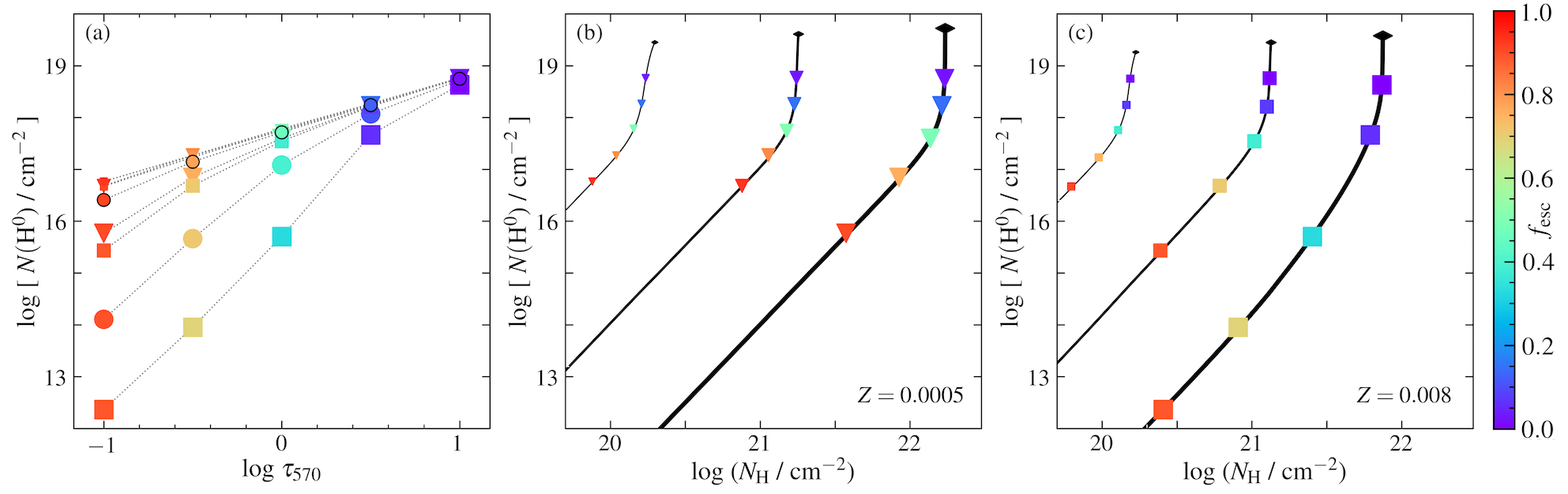}}
\end{center}
\caption{Relationship between optical depth to LyC photons at $\lambda=570\,$\AA, \tauref, neutral-H column density, $N(\mathrm{H}^0)$, and total H-column density, \Nh, at age $\tprime=0$ in the density-bounded models of Section~\ref{sec:fesc}. (a)  $N(\mathrm{H}^0)$ plotted against \tauref\ for the same models as in Fig.~\ref{fig:fesc}c. (b) $N(\mathrm{H}^0)$ plotted against \Nh\ for the subset of models in (a) with metallicity $Z=0.0005$. Lines join models of fixed $\log\Uav=-3.0$, $-2.0$ and $-1.0$ (in order of increasing thickness). At the top of each line, a black diamond indicates the location of the ionization-bounded model. (c) Same as (b), but for $Z=0.008$. }
\label{fig:NH0}
\end{figure*}

It is also interesting to examine the dependence of the neutral-H column density, \Nhzero, on \tauref\ and \Nh\ in the density-bounded models of Fig.~\ref{fig:fesc}. Fig.~\ref{fig:NH0}a shows \Nhzero\ against \tauref\ for the same zero-age models with various metallicities and ionization parameters as in Fig.~\ref{fig:fesc}c. At fixed \tauref, the drop in \Nhzero\ mentioned above to compensate the enhanced opacity from metals and dust when increasing $Z$ and \Uav\ is clearly apparent in this diagram, especially at low \tauref, when the outer H$^0$ layer of the density-bounded \hii\ regions is very thin [i.e., $\Nhzero\lesssim1.6\times10^{17}\rm{cm}^{-2}$, the column density required to produce unit optical depth at the Lyman edge]. In Figs~\ref{fig:NH0}b and \ref{fig:NH0}c, we plot \Nhzero\ against total H-column density \Nh, for $Z=0.0005$ and 0.008, respectively, and in each case for the same models as in Fig.~\ref{fig:NH0}a with $\log\Uav=-3.0$, $-2.0$ and $-1.0$ (the lines join models of fixed \Uav). As noted previously, \Nh\ increases together with \Uav. Also, at fixed \Uav\ and \tauref, \Nh\ is smaller for $Z=0.008$ than for $Z=0.0005$, because more ionizing photons are absorbed by metals and dust relative to hydrogen at higher $Z$. At fixed ionization parameter, decreasing \tauref\ relative to the ionization-bounded model firstly amounts to making \Nhzero\ decrease at nearly fixed \Nh, until the outer H$^0$ layer of the \hii\ region is nearly peeled off [i.e., around $\Nhzero\sim1.6\times10^{17}\rm{cm}^{-2}$]. Further reducing \tauref\ requires a drop in the optical depth to LyC photons arising from metals and dust, and hence smaller \Nh. The transition between the two regimes occurs at smaller \fesc\ for $Z=0.008$ (Fig.~\ref{fig:NH0}c) than for $Z=0.0005$ (Fig.~\ref{fig:NH0}b), because of the larger metal and dust optical depths at higher $Z$. 

\begin{figure*}
\begin{center}
\resizebox{\hsize}{!}{\includegraphics{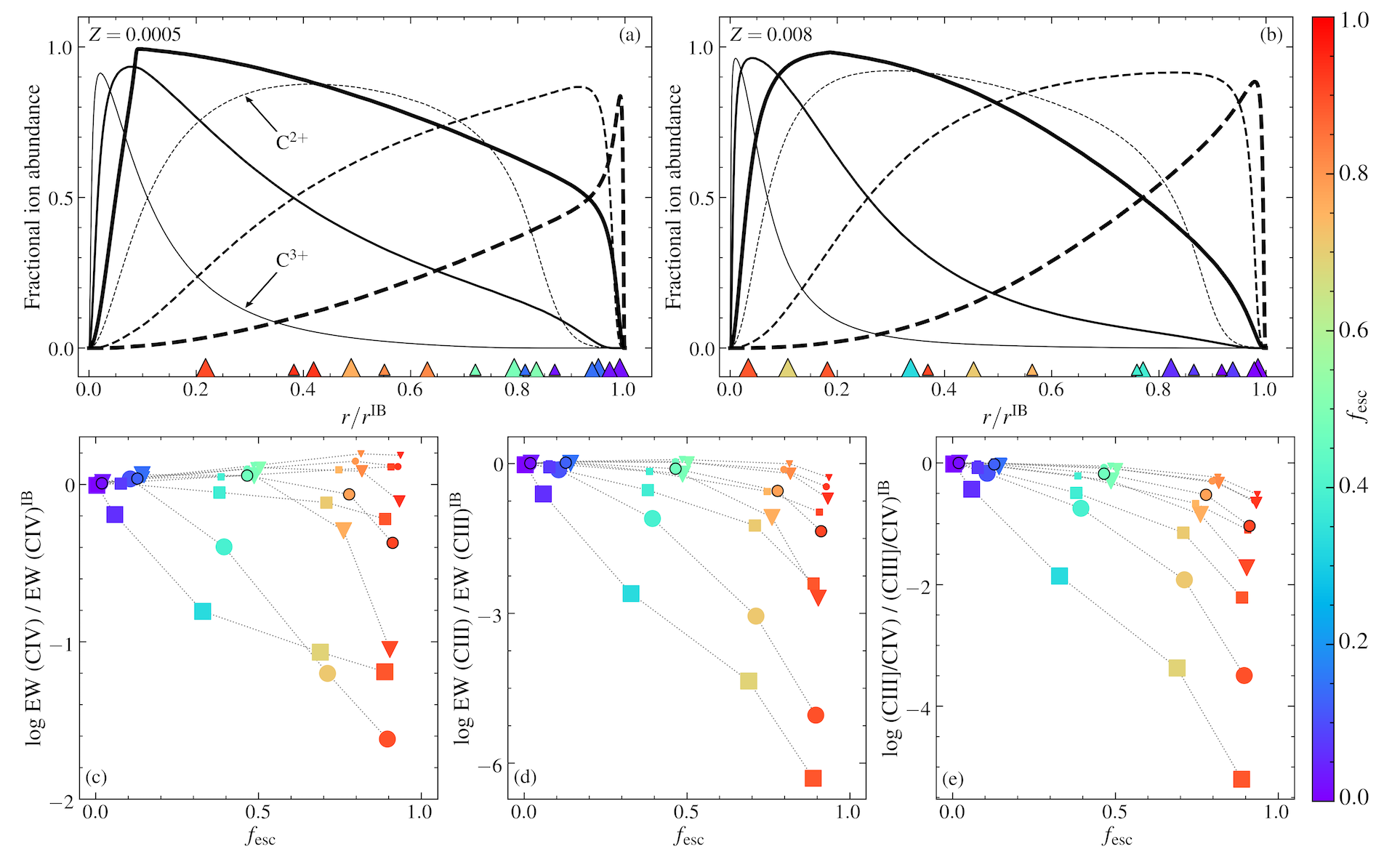}}
\end{center}
\caption{Carbon emission-line properties at age $\tprime=0$ in the density-bounded models of Section~\ref{sec:fesc}. (a) Fractional abundances of C$^{2+}$ (dashed lines) and C$^{3+}$ (solid lines) plotted against radius $r$ (in units of the ionization-bounded \hii-region radius, \rIB), for reference ionization-bounded models with $Z=0.0005$, $\xid=0.3$, $\mup=300\,$\Msun\ and 3 values of the ionization parameter, $\log\Uav=-3.0$, $-2.0$ and $-1.0$ (in order of increasing line thickness). Triangles at the bottom locate the cutoff radii of the models with different \tauref\ and $\log\Uav$ (in order of increasing symbol size) of Fig.~\ref{fig:NH0}b. (b) Same as (a), but for $Z=0.008$. (c) Equivalent width of the \civ\ nebular emission line (in units of the equivalent width in the ionization-bounded case) plotted against \fesc, for the same density-bounded models as in Fig.~\ref{fig:fesc}c. (d) Same as (c), but for the equivalent width of \ciii. (e) Same as (c), but for the \ciii/\civ\ emission-line luminosity ratio. }
\label{fig:struction_carbon}
\end{figure*}

\subsubsection{Implications for emission-line properties}

We now turn to the emission-line properties of these density-bounded models. Figs~\ref{fig:struction_carbon}a and \ref{fig:struction_carbon}b show the fractions of total C abundance in the form of C$^{2+}$ (dashed lines) and C$^{3+}$ (solid lines), as a function of radius, in three reference ionization-bounded models with $\log\Uav=-3.0$, $-2.0$ and $-1.0$ (in order of increasing line thickness), for $Z=0.0005$ and 0.008, respectively, at age $\tprime=0$.  The cutoff radii of the density-bounded models with different \tauref\ and $\log\Uav$ from Fig.~\ref{fig:NH0}b are indicated by triangles at the bottom of each panel. At fixed \Uav, the fractional abundance of C$^{3+}$ is largest in the inner, highly-ionized parts of the nebula, while C$^{2+}$ dominates on the outer, lower-ionization parts. Increasing \Uav\ (which can be achieved by raising $\epsilon$ at fixed $Q$ in equation~\ref{eq:Udef}) increases the probability for carbon to be multiply ionized in the inner parts of the nebula, causing an inner C$^{4+}$ zone (not shown) to develop, while the C$^{3+}$ zone thickens to the detriment of the C$^{2+}$ zone.  At fixed \tauref, the cutoff radii corresponding to density-bounded models with  different \Uav\ and $Z$ sample different global abundances of C$^{2+}$ and C$^{3+}$ in Figs~\ref{fig:struction_carbon}a and \ref{fig:struction_carbon}b.

\begin{figure*}
\begin{center}
\resizebox{\hsize}{!}{\includegraphics{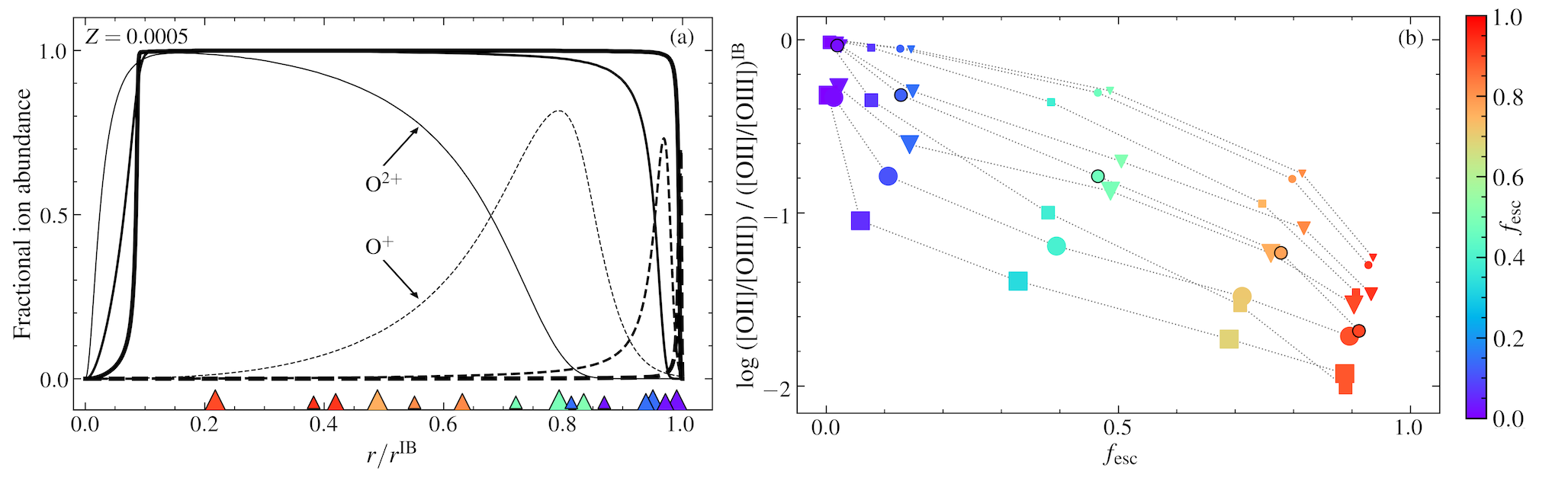}}
\end{center}
\caption{Oxygen emission-line properties at age $\tprime=0$ in the density-bounded models of Section~\ref{sec:fesc}. (a) Same as Fig.~\ref{fig:struction_carbon}a, but for the fractions of O$^{+}$ (dashed lines) and O$^{2+}$ (solid lines). (b) Same as Fig.~\ref{fig:struction_carbon}e, but for the \oiiopt/\oiiiopt\ emission-line luminosity ratio.}
\label{fig:struction_oxygen}
\end{figure*}

The implications for the \ciii\ and \civ\ emission-line properties of models with different \fesc\ are shown in the bottom panels of Fig.~\ref{fig:struction_carbon}. Figs~\ref{fig:struction_carbon}c and \ref{fig:struction_carbon}d show the equivalent widths EW(\ciii) and EW(\civ) (in units of the equivalent widths in the ionization-bounded case), respectively, as a function of \fesc, for the same zero-age models with different \Uav\ and $Z$ as in Fig.~\ref{fig:fesc}c above. Fig.~\ref{fig:struction_carbon}e shows the \ciii/\civ\ line-luminosity ratio. As expected from Figs~\ref{fig:struction_carbon}a and \ref{fig:struction_carbon}b, the gradual removal of the outer low-ionization zone when \fesc\ rises makes EW(\ciii) decrease more rapidly than EW(\civ), and the \ciii/\civ\ ratio drop, the strengths of these effects increasing with both \Uav\ and $Z$. We note that, for low \Uav\ and $Z$, the rise in EW(\civ) when \fesc\ increases in Fig.~\ref{fig:struction_carbon}c is caused by the drop in recombination-continuum flux at nearly constant line luminosity, since the \civ\ zone (thin solid lines in Figs~\ref{fig:struction_carbon}a and \ref{fig:struction_carbon}b) is unaffected by the cuts in \Nh\  \citep[small coloured triangles; see also][]{raiter10,jaskot16}. In Fig.~\ref{fig:struction_oxygen}, we show the analogues of Figs~\ref{fig:struction_carbon}a and \ref{fig:struction_carbon}e for the \oiiopt\ and \oiiiopt\ lines. The fractional abundances of O$^{2+}$ and O$^+$ (Fig.~\ref{fig:struction_oxygen}a) exhibit a dependence on radius similar to that of C$^{2+}$ and C$^{3+}$ (Fig.~\ref{fig:struction_carbon}a), except that the outer low-ionization O$^+$ zone is thinner than the outer C$^{2+}$ zone at all ionization parameters. This causes the \oiiopt/\oiiiopt\ ratio (Fig.~\ref{fig:struction_oxygen}b) to drop more steeply than the \ciii/\civ\ ratio (Figs~\ref{fig:struction_carbon}e) when \fesc\ increases, until the O$^+$ zone disappears.

\begin{figure*}
\begin{center}
\resizebox{\hsize}{!}{\includegraphics{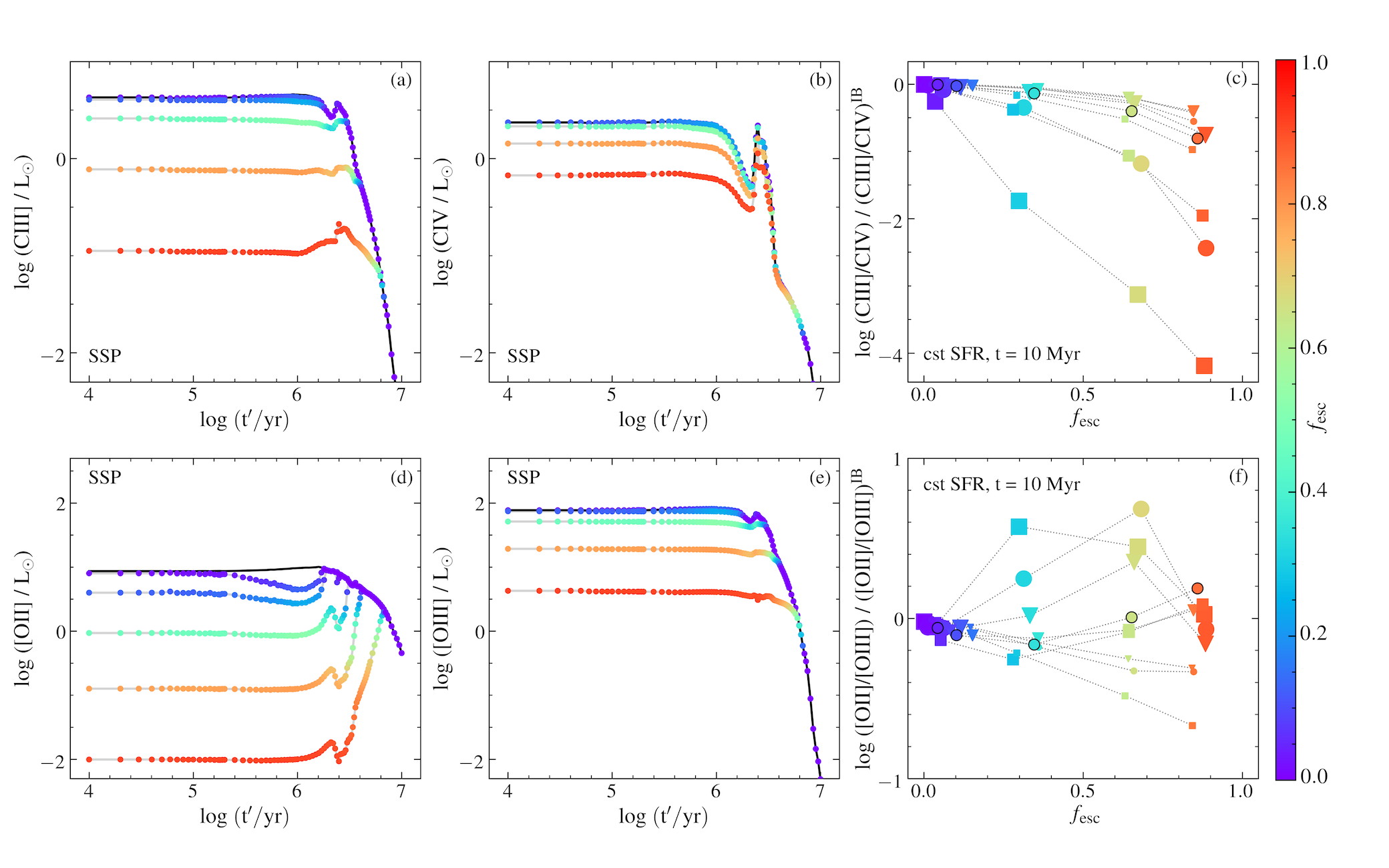}}
\end{center}
\caption{Emission-line properties of the density-bounded models of Section~\ref{sec:fesc}. (a) \ciii\ emission-line luminosity plotted against \hii-region age \tprime\ for the same models with $Z=0.002$, $\log\Uav=-2.0$, $\xid=0.3$, $\mup=300\,$\Msun\ and 5 choices of \tauref\ as in Fig.~\ref{fig:fesc}d. (b) Same as (a), but for \civ. (c) \ciii/\civ\ emission-line luminosity ratio for the same models with different $Z$ and $\log\Uav$ as in Fig.~\ref{fig:struction_carbon}e, but at age $t=10\,$Myr for a galaxy with constant star formation rate (i.e., adopting $\psi=\,$cst in equation~\ref{eq:flux_gal}). (d) Same as (a), but for \oiiopt. (e) Same as (a), but for \oiiiopt. (f) Same as (c), but for the \oiiopt/\oiiiopt\ ratio.}
\label{fig:convol}
\end{figure*}

So far, we have described the emission-line properties of density-bounded \hii-region models at age $\tprime=0$ only. Fig.~\ref{fig:convol} shows the evolution of the \ciii\ (Fig.~\ref{fig:convol}a), \civ\ (Fig.~\ref{fig:convol}b), \oiiopt\ (Fig.~\ref{fig:convol}d) and \oiiiopt\ (Fig.~\ref{fig:convol}e) emission-line luminosities as a function of \tprime\ for the same models with $Z=0.002$, $\log\Uav=-2.0$, $\xid=0.3$, $\mup=300\,$\Msun\ and 5 choices of \tauref\ as in Fig.~\ref{fig:fesc}d. As in the case of \Nh\ in Fig.~\ref{fig:fesc}d, the luminosity of emission lines in the reference ionization-bounded model (black curve in Figs~\ref{fig:convol}a--\ref{fig:convol}b and \ref{fig:convol}d--\ref{fig:convol}e) reaches a maximum at ages around 1\,Myr and exhibits a secondary peak when the hard ionizing radiation from hot WR stars kicks in, around 2.5\,Myr. Then, after the most massive stars have died, line emission fades. The secondary peak is most prominent in the evolution of the \civ\ luminosity, since \lciv\ requires the most energetic photons to be produced ($E_\mathrm{ion}>47.9\,$eV, compared to $35.1\,$eV for O\,\textsc{iii}). In contrast, the \oiiopt\ luminosity, which requires the least energetic photons ($E_\mathrm{ion}>13.6\,$eV to produce O\,\textsc{ii}, compared to $24.4\,$eV for C\,\textsc{iii}) does not drop as sharply as that of the other three lines at ages greater than a few Myr. For density-bounded models,  the gradual removal of the outer low-ionization envelope when \tauref\ decreases reduces the \oiiopt\ luminosity more strongly than the \ciii, \oiii\ and \civ\ ones at early ages, until the ionizing flux has dropped low enough for the nebula to become ionization-bounded. The low-ionization zone reappears, causing a sharp rise in \oiiopt\ luminosity at ages $\tprime\gtrsim3\,$Myr.

In Figs~\ref{fig:convol}c and \ref{fig:convol}f, we show the resulting dependence on \fesc\ of the \ciii/\civ\ and \oiiopt/\oiiiopt\ emission-line luminosity ratios, respectively, at age $t=10\,$Myr for a galaxy with constant star formation rate (i.e., adopting $\psi=\,$cst in equation~\ref{eq:flux_gal}), for the same models with different $Z$ and $\log\Uav$ as in Fig.~\ref{fig:struction_carbon}e. The results for \ciii/\civ\ are very similar to those described above for single zero-age \hii\ regions (Fig.~\ref{fig:struction_carbon}e), as expected from the similar effect of reducing \tauref\ on the evolution of the \ciii\ and \civ\ luminosities (Figs~\ref{fig:convol}a and \ref{fig:convol}b). Instead, the dependence of \oiiopt/\oiiiopt\ on \fesc\ in Fig.~\ref{fig:convol}f differs from that found for zero-age \hii\ regions in Fig.~\ref{fig:struction_oxygen}b (for $ \log\Uav\geq-2.0$). For $\tauref=1$, for example, corresponding roughly to $\fesc\approx0.3$ in the different models of Fig.~\ref{fig:convol}f (and $\fesc\approx0.5$ at $\tprime=0$ in the single \hii-region model of Fig.~\ref{fig:convol}d), the rise in \oiiopt\ luminosity (Fig.~\ref{fig:convol}d) and corresponding drop in \oiiiopt\ luminosity (Fig.~\ref{fig:convol}e) at ages $\tprime\gtrsim3\,$Myr in the evolution of single \hii\ regions can cause \oiiopt/\oiiiopt\ to exceed the ionization-bounded value at $t=10\,$Myr for a galaxy with constant star formation rate, especially for large \Uav\ (Fig.~\ref{fig:convol}f).

Hence, while a small \oiiopt/\oiiiopt\ ratio can be a clue of significant LyC leakage in models of individual density-bounded \hii\ regions, this is not the case for model galaxies containing several generations of \hii\ regions \citep[see also][]{jaskot13}. It is worth noting that the models presented here are highly idealized, and that, in practice, a galaxy will contain different types of \hii\ regions with different optical depths to LyC photons, metallicities and ionization parameters. In any case, the complex dependence of the \ciii/\civ\ and \oiiopt/\oiiiopt\ ratios on \fesc\ identified in Figs~\ref{fig:struction_carbon}--\ref{fig:convol} above illustrates the difficulty of observationally tracking galaxies which lose significant amounts of LyC photons. We will return to this point in Section~\ref{sec:constraints}.

\subsection{Nebular emission from AGN}\label{sec:agn}

To explore the influence of an AGN on the nebular emission from a young star-forming galaxy, we use photoionization calculations of AGN narrow-line regions based on the approach of \citet{Feltre2016}. This relies on a parametrization of \cloudy\ similar to that described above in terms of hydrogen density \nh, gas metallicity $Z$, C/O ratio, dust-to-metal mass ratio \xid\ and volume-averaged ionization parameter \Uav, but using the emission from an accretion disc in place of equation~\eqref{eq:flux_gal} as input radiation. The spectral energy distribution of the accretion disc is parametrized as
\begin{equation}
S_\nu \propto
\left \{
\begin{array}{ll}
\nu^{\alpha} & \mbox{at wavelengths} \quad  0.001\leq  \lambda/\micron \leq 0.25\,,\\
\nu^{-0.5} & \mbox{at wavelengths} \quad  0.25< \lambda/\micron \leq 10.0\,,\\ 
\nu^{2} & \mbox{at wavelengths} \quad  \lambda/\micron >10.0\,.\\
\end {array}
\right.
\label{eq:Lagn}
\end{equation}
We adopt here for simplicity a fixed slope $\alpha=-1.7$ at high energies \citep{Zheng97,Lusso15} and a fixed gas density $\nh=10^3\,$cm$^{-3}$ in the narrow-line region \citep[e.g.,][]{Osterbrock2006}. The \cloudy\ calculations are performed in `open geometry' \citep[see][for more details]{Feltre2016}.

The models adopted here differ from those originally published by \citet{Feltre2016} in that they are computed using version c17.00 of \cloudy\ \citep{cloudyc17} and include dissipative microturbulence in the gas clouds in the narrow-line region (with a microturbulence velocity of 100\,km\,s$^{-1}$) and a smaller inner radius of this region (90\,pc instead of 300\,pc, for an AGN luminosity of $10^{45}$\,erg\,s$^{-1}$). These parameters were found to better reproduce the observed ultraviolet emission-line spectra (in particular, the \nv\ emission) of a sample of 90 type-2 AGN at redshifts $z=1.5$--3.0 \citep{Mignoli19}.

\subsection{Nebular emission from radiative shocks}\label{sec:shocks}

We are also interested in the effects of a contribution by shock-ionized gas to the nebular emission from actively star-forming galaxies. We appeal to the 3MdBs database\footnote{See \url{http://3mdb.astro.unam.mx:3686}} of fully radiative shock models recently computed by \citet{3MdBs19} using the \mappings\ shock and photoionization code \citep{MappingsV2017}. The models (computed in plane-parallel geometry) are available for the same sets of element abundances as adopted in the stellar and AGN photoionization models described in Sections~\ref{sec:ionb}--\ref{sec:agn} above (albeit for only two values of the C/O ratio: 0.11 and 0.44). Metal depletion on to dust grains is not included in this case, as in fast shocks, dust can be efficiently destroyed by grain-grain collisions, through both shattering and spallation, and by thermal sputtering \citep{allen08}. The other main adjustable parameters defining the model grid are the shock velocity (from $10^2$ to $10^3$\,km\,s$^{-1}$), pre-shock density (from 1 to $10^4$\,cm$^{-2}$) and transverse magnetic field (from $10^{-4}$ to 10\,$\mu$G). The pre-shock density (\nh) and transverse magnetic field (noted $B$) have a much weaker influence than shock velocity on most emission lines of interest to us (see Section~\ref{sec:params}). Thus, in the following, to probe global trends in the influence of radiative shocks on the nebular emission from star-forming galaxies, we consider for simplicity models with fixed $\nh=10^2\,$cm$^{-3}$ and $B=1\,\mu$G in the full available range of shock velocities. We focus here on the predictions of models including nebular emission from both shocked and shock-precursor (i.e., pre-shock gas photoionized by the shock) gas (see \citealt{3MdBs19} for more details).


\afterpage{%
\begin{landscape}
\begin{table}
\begin{threeparttable}
\centering
\caption{Published ultraviolet (/optical) spectroscopic analyses of metal-poor star-forming galaxies at redshifts between 0.003 and 7.1.}
\begin{tabular}{lclllcl}
\toprule
Reference & $z$ & Sample &  \logohgas\ & Modelling & Physical parameters\tnote{\it a} & Comment \\
\midrule
\citet{senchyna17,senchyna19} &  $<0.05$ & 10 extreme SF regions and & 7.5--8.5 & \beagle\             & $-1.4<\log\CO<-0.7$\tnote{\it b}           & Transition from primarily stellar to purely nebular \heii\  \\
                              &                 & 6 extremely metal-poor SF             &              & \cloudy\ 13.03   & $-3.6<\log \Uav<-1.9$\tnote{\it b} & near $\logoh\lesssim 8.0$; no evidence for shocks nor XRBs;\tnote{\it b}  \\
                              &                 &  galaxies from SDSS         &              &  (dust physics)  &   sSFR$\,\sim\,$2--300\,Gyr$^{-1}$   & strong \civ\ traces young ages, extremely low $Z$ and $\alpha$/Fe  \\
                              
\rule{-2pt}{3ex}                            
\citet{berg16,berg19} &  $<0.14$ & 32 compact SF regions with  & 7.4--8.0 & \bpass\ v2.14\tnote{\it c}       & $-1.0<\log\CO<-0.3$   & No obvious AGN ($\civ/\ciii < 1$) nor shock \\
                       &                & EW$(\oiiiopt)>50\,$\AA\     &               &    \cloudy\ 17.00    & $-2.8<\log U_0<-1.8$  & ($\oiopt/\oiiiopt < 0.01$; see Fig.~\ref{fig:OI}) contribution \\
                       &                       &                                      &              & (no dust)              & sSFR$\,\sim\,$1--40\,Gyr$^{-1}$     & \\
                       
\rule{-2pt}{3ex}       
\citet{berg18}  & 1.8  & Lensed, extreme-SF galaxy & $\sim7.5$ & \bpass\ v2.14          & $\log\CO\sim-0.8$          & Strong nebular \heii\ not reproducible by models; no  \\
                       &        &  SL2SJ021737-051329      &              & \cloudy\ 17.00   & $\log U_0\sim-1.5$ &  obvious AGN nor shock contribution (\citealt{groves04}-AGN \\
                       &        &                                            &              &  (no dust)  &   sSFR$\,\sim\,$100\,Gyr$^{-1}$       & and \citealt{allen08}-shock prescriptions deteriorate fits) \\  

\rule{-2pt}{3ex}       
\citet{stark14} &  1.5--3.0 & 17 lensed, dwarf SF galaxies & 7.3--7.8 & \beagle-like            & $-1.0<\log\CO<-0.3$           & \ciii\ detected in 16 systems; strongest (EW>10\AA) \\
                              &                 &                                                        &              & \cloudy\ 13.03   & $-1.8<\log \Uav<-1.5$\tnote{\it d}  & emitters show \civ\ emission, while nebular \heii\ \\
                              &                 &                                                        &              &  (dust physics)  &   sSFR$\,\sim\,$2--40\,Gyr$^{-1}$       & is weak or non-detected in many systems \\

\rule{-2pt}{3ex}       
\citet{erb10}  & 2.3  & Low-mass SF galaxy & $\sim7.8$ & \starburst\tnote{\it e}          & $\log\CO\sim-0.6$          & Unlikely AGN contribution ($\civ/\ciii\sim0.3$)\\
                       &        &  Q2343-BX418      &              & \cloudy\ 08.00   & $\log U_0\sim-1.0$ &  \\
                       &        &                                            &              &  (dust physics)  &   sSFR$\,\sim\,$16\,Gyr$^{-1}$       & \\  
                       
\rule{-2pt}{3ex}   
\citet{amorin17} &  2.4--3.5 & 10 VUDS SF galaxies with & 7.4--7.7 & \popstar\tnote{\it f}             & $-1.0<\log\CO<-0.4$           & No obvious AGN contributions (from \civ, \heii,\\
                              &                 &  \lya, \oiii\  and  &              & \cloudy\ 13.03   & $-2.3<\log U_0<-1.7$ & \ciii, \oiii; no X-ray emission nor broad lines) \\
                              &                 & \ciii\ emission    &              &  (dust physics)  &   sSFR$\,\sim\,$5--50\,Gyr$^{-1}$   &   \\
                              
\rule{-2pt}{3ex}   
\citet{nakajima18} &  2--4 & 450 VUDS SF galaxies with & $\sim8.3$ (C) & \popstar\             & $\log\CO\sim-0.5$ (C),           & AGN contribution to ionizing radiation required in strongest \\
                              &                 & \ciii\ emission (C)\tnote{\it g}    &  $\sim7.6$ (B)        & \bpass\ v2.0   &$-0.3$ (B) and $-0.4$ (A) & \ciii\ emitters, irrespective of the inclusion of binary \\
                              &                 & 43 with EW$\,=\,$10--20\,\AA\ (B)\tnote{\it g} &   $\sim7.8$ (A)   &   \cloudy\ 13.03   &   $\log U_0\sim-2.9$ (C),      & stars in the stellar population modelling \\
                              &                 & 16 with EW$\,>\,$20\,\AA\  (A)\tnote{\it g}  &              &  (dust physics)  &   $-1.7$ (B) and $-1.6$ (A)    &   \\
                              
\rule{-2pt}{3ex}   
\citet{nanayakkara19} &  2--4 & 12 MUSE SF galaxies with & 7.9--8.6 & \bpass\ v2.1  & $\log\CO\sim-0.4$        & Rest-frame \heii\ equivalent widths of 0.2--10\,\AA\ not \\
                              &                 & \heii\ emission                  &              & \cloudy\ 13.03   & $-2.5<\log U_0<-1.5$ & reproducible by models with or without binary stars \\
                              &                 &                                       &              &   (no dust)  &     &   \\
                              
\rule{-2pt}{3ex}   
\citet{vanzella17} &  3.2 & Lensed double-super star & $\sim7.7$ &  &   sSFR$\,\sim\,$20\,Gyr$^{-1}$   & Emission-line spectrum from \civ\  through \\
                              &            & cluster ID14                     &              &   & & \ciii\ consistent with photoionization by stars  \\
                              
\rule{-2pt}{3ex}   
\citet{fosbury03} &  3.4 & Lensed \hii\ galaxy  & $\sim7.6$ & Pure blackbody  &     $\log U_0\sim-1.0$    & Absence of \nv\ and weakness  of \niii\ \\
                              &              & RX J0848+4456                    &              & \mappings\  Ic &   & taken as evidence against photoionization by an AGN \\
                              &                 &                                       &              &   (no dust)  &     &   \\

\rule{-2pt}{3ex}   
\citet{schmidt17} &  6.1 & Lensed Ly$\alpha$ galaxy  & $<8.3$ &   &    sSFR$\,\sim\,$40\,Gyr$^{-1}$     & Emission-line spectrum from \civ\  through\\
                              &              &                    &              &  &   &  \ciii\ consistent with photoionization by stars \\
                              &                 &                                       &              &    &     &   \\
                              
\rule{-2pt}{3ex}       
\citet{stark15} &  7.0 & Lensed Ly$\alpha$ galaxy & $\sim7.0$ & \beagle-like            & $\log \Uav\sim1.0$      & \civ\ emission and upper limits on \heii\  \\
                              &                 &                                                        &              & \cloudy\ 13.03   &   & and \oiii\ emission consistent with photoionization \\
                              &                 &                                                        &              &  (dust physics)  &          & by both an AGN and stars \\

\rule{-2pt}{3ex}       
\citet{laporte17} &  7.1 & \oiiiopt-strong SF  &  &         &   & Prominent \nv\ (and \heii) emission supports  \\
                              &                 &  galaxy COSY                                                    &              &   &   & photoionization by an AGN \\
\bottomrule
\end{tabular}
\label{tab:analogs}
\begin{tablenotes}
\item [{\it a}] $U_0$ refers to the ionization parameter at the inner edge of the gas cloud, which, in models with spherical geometry, corresponds to the inner radius of the \hii\ region and is roughly 3 times larger than the volume-averaged ionization parameter \Uav\ described in Section~\ref{sec:ionb}.
$^{\it b}$ Pertains to the 10 extreme SF regions studied by \citet{senchyna17}.
$^{\it c}$ \citet{berg16} used \starburst.
$^{\it d}$ Range spanned by the 4 most extreme \ciii-emitting galaxies in the sample.
$^{\it e}$ \citep{popstar}.
$^{\it f}$ \citep{starburst99,Leitherer14}.
$^{\it g}$ Letter referring to te corresponding sample in \citet{nakajima18}.
\end{tablenotes}
\end{threeparttable}
\end{table}
\end{landscape}
}
%
\section{Observational constraints}\label{sec:obs}

In this section, we build a reference sample of the nebular emission from metal-poor star-forming galaxies and LyC leakers at various redshifts, including also other star-forming galaxies and AGN, which we will use in Sections~\ref{sec:params} and \ref{sec:constraints} to explore potentially discriminating signatures of the different adjustable parameters of the versatile models presented in Section~\ref{sec:models}. To this end, we wish to assemble a large homogeneous sample of observations of metal-poor star-forming galaxies at ultraviolet and optical wavelengths, by gathering from the literature data often analysed in independent ways using different models and assumptions. In the following, we assemble observations of such galaxies (Section~\ref{obs:analogs}), as well as of confirmed and candidate LyC leakers (Section~\ref{obs:leakers}) and other star-forming galaxies and AGN (Section~\ref{obs:normal}) in a wide redshift range. We consider here only observational studies which gathered enough nebular emission-line properties to be plotted in at least one of the diagrams we investigate. Observations involving ultraviolet lines are presented in Fig.\ref{fig:obs_uv}, and those involving optical lines in Fig.~\ref{fig:obs_opt}. We comment on the general properties of this reference sample in Section~\ref{sec:obsprop}.

\subsection{Metal-poor star-forming galaxies}\label{obs:analogs}

We list in Table~\ref{tab:analogs} the main characteristics of 13 samples of low-metallicity, actively star-forming galaxies. The samples are arranged in order of increasing redshift. In each case, we indicate the nature of the sample; the published gas-phase oxygen abundances of galaxies; the modelling tools used to interpret the observations (ionizing stellar population spectra and photoionization model); the constraints derived on physical parameters such as \CO\ ratio, ionization parameter and specific star formation rate; and the main conclusions drawn in the original studies. The gas-phase oxygen abundance, \logohgas, is usually estimated using the direct-$T_\mathrm{e}$ method (see section~5.1 of \citealt{gutkin2016} for potential caveats of this method), and otherwise through photoionization calculations, including or not (in which case the total gas+dust-phase O abundance is not computed) depletion of oxygen onto dust grains. Also, we note that the differences in model analyses between the different studies in Table~1 go beyond the listed details. For example, the ionizing stellar population spectra can refer to different IMF shapes and upper-mass limits, star formation histories and ages. We do not focus on such differences here, as our main goal is to provide rough estimates of the characteristics of the various sample, which we will compare globally with a homogeneous set of models in Section~\ref{sec:params}. We now briefly describe these samples.

In the nearby Universe ($z\lesssim0.1$), \HST/COS observations have brought valuable insight into the rest-ultraviolet properties of metal-poor star-forming galaxies with hard ionizing spectra. \citet{senchyna17} observed 10 galaxies from the sample of \heiiopt-emitting, Sloan Digital Sky Survey (SDSS) star-forming galaxies of \citet{shirazi12}, \citet{senchyna19} 6 extremely metal-poor ($Z/\zsun\lesssim 0.1$) galaxies from the SDSS sample of \citet{Morales11} and \citet{berg16,berg19} 32 compact, ultraviolet-bright, SDSS star-forming galaxies with \oiiiopt\ emission equivalent widths larger than 50\,\AA. Objects in these samples show no sign of AGN activity and range from high-ionization \hii\ regions embedded in larger galaxies to blue compact dwarf galaxies. They can reach \ciii\ equivalent widths as large as $\sim15\,$\AA, similar to those found in galaxies at redshifts $z>6$ (see below). \citet{senchyna17} identify a marked transition with decreasing metallicity around $\logoh\approx8.0$, from stellar-wind dominated to nebular-dominated \heii\ and \civ\ emission. Analysis with the \beagle\ code \citep{Chevallard2016}, allows them to reproduce all the stellar (e.g., \civ\ P-Cygni and broad-\heii\ wind features) and nebular ultraviolet/optical emission-line properties of their sample \citep[see also][]{Chevallard18}, except for the strong nebular \heii\ emission in the most metal-poor systems, which does not seem to be reproducible by any current stellar population synthesis prescription. Like \citet{berg16,berg19}, they do not find any strong evidence for a contribution by radiative shocks to the nebular emission of galaxies in their sample.

At intermediate redshifts ($2\lesssim z\lesssim4$), spectroscopic observations with large optical telescopes have allowed detailed studies of the rest-ultraviolet spectra of unusually bright or lensed, dwarf star-forming galaxies (Table~\ref{tab:analogs}). The galaxies in these sample exhibit spectral characteristics typical of high-ionization, metal-poor star-forming galaxies. Remarkably, they also often show strong nebular \heii\ emission, which cannot be reproduced by any current stellar population synthesis models, even when including enhanced production of hard ionizing radiation via binary mass transfer \citep{berg18,nakajima18,nanayakkara19}. Significant contribution from a luminous AGN is disfavoured in most cases, based on the weakness of \nv, the low observed \civ/\ciii\ ratio and the lack of X-ray detection and broad emission lines, standard optical \citep[e.g.][hereafter BPT]{BPT} diagnostic diagrams being generally not available in this redshift range \citep{stark14,erb10,amorin17,vanzella17,berg18,nanayakkara19}. This is not the case for \citet[][we adopt here the 3 composite spectra representative of classes A, B and C from this sample, with additional data from \citealt{lefevre19} for classes A and B; see Table~\ref{tab:analogs}]{nakajima18}, who argue that an AGN contribution is required to account for the \heii, \ciii\ and \civ\ properties of the galaxies with $\mathrm{EW(\ciii)}>20\,$\AA\ in their sample. 

At the highest redshifts ($z>6$), we report in Table~\ref{tab:analogs} the constraints from near-infrared spectroscopy on the rest-ultraviolet emission of three galaxies probing the reionization era. In one of these, the emission-line spectrum from \civ\ through \ciii\ favours photoionization by stars rather than by an AGN \citep{schmidt17}. In another, photoionization could arise from an AGN or hot stars \citep{stark15}. In the latter, prominent \nv\ emission and the low \ciii/\heii\ ratio both support photoionization by an AGN \citep{laporte17}. Remarkably, all three galaxies show strong Ly$\alpha$ emission, suggesting intense radiation fields capable of creating early ionized bubbles in the surrounding hydrogen distribution.

Overall, the published spectral analyses of the observations listed in Table~\ref{tab:analogs} consistently point toward galaxies with low metallicities, typically $7.5\lesssim\logoh\lesssim8.0$, low C/O ratios, $-1.0\lesssim\log\CO\lesssim-0.3$, high ionization parameters, $-3.0\lesssim\log \Uav\lesssim-1.5$ and large specific star formation rates, from $\sim10$ to a few $\times100$\,Gyr$^{-1}$, across a wide redshift range. The properties of these galaxies approach those expected for primeval galaxies near the reionization epoch \citep[e.g.,][]{stark16}.


%
\afterpage{%
\begin{landscape}
\begin{table}
\begin{threeparttable}
\centering
\caption{Published ultraviolet (/optical) spectroscopic analyses of confirmed and candidate LyC leakers at redshifts between 0.02 and 3.2.}
\begin{tabular}{lclllcl}
\toprule
Reference & $z$ & Sample &  \logohgas\ & Modelling & Physical parameters\tnote{\it a} & Comment \\
\midrule
\citet{leitet11} &  $0.02$ & {\it FUSE} rest-900\,\AA\ detection of & $\sim7.9$ &          & $\fesc\sim3\%$           & Consistent with 2-dimensional data, count rates, and limits on \\
                              &                 & blue compact galaxy Haro~11            &              &   & & residual flux in \ciia\ interstellar absorption line \\
                              
\rule{-2pt}{3ex}                            
\citet{izotov17oct} &  $<0.14$ & 5 SDSS compact SF galaxies   & 7.5--7.8 & \starburst               & sSFR$\,\sim\,$50--400\,Gyr$^{-1}$   &  $\oratio$ alone not good indicator of LyC leakage because depends \\
                       &                &  with $\oratio>20$ and no AGN  &               &    \cloudy\ 13.04    & $\Nh<\NhIB$ in 3 galaxies\tnote{\it b} &  on details of ionizing spectrum;  \heiopta, \heioptb\ and  \\
                       &                       & spectral feature\tnote{\it c}             &              & (dust physics)              &  (potentially $\fesc>20$\%)    & \heioptc\ more promising tracers of density-bounded regions\\
                       
\rule{-2pt}{3ex}       
\citet{chisholm17}  & 0.04--0.2  & {\it HST}/COS rest-900\,\AA\  archival & 8.1--8.7 &            & $\fesc\sim0.4$--1.9\%        & Weak \siliia\ and \siliiia\ absorption consistent with  \\
                       &        & data of 3 confirmed leakers &              &          &            & density-bounded regions, although gas covering factor may vary \\

\rule{-2pt}{3ex}       
\citet{jaskot13} &  0.1--0.2 & 6  SDSS Green-Pea galaxies & 7.9--8.0 & \starburst             & sSFR$\,\sim\,$70--200\,Gyr$^{-1}$           & Large \heii/\hb\ ratio not reproducible by standard stellar \\
                              &                 & with $\oratio>7$ and no AGN    &              & \cloudy\ 10.00   & burst age$\,\sim\,$3--5\,Myr & population models; if arising from a contribution by shocks, the   \\
                              &                 & spectral feature\tnote{\it c}        &              &  (dust physics)  &        & associated large $\oratio$ may indicate $\Nh<\NhIB$ \tnote{\it b} \\

\rule{-2pt}{3ex}       
\citet{izotov16oct,izotov16jan}  & 0.3--0.4  & {\it HST}/COS rest-900\,\AA\ detection & 7.8--8.0 &   \starburst   &  sSFR$\,\sim\,$10--200\,Gyr$^{-1}$         & Compact SF galaxies with large \oratio\  appear to pick up \\
                        &                      & of 5 compact galaxies with     &              &   &  burst age$\,<\,$4\,Myr  & efficiently LyC leakers \\
                        &                      & $\oratio>5$ and no AGN feature\tnote{\it c}      &              & & $\fesc\sim6$--13\%       & \\  
                       
\rule{-2pt}{3ex}   
\citet{izotov18mar,izotov18aug} & 0.3--0.4 &  {\it HST}/COS rest-900\,\AA\ detection  & 7.6--8.2 &  \starburst  & sSFR$\,\sim\,$3--1,000\,Gyr$^{-1}$ & General increase of \fesc\ with increasing \oratio\ (and decreasing \\
&                 & of 6 compact galaxies with   &              &          &   burst age$\,\sim\,$2--3\,Myr      & velocity spread of Ly$\alpha$ double-peaked emission) \\
                              &                 & $\oratio>8$ and no AGN feature\tnote{\it c}      &              &        &  $\fesc\sim2$--70\%  &   \\
                              
\rule{-2pt}{3ex}   
\citet{nanayakkara19} &   2.2 & MUSE SF galaxy 1273, also & $\sim8.3$ & \fast\tnote{\it d}   & sSFR$\,\sim\,$1.5\,Gyr$^{-1}$        & $\mathrm{EW(\oiiiopttot)\sim1200}$\,\AA \\
                              &                 & candidate LyC leaker from &              &  &  (potentially $\fesc\sim60$\%)  & \\
                              &                 &\citet[][GS 30668]{Naidu17}        &              &     &    &  \\
                              
\rule{-2pt}{3ex}   
\citet{nakajima16} &  3.0--3.7 & 13 candidate LyC-leaker Ly$\alpha$  & $\sim8.1$ & \starburst             & sSFR$\,\sim\,$1--50\,Gyr$^{-1}$  & Ly$\alpha$ emitters have larger \oratio\ than Lyman-break galaxies \\
                              &                 & emitters (including 1 AGN)  &          &\cloudy\ 13.02   & (Ly$\alpha$ emitters have & at similar metallicity, suggesting that the ionized regions \\
                              &                 & and 2 Lyman-break galaxies &      &   (dust physics)   &  potentially $\fesc>0$)  &  are, at least in part, density bounded \\
                              
\rule{-2pt}{3ex}   
\citet{vanzella16} &  3.1 & Lensed compact Ly$\alpha$ emitter & <7.8 &  & $\fesc>0$ expected      & $\oratio>10$, small velocity spread of Ly$\alpha$ double-peaked emission \\
                              &                 & ID11                 &              &   &  &  and \civ\ emission suggest low \Nh\ and LyC leakage \\
                              &                 &                                       &              &    &     &   \\
                              
\rule{-2pt}{3ex}   
\citet{debarros16} &  3.2 & VLT/VIMOS rest-900\,\AA\  & $\sim8.1$ & \popstar &   sSFR$\,\sim\,$10\,Gyr$^{-1}$ & LyC leakage, Ly$\alpha$ emission, $\oratio>10$ and weak \ciib\ and  \\
                              &              &  detection of candidate LyC      &              & \cloudy\ 13.03    & $\log\CO\sim-0.8$ & \siliia\ absorption suggest density-bounded \hii\ region;  \\
                              &                 &  leaker Ion2  &      &   (dust physics)   &  $\fesc\sim64$\% &  a faint AGN not excluded \\

\bottomrule
\end{tabular}
\label{tab:leakers}
\begin{tablenotes}
\item [{\it a}] sSFR typically quoted under the assumption $\fesc=0$.
$^{\it b}$ $\NhIB$ is the H-column density required to produce an ionization-bounded nebula.
$^{\it c}$ $\oratio=\oiiiopt/\oiiopt$.
$^{\it d}$ \citep{Kriek09}.
\end{tablenotes}
\end{threeparttable}
\end{table}
\end{landscape}
}
%


\subsection{LyC leakers}\label{obs:leakers}

We are also interested in existing ultraviolet and optical observations of confirmed or potential LyC-leaking galaxies at any redshift. In Table~\ref{tab:leakers}, we list the main characteristics of 10 such samples, arranged as before in order of increasing redshift. These include 5 samples of confirmed LyC leakers, in which the rest-ultraviolet emission around 880--912\,\AA\ (i.e. just blueward of the Lyman limit) has been directly observed, typically with {\it FUSE} and \HST\ at low redshift and by means of deep optical spectroscopy at higher redshift \citep{leitet11,debarros16,izotov16oct,izotov16jan,chisholm17,izotov18mar,izotov18aug}. The reported fractions of escaping LyC photons span a wide range, $1\lesssim\fesc\lesssim70$ per cent. As mentioned in Section~\ref{sec:fesc} above, this leakage is generally thought to occur either through holes carved into the neutral ISM by extreme galactic outflows or through density-bounded (i.e. optically thin to LyC photons) \hii\ regions, which are expected to lead to weak low-ionization emission lines, a small velocity spread of the Ly$\alpha$ double-peaked emission and a large \oiiiopt/\oiiopt\ ratio. 

The observations reported in Table~\ref{tab:leakers} seem to support the occurrence of enhanced \oiiiopt/\oiiopt\ ratios in confirmed LyC leakers (along with, in some cases, weak low-ionization emission lines and a small velocity spread of the \lya\ double-peaked profile), which favours density-bounded \hii\ regions as the main leakage mechanism. In Table~\ref{tab:leakers}, we also list 5 samples of candidate LyC leakers, selected on the basis of large observed \oiiiopt/\oiiopt\ ratios, the presence of \lya\ emission or deep ultraviolet imaging \citep{jaskot13, nakajima16, vanzella16, izotov17oct,nanayakkara19}. Among these studies, \citet{izotov17oct} caution that the \oiiiopt/\oiiopt\  ratio alone is not a certain indicator of LyC leakage, as it depends also on other parameters, such as the ionization parameter, the hardness of ionizing radiation and metallicity. These authors propose an alternative spectral diagnostic of density-bounded \hii\ regions, based on the \heiopta, \heioptb\ and \heioptc\ lines. 

A comparison between Tables~\ref{tab:analogs} and \ref{tab:leakers} reveals that LyC leakers have, typically, gas-phase oxygen abundances similar to those of (presumably ionization-bounded) metal-poor star-forming, but specific star formation rates several times larger. The implied extreme radiation fields of these intensively star-forming galaxies likely contribute to the escape of ionizing photons, as suggested by the apparent trend of increasing \fesc\ with increasing strength of stellar-wind features in the sample of \citet{izotov18aug}.

\subsection{Other star-forming galaxies and AGN}\label{obs:normal}

To complement our reference observational sample, we also include constraints on the ultraviolet and optical nebular emission of star-forming galaxies and AGN at various redshifts from more heterogeneous surveys. In the local Universe, we appeal to the samples of 21 low-metallicity starburst galaxies from \citet{giavalisco96}, 20 Wolf-Rayet galaxies from \citet[][see also \citealt{lopezsanchez10}]{lopezsanchez08} and 28 star-forming galaxies from \citet[][including \hii, Seyfert-2 and LINER galaxies]{leitherer11}. These samples span wide ranges of metallicities, $7.2\lesssim\logoh\lesssim9.2$. We also gather different samples of more distant galaxies: 9 lensed star-forming galaxies at redshifts $z=1.0$--3.5 with optical (3 with ultraviolet) emission-line measurements from \citet[][with complementary data from \citealt{patricio16}]{christensen12} and metallicities $7.6\lesssim\logoh\lesssim8.9$; a composite spectrum of 30 star-forming galaxies with median redshift $z\approx2.4$ and metallicity $\logoh\approx8.14$ from \citet{steidel16}; 20 Lyman-break galaxies at redshifts $z=3.0$--3.8 from \citet[][we retain 15 galaxies with H$\beta$ signal-to-noise ratio greater than 2]{schenker13}; and 24 Lyman-break galaxies at redshifts $z=3.2$--3.7 from \citet[][we retain 19 galaxies with H$\beta$ signal-to-noise ratio greater than 2]{holden16}. We note that the above samples span wide ranges of metallicities, stellar masses and specific star formation rates, which can overlap in part with those of the samples in Table~\ref{tab:analogs}. Finally, we include constraints on the ultraviolet and optical nebular emission of a sample of 12 nearby ($z\lesssim0.04$) Seyfert-2 galaxies and 59 radio galaxies and 10 type-2  (i.e. obscured) quasars at redshifts $1\lesssim z\lesssim4$ from \citet[][with complementary data from  \citealt{diaz88} and  \citealt{kraemer94}]{dors14}, sampling metallicities in the range from roughly 0.1 to 1.0 times solar. We also report the emission-line properties measured by \citet{Mignoli19} in a composite spectrum of 92 type-2 AGN in massive galaxies at $1.45<z<3.05$ from the zCOSMOS-deep survey \citep{Lilly07}.


\begin{figure*}
\begin{center}
\resizebox{0.95\hsize}{!}{\includegraphics{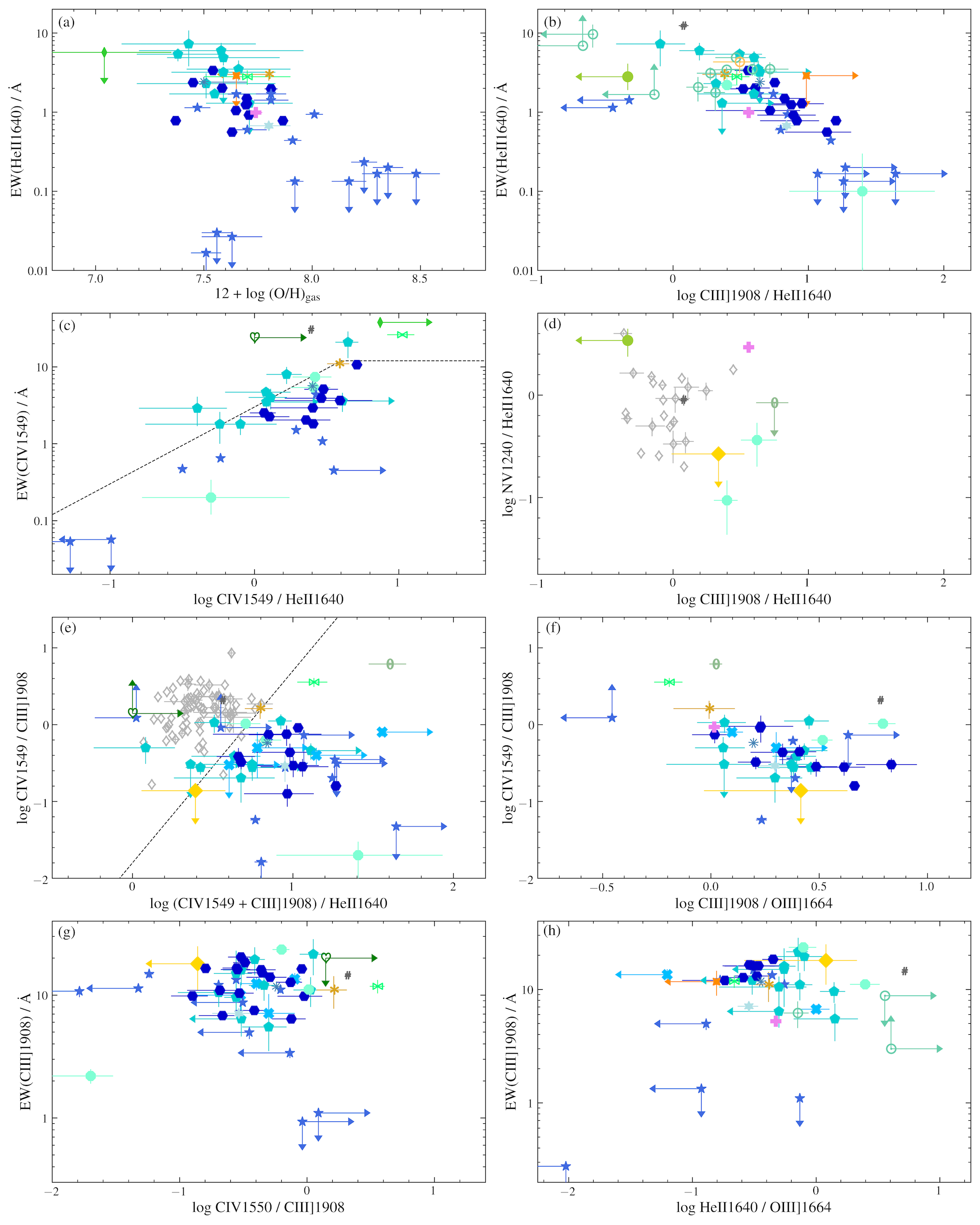}}
\end{center}
\caption{Ultraviolet emission-line properties of the reference observational sample of star-forming galaxies and AGN described in Section~\ref{sec:obs}. Different symbols refer to different samples, as indicated at the bottom of Fig.~\ref{fig:obs_opt}, with blue-like colours corresponding to metal-poor star-forming galaxies, orange-like colours to LyC leakers, purple-like colours to other star-forming galaxies and grey to AGN. The diagrams show different combinations of equivalent widths and ratios of the \nv,  \civ, \heii, \oiii\ and \ciii\ nebular emission lines [and the gas-phase oxygen abundance in (a)]. In (c) and (e), the dashed lines show the criteria proposed by \citet{nakajima18} to separate AGN-dominated from star-forming galaxies. All line fluxes are corrected for attenuation by dust, as prescribed in the original studies. Arrows show 1$\sigma$ upper limits. See description in Section~\ref{sec:obsprop}.}
\label{fig:obs_uv}
\end{figure*}

\begin{figure*}
\begin{center}
\resizebox{0.95\hsize}{!}{\includegraphics{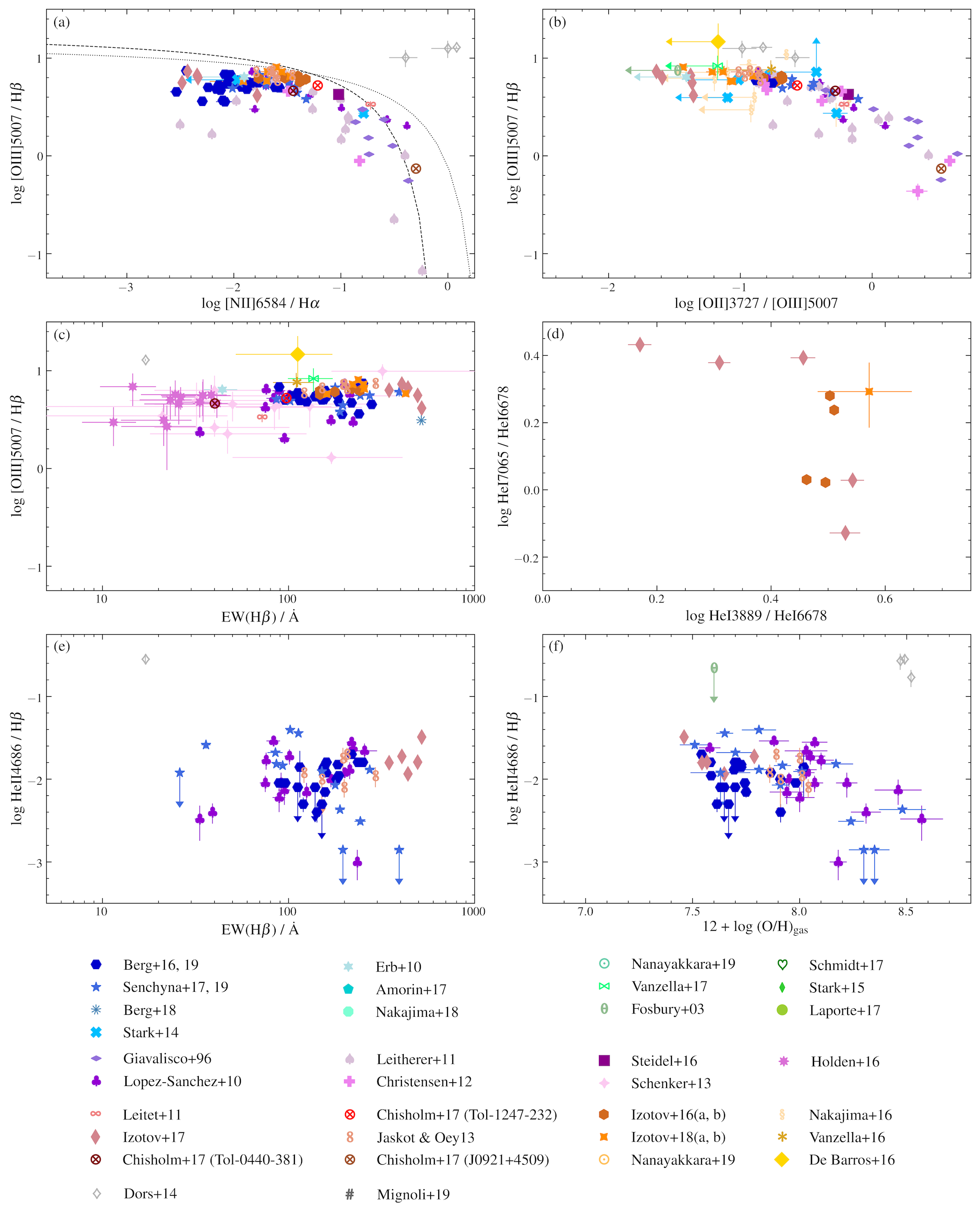}}
\end{center}
\caption{Optical emission-line properties of the reference observational sample of star-forming galaxies and AGN described in Section~\ref{sec:obs}. Different symbols refer to different samples, as indicated at the bottom of the figure, with blue-like colours corresponding to metal-poor star-forming galaxies, orange-like colours to LyC leakers, purple-like colours to other star-forming galaxies and grey to AGN. The diagrams show different ratios of the \oiiopt, \heiopta, \heiiopt, \hb, \oiiiopt, \ha, \niiopt, \heioptb\ and \heioptc\ nebular emission lines (and the \hb\ equivalent width). In (a), the dotted and dashed lines show the criteria of \citet{Kewley01} and \citet{Kauffmann03}, respectively, to separate AGN-dominated from star-forming galaxies. All line fluxes are corrected for attenuation by dust, as prescribed in the original studies. See description in Section~\ref{sec:obsprop}.}
\label{fig:obs_opt}
\end{figure*}


\subsection{Global observational properties of the full sample}\label{sec:obsprop}

We plot in Fig.~\ref{fig:obs_uv} different ultraviolet properties of the full reference sample of metal-poor star-forming galaxies (blue-like colours; see coding at the bottom of Fig.~\ref{fig:obs_opt}), LyC leakers (orange-like colours) and other star-forming galaxies (purple-like colours) and AGN (grey). After exclusion of a few galaxies with incomplete data, the final sample includes 68 metal-poor star-forming galaxies, 16 confirmed and 23 candidate LyC leakers, 75 other star-forming galaxies and 73 AGN. The diagrams in Fig.~\ref{fig:obs_uv} include several combinations of equivalent widths and ratios of the \nv\ (hereafter simply \lnv), \civd\ (hereafter simply \lciv), \heii, \oiiid\ (hereafter simply \loiii) and \ciiid\ (hereafter simply \lciii) nebular emission lines (and the gas-phase oxygen abundance in Fig.~\ref{fig:obs_uv}a). All but one diagram (Fig.~\ref{fig:obs_uv}g) involve the \heii\ line, as a main goal of our analysis in the next section will be to assess the specific influence of a comprehensive set of model parameters on predictions of this relative to other line intensities. In Fig.~\ref{fig:obs_opt}, we show the corresponding optical properties of this reference sample, through different ratios involving the \oiid\ (hereafter simply \loiiopt), \heiopta, \heiiopt, \hb, \oiiiopt\ (hereafter simply \loiiiopt), \ha, \niiopt\ (hereafter simply \lniiopt), \heioptb\ and \heioptc\ nebular emission lines (and the \hb\ equivalent width). All line fluxes in Figs~\ref{fig:obs_uv} and \ref{fig:obs_opt} are corrected for attenuation by dust, as prescribed in the original studies. We note that the line ratios in these figures are subject to uncertainties linked to the different apertures used to observe different galaxy samples, as high- and low-ionization lines are not necessarily co-spatial \citep[see, e.g.,][]{kehrig18}.

The fact that different symbols can populate different diagrams in Figs~\ref{fig:obs_uv} and \ref{fig:obs_opt} illustrate how the many spectral properties we consider are not always available homogeneously for all galaxy samples. This highlights, by itself, the value of the reference sample we have assembled, which allows one to grasp at once the broad ultraviolet and optical properties of the closest known analogues to primeval galaxies, LyC leakers and other star-forming galaxies and AGN. In Fig.~\ref{fig:obs_uv}, over three quarters of all galaxies with ultraviolet data have \lciii\ measurements, sometimes by requirement \citep{amorin17,nakajima18}. The line is unavailable for the highest-redshift galaxies, because in part of the limitations affecting ground-based infrared spectroscopy \citep{schmidt17,stark15,laporte17}. Many galaxies with \lciii\  measurements also have \loiii\ ones. Over half of all galaxies have \heii\ and/or \lciv\ measurements, the availability of one line relative to the other being independent of redshift. Finally, in Fig.~\ref{fig:obs_uv}d,  \lnv\ is available for only a few galaxies (a young star-forming galaxy with extended \lya\ halo from the \citealt{christensen12} sample, observed by \citealt{patricio16}, which also appears in Figs~\ref{fig:obs_uv}a,b,d and f; the \citealt{laporte17} galaxy; and samples A and B of \citealt{nakajima18}). 

The dashed lines in Figs~\ref{fig:obs_uv}c and \ref{fig:obs_uv}e show the criteria proposed by \citet{nakajima18} to separate AGN-dominated from star-forming galaxies, based on the \heii, \lciii\ and \lciv\ emission lines. Aside from the \citet{dors14} AGN sample, only a few galaxies with strong \lciv\ emission lie above the \citet{nakajima18} AGN criterion: a few galaxies from the \citet{amorin17} sample (although within 2$\sigma$ of the criterion); the \citet{stark15} lensed \lya\ galaxy, which could be powered by an AGN (Table~\ref{tab:analogs}); the \citet{schmidt17} lensed \lya\ galaxy; in one diagram only (Fig.~\ref{fig:obs_uv}e), the most extreme star-forming galaxy SB~111 in the \citet{senchyna17} sample (with an upper limit on \lciii\ emission, accounted for in the horizontal error bar); and, again in one diagram only (Fig.~\ref{fig:obs_uv}c), the \citet{vanzella17} lensed double-super star cluster. We note that \citet{schmidt17}, \citet{senchyna17} and \citet{vanzella17} find these last three galaxies to be more likely powered by star formation than by an AGN, based on the \citet{gutkin2016} and \citet{Feltre2016} photoionization models.

A particularly notable feature of Fig.~\ref{fig:obs_uv} is the trend of increasing \heii\ equivalent width with decreasing gas-phase oxygen abundance (Fig.~\ref{fig:obs_uv}a). High \heii\ equivalent widths ($\gtrsim1\,$\AA) correspond typically to galaxies with low \lciii/\heii\ ratios ($\lesssim10$; Fig.~\ref{fig:obs_uv}b) and potentially \lnv\ emission (according to Fig.~\ref{fig:obs_uv}d). At low EW(\heii), we find more metal-rich galaxies, with generally lower-ionization gas (e.g., larger \lciii/\heii\ and lower \lciv/\lciii\ ratios). This is the case for the few `normal' star-forming galaxies of our sample with enough ultraviolet data to appear in at least some diagrams of Fig.~\ref{fig:obs_uv} (e.g.; sample C of \citealt{nakajima18}; one of the  \citealt{christensen12} galaxies, appearing only in Fig.~\ref{fig:obs_uv}h). 

Remarkably, galaxies in wide ranges of redshift populate similar regions of the diagrams in Fig.~\ref{fig:obs_uv}, which confirms that modelling low-redshift, metal-poor star-forming galaxies is a useful step toward understanding the physical properties of reionization-era galaxies. A potential shortcoming of such studies is the occurrence at redshifts $z\gtrsim6$ of \lciv\ equivalent widths well in excess of those found in the nearby Universe (Fig.~\ref{fig:obs_uv}c), which could be attributable to enhanced $\alpha$/Fe abundance ratios in high-redshift galaxies \citep{senchyna19}. As expected from the comparison between Tables~\ref{tab:analogs} and \ref{tab:leakers} (Section~\ref{obs:leakers}), the handful of LyC leakers with available ultraviolet data in our sample tend to overlap with the most extreme star-forming galaxies in Fig.~\ref{fig:obs_uv} (but notice the low \lciv/\lciii\ ratio of the \citealt{debarros16} galaxy, in which the \heii\ emission could also be dominated by winds of WR stars; see \citealt{Vanzella19}).

This is even more apparent in Fig.~\ref{fig:obs_opt}, as more optical than ultraviolet data are available for LyC leakers and quiescent star-forming galaxies in our sample. In Figs~\ref{fig:obs_opt}a and \ref{fig:obs_opt}b, for example, all but a few LyC leakers are concentrated -- on top of the general galaxy population -- in the high-ionization parts (low \lniiopt/\ha\ and \loiiopt/\loiiiopt) of the standard BPT diagrams defined by the \loiiiopt/\hb, \lniiopt/\ha\ and $\loiiopt/\loiiiopt$ ratios. In Figs~\ref{fig:obs_opt}c, \ref{fig:obs_opt}e and  \ref{fig:obs_opt}f, they are concentrated in the regions of highest \hb\ equivalent width, highest \heiiopt/\hb\ ratio and lowest gas-phase oxygen abundance. The exceptions are a few weak ($\fesc\lesssim\;$a few per cent) nearby leakers from \citet{leitet11} and \citet[][Fig.~\ref{fig:obs_opt}c]{chisholm17}. The dotted and dashed lines in Fig.~\ref{fig:obs_opt}a show the criteria of \citet{Kewley01} and \citet{Kauffmann03}, respectively, to separate AGN-dominated from star-forming galaxies. Only the \citet{dors14} AGN lie above these lines. Finally, Fig.~\ref{fig:obs_opt}d shows the diagram advertised by \citet{izotov17oct} to diagnose density-bounded \hii\ regions, based on the \heiopta, \heioptb\ and \heioptc\ lines (Section~\ref{obs:leakers}). 

\section{Emission-line signatures of galaxy physical parameters}\label{sec:params}

In this section, we use the models introduced in Section~\ref{sec:models} to investigate the ultraviolet and optical emission-line signatures of a wide range of physical parameters of metal-poor star-forming galaxies in the reference observational diagrams assembled in Section~\ref{sec:obs}.  We consider physical parameters pertaining to the interstellar gas, stellar populations, LyC-photon leakage, AGN narrow-line regions and radiative shocks. As described in Section~\ref{sec:models}, a main attribute of our approach is the adoption of a common parametrization of nebular-gas abundances in all calculations, allowing direct comparisons between models powered by different sources.

To explore the observable signatures of any specific physical parameters in the emission-line diagrams of Figs~\ref{fig:obs_uv} and \ref{fig:obs_opt}, it is convenient to examine the offsets implied by changes in this parameter with respect to a `standard' model with stellar and interstellar properties typical of those expected for young, metal-poor star-forming galaxies. Referring to Table~\ref{tab:analogs}, we take this model to correspond to an ionization-bounded galaxy with constant star formation rate [$\psi(t)=\mathrm{constant}$] and the following parameters (see Section~\ref{sec:ionb}):
\begin{itemize}
\item[] $\fesc=0$\;;
\item[] $\nh=10^2\,{\rm cm}^{-3}$\;;
\item[] $Z=0.002$\;;
\item[] $\CO=0.38\COsol\approx0.17$\;;
\item[] $\xid=0.3$\;;
\item[] $\log\Uav=-2$\;;
\item[] $\mup=300\,\Msun$\;;
\item[] $t=3\,$Myr.
\end{itemize}
The gas-phase oxygen abundance corresponding to these choices of $Z$ and \xid\ is $\logohgas\approx7.83$ \citep[see table~1 of][]{gutkin2016}. In this standard model, we do not include interstellar-line absorption in the \hii\ interiors and \hi\ envelopes of stellar birth clouds, nor any contribution by an AGN or radiative shocks to nebular emission. We compute the emission-line properties of the model using the \CB\ stellar population synthesis code.

Figs~\ref{fig:baton3z_uv} and \ref{fig:baton3z_opt} show the ultraviolet and optical emission-line properties of the standard model (black circle) in the same diagrams as in Figs~\ref{fig:obs_uv} and \ref{fig:obs_opt}, where all the observations have been greyed for clarity. Also shown are a more metal-poor model with $Z=0.0005$ (black upside-down triangle) and a more metal-rich one with $Z=0.008$ (black square), with all other parameters fixed. Figs~\ref{fig:baton3u_uv} and \ref{fig:baton3u_opt} show the standard model again, along with a lower-ionization model with $\log\Uav=-3$ (small black circle) and a higher-ionization one with $\log\Uav=-1$ (large black circle), at fixed other parameters. Overall, Figs~\ref{fig:baton3z_uv}--\ref{fig:baton3u_opt} indicate that these five models sample reasonably well the observed ultraviolet and optical emission-line properties of metal-poor, actively star-forming galaxies [i.e., with $\lniiopt/\ha\lesssim0.1$ and EW(\hb)$\,\gtrsim200\,$\AA\ in Figs~\ref{fig:baton3z_opt} and \ref{fig:baton3u_opt}], except in diagrams involving the \heii\ and \heiiopt\ lines. In such diagrams, the data for low oxygen abundances [$\logohgas\lesssim8.0$; see Figs~\ref{fig:baton3z_uv}a, \ref{fig:baton3z_opt}f, \ref{fig:baton3u_uv}a and \ref{fig:baton3u_opt}f] tend to exhibit much stronger \lheii\ emission than predicted by models with classical stellar populations, as pointed out in several previous studies (Section~\ref{sec:intro}). The five benchmark models also do not quite reach the highest observed equivalent widths of \lciv\ ($\gtrsim10\,$\AA; Figs~\ref{fig:baton3z_uv}c and \ref{fig:baton3u_uv}c) and \lciii\ ($\gtrsim20\,$\AA; e.g., Figs~\ref{fig:baton3z_uv}g and \ref{fig:baton3u_uv}g), nor the highest \lciii/\loiii\ ratios ($\gtrsim3$; Figs~\ref{fig:baton3z_uv}f and \ref{fig:baton3u_uv}f).

We now examine the observable signatures of each adjustable parameter of the models in the emission-line diagrams of Figs~\ref{fig:baton3z_uv}--\ref{fig:baton3u_opt}. We describe the effects of altering a single parameter at a time, keeping all other parameters fixed:

\begin{figure*}
\begin{center}
\resizebox{0.95\hsize}{!}{\includegraphics{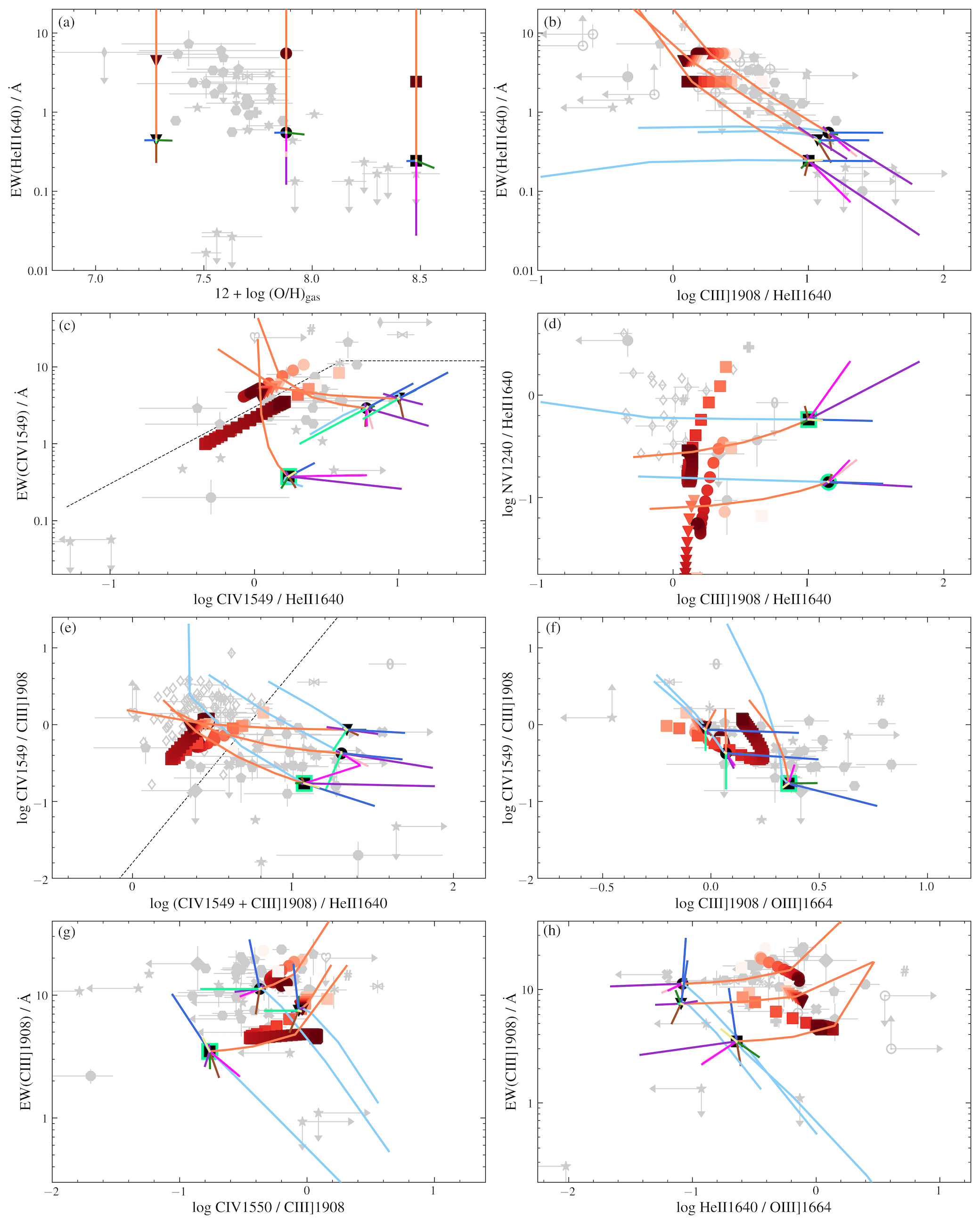}}
\end{center}
\caption{Same diagrams as in Fig.~\ref{fig:obs_uv}, but where the observations have been greyed for clarity. The black circle corresponds to the `standard' model (with metallicity $Z=0.002$ and $\log\Uav=-2$) described in Section~\ref{sec:params}, while the black upside-down triangle and black square are benchmark models with the same parameters, but with $Z=0.0005$ and 0.008, respectively. Segments of different colours show the effect of altering a single parameter at the time (as summarized at the bottom of Fig.~\ref{fig:baton3z_opt}): rise in \CO\ ratio from 0.17 to $\COsol=0.44$ (blue); drop in dust-to-mass ratio from $\xid=0.3$ to 0.1 (dark green); rise in \nh\ from $10^2$ to $10^3\,$cm$^{-3}$ (yellow); inclusion of interstellar-line absorption in the \hii\ interiors and outer \hi\ envelopes of stellar birth clouds (light green); increase in stellar population age from 3 to 10\,Myr (brown); rise in \mup\ from 100, to 300, to 600\,\Msun\ (dark purple); adopting the \bpass\ single- (light purple) and binary-star (magenta) models in place of the \CB\ model (\bpass\ models are not available for $Z=0.0005$); drop in the optical depth \tauref\ from +1.0 to $-1.0$ (light blue); inclusion of an AGN component contributing from 0 to 99 per cent of the total \heii\ emission (orange); and inclusion of a radiative-shock component contributing 90 per cent of the total \heii\ emission [red symbols, with shape corresponding to the metallicity of the associated benchmark model, and darkness to the shock velocity, from $10^2\,$km\,s$^{-1}$ (light) to $10^3\,$km\,s$^{-1}$ (dark)].}
\label{fig:baton3z_uv}
\end{figure*}

\begin{figure*}
\begin{center}
\resizebox{0.95\hsize}{!}{\includegraphics{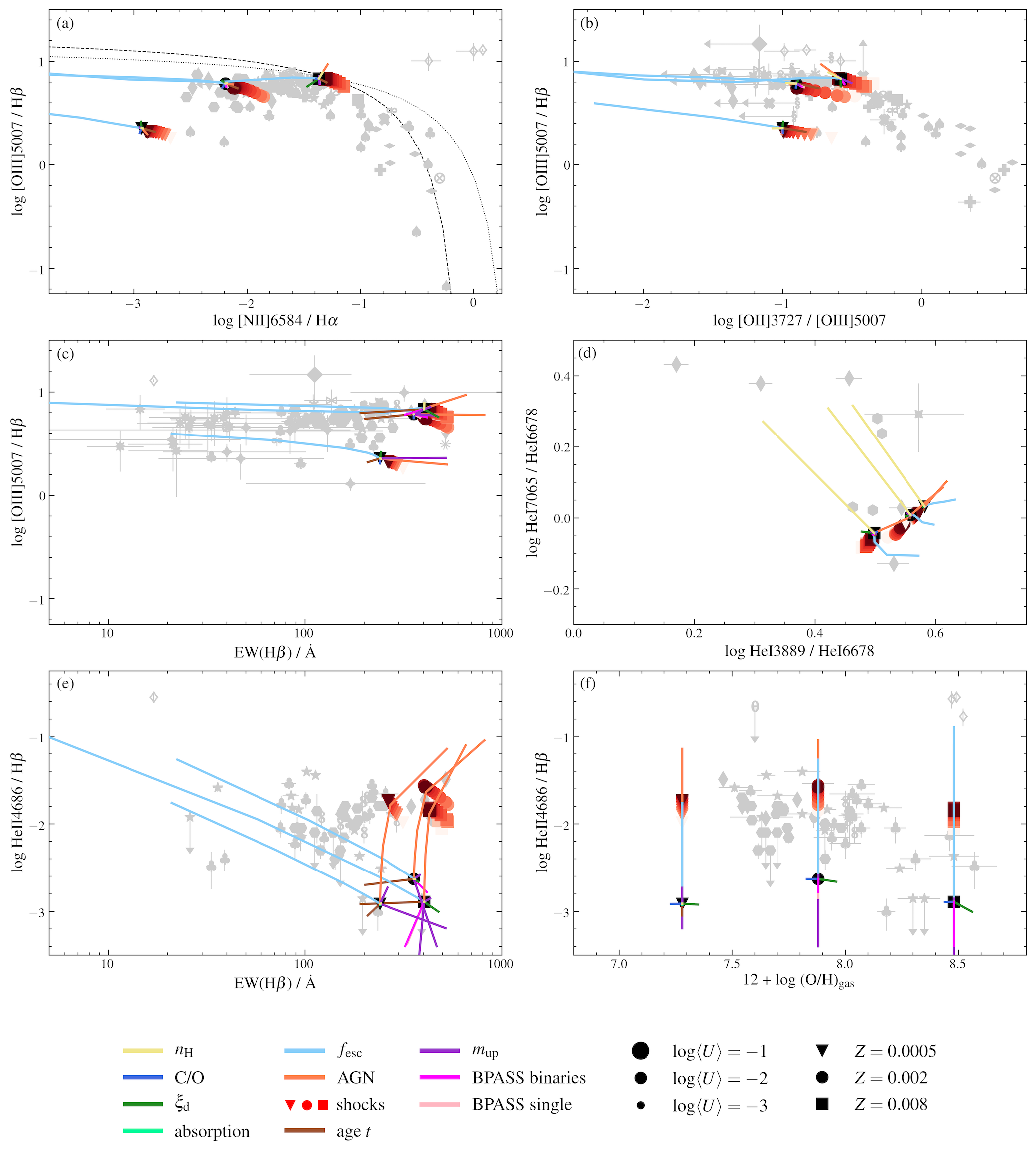}}
\end{center}
\caption{Same diagrams as in Fig.~\ref{fig:obs_opt}, but where the observations have been greyed for clarity. The models are the same as in Fig.~\ref{fig:baton3z_uv}.}
\label{fig:baton3z_opt}
\end{figure*}

\begin{figure*}
\begin{center}
\resizebox{0.95\hsize}{!}{\includegraphics{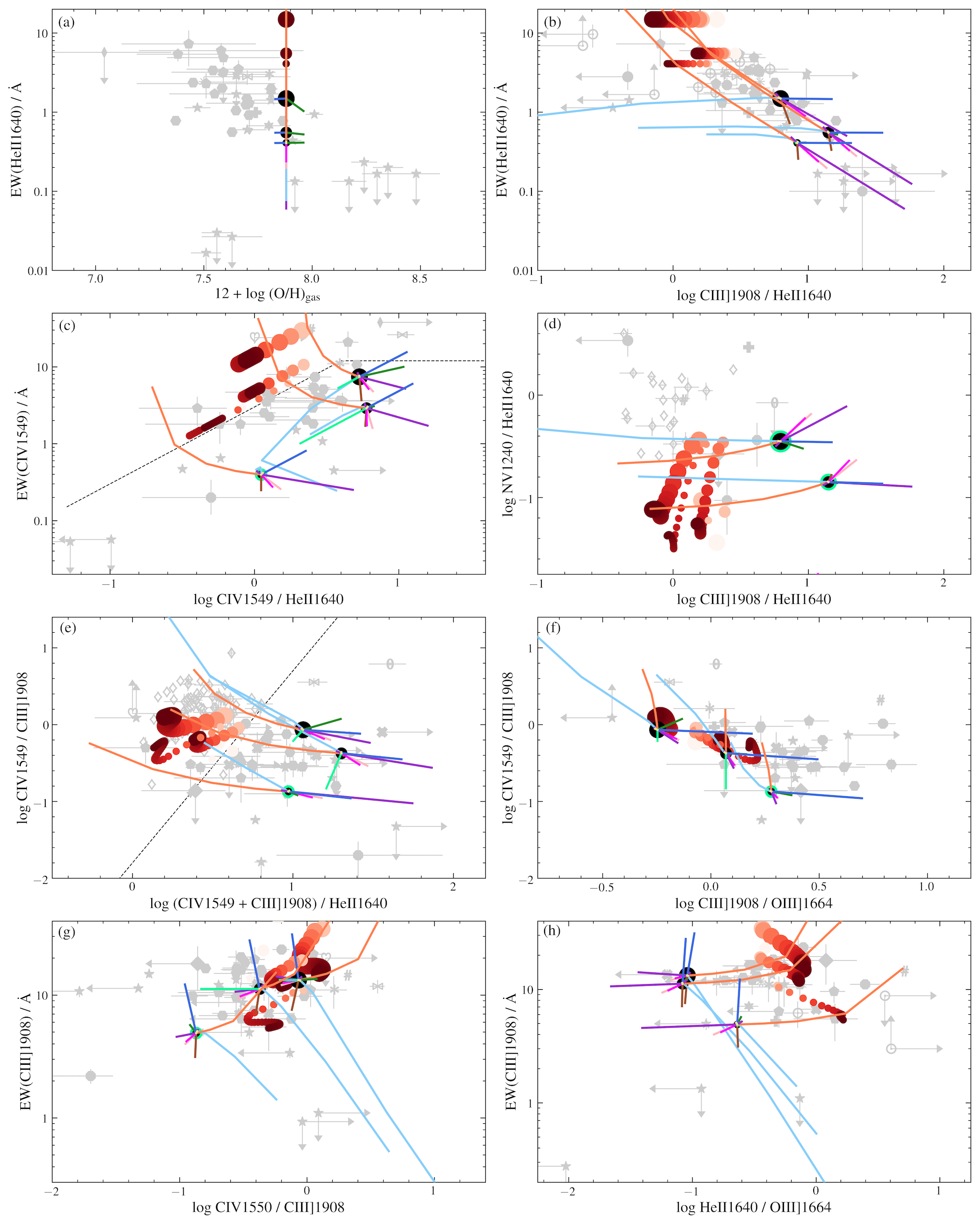}}
\end{center}
\caption{Same as Fig.~\ref{fig:baton3z_uv}, but for models with the metallicity $Z= 0.002$ only, and for three values of the zero-age volume-averaged ionisation parameter, $\log\Uav=-3$, $-2$ and $-1$ (in order of increasing symbol size).}
\label{fig:baton3u_uv}
\end{figure*}

\begin{figure*}
\begin{center}
\resizebox{0.95\hsize}{!}{\includegraphics{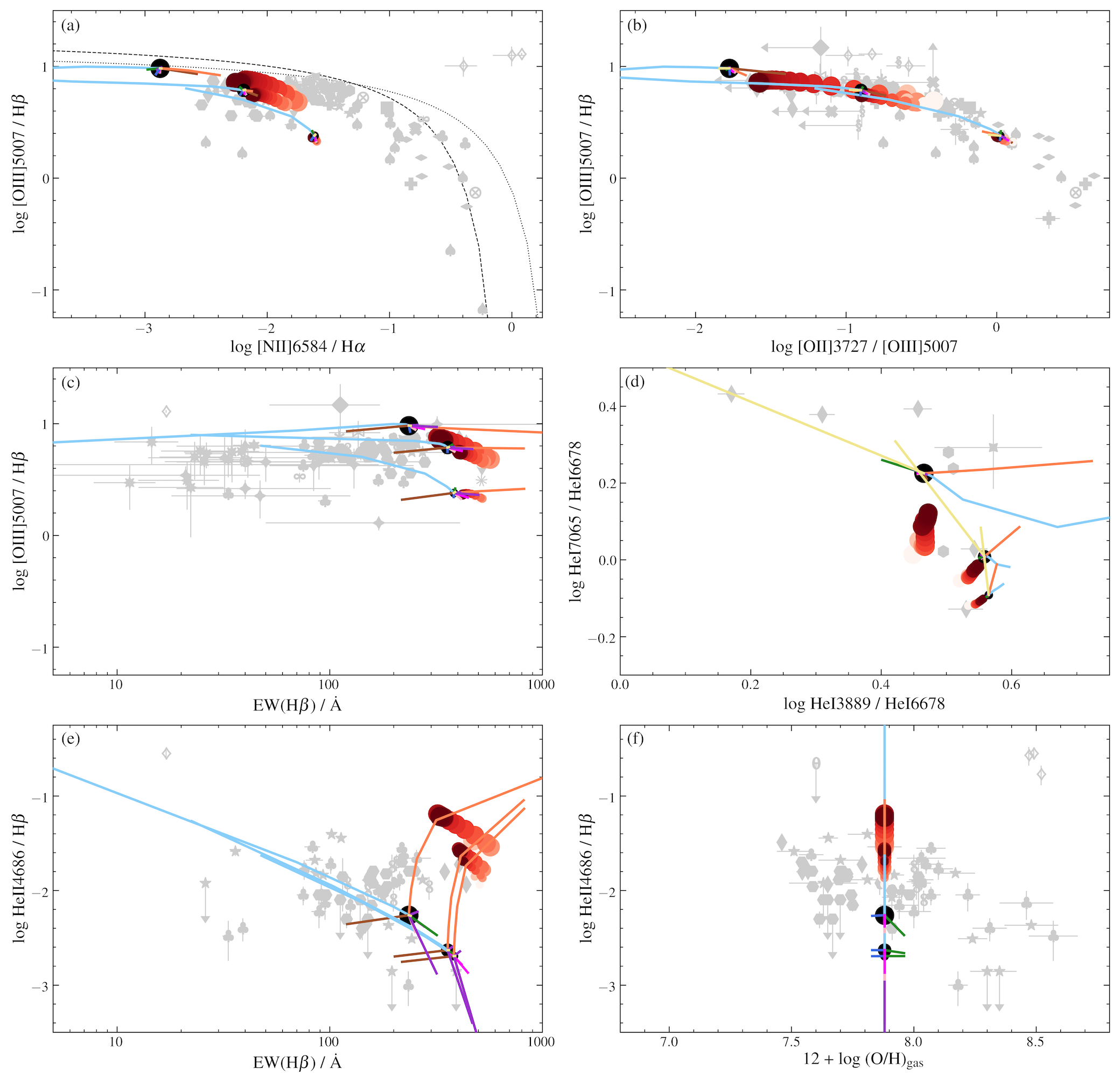}}
\end{center}
\caption{Same as Fig.~\ref{fig:baton3z_opt}, but for the same models as in Fig.~\ref{fig:baton3u_uv}.}
\label{fig:baton3u_opt}
\end{figure*}

\begin{itemize}
\item[] \textit{Metallicity, $Z$}. Increasing metallicity from $Z=0.0005$ (black upside-down triangle), to 0.002 (circle), to 0.008 (square) in Figs~\ref{fig:baton3z_uv} and \ref{fig:baton3z_opt} raises cooling through collisionally-excited metal transitions, causing the electronic temperature, \Te, to drop. The luminosity ratios of metal-to-H and He lines tend to increase at first, and then drop when \Te\ is low enough for cooling to shift from ultraviolet and optical to infrared transitions \citep[e.g.][]{Spitzer78}. This is particularly visible for \lciii/\heii\ [Fig.~\ref{fig:baton3z_uv}b; note also the behaviour of EW(\lciii) in Fig.~\ref{fig:baton3z_uv}g], while for \loiiiopt/\hb\ (Fig.~\ref{fig:baton3z_opt}a) the maximum is reached around $Z\approx0.006$ \citep[see fig.~2 of][]{gutkin2016}. In contrast, \lnv/\heii\ (Fig.~\ref{fig:baton3z_uv}d) and \lniiopt/\ha\ (Fig.~\ref{fig:baton3z_opt}a) keep rising as $Z$ increases, because of the inclusion of secondary nitrogen production in our model. Meanwhile, \heiopta/\heioptb\ and \heioptc/\heioptb\ both decrease as $Z$ rises and \Te\ declines \citep{izotov17oct}. At fixed ionization parameter, a rise in $Z$ also implies lower ratios of high- to low-ionization lines, such as smaller \lciv/\lciii\ [Fig.~\ref{fig:baton3z_uv}c;  note also the behaviour of EW(\lciv) in Fig.~\ref{fig:baton3z_uv}c] and larger \lciii/\loiii\ (Fig.~\ref{fig:baton3z_uv}f) and \loiiopt/\loiiiopt\ (Fig.~\ref{fig:baton3z_opt}b), because the inner high-ionization parts of \hii\ regions shrink (e.g., Fig.~\ref{fig:struction_carbon} above; see also \citealt{stasinska80}). 

The behaviour of the recombination-line ratio \heiiopt/\hb\ in Fig.~\ref{fig:baton3z_opt}e is linked to the evolution of very massive stars ($>100\,\Msun$) in the models. As $Z$ increases, the competing effects of the weakening of ionizing radiation caused by the lower effective temperatures of metal-rich relative to metal-poor stars \citep[e.g., fig.~15 of][]{Bressan2012} and the hardening of ionizing radiation caused by an increase in mass-loss rate \citep[e.g.,][]{Vink2001,Crowther06} conspire in making \heiiopt/\hb\ larger at $Z=0.002$ than at 0.0005 and 0.008 (see also Fig.~\ref{fig:HeII_ssp_age} of Section~\ref{sec:constraints}; we note that this is not the case for $\mup=100\,$\Msun; see below). The rise in EW(\hb) as $Z$ increases in this diagram results from the slower evolution of metal-rich relative to metal-poor stars, delaying the appearance of evolved stars with strong optical continua.

\item[] \textit{Zero-age volume-averaged ionisation parameter, \Uav}. Increasing $\log\Uav$ from $-3$ (small black circle), to $-2$ (medium-size circle), to $-1$ (large circle) in Figs~\ref{fig:baton3u_uv} and \ref{fig:baton3u_opt}, which can be achieved in our model by raising the gas-filling factor $\epsilon$ at fixed ionizing-photon rate $Q$ and H-density \nh\ (equation~\ref{eq:Udef}), increases the probability of multiple ionization. This causes \lnv/\heii\ (Fig.~\ref{fig:baton3u_uv}d), \lciv/\lciii\ (Fig.~\ref{fig:baton3u_uv}e), \loiiiopt/\hb\ (Fig.~\ref{fig:baton3u_opt}a), \loiiiopt/\loiiopt\ (inverse abscissa of Fig.~\ref{fig:baton3u_opt}b) and \heiiopt/\hb\ (Fig.~\ref{fig:baton3u_opt}f) to rise. The equivalent widths of \heii\ (Fig.~\ref{fig:baton3u_uv}a), \lciv\ (Fig.~\ref{fig:baton3u_uv}c) and \lciii\ (Fig.~\ref{fig:baton3u_uv}h) also increase. Instead, \lciii/\heii\ (Fig.~\ref{fig:baton3u_uv}b), \lciv/\heii\ (Fig.~\ref{fig:baton3u_uv}c) and \loiii/\heii\ (inverse abscissa of Fig.~\ref{fig:baton3u_uv}h) first increase, and then decrease when \Uav\ rises. This is because while the \heii\ luminosity continues to rise, the rise of \lciii, \lciv\ and \loiii\ is slowed down by the conversion of C$^+$ into C$^{2+}$, C$^{2+}$ into C$^{3+}$ and O$^{2+}$ into O$^{3+}$. Similarly, \niiopt/\ha\ drops (Fig.~\ref{fig:baton3u_opt}a) because of the conversion of N$^+$ into N$^{2+}$. 

Increasing \Uav\ at fixed other parameters also makes the H-column density, and hence the dust optical depth, larger (Section~\ref{sec:fesc}). The enhanced absorption of ionizing photons by dust is the reason for the drop in EW(\hb) in Fig.~\ref{fig:baton3u_opt}e. Since the grain opacity peaks near 912\,\AA\ and declines toward shorter wavelengths \citep[e.g.,][]{Bottorff98}, \heiiopt-ionizing photons are less absorbed than H-ionizing ones, an effect amplified by the fact that the \hb\ line is produced further out in \hii\ regions than the \heiiopt\ line. This effect contributes to the rise in \heiiopt/\hb\ from $\log\Uav=-3$ to $-1$ in this diagram \citep[see also][]{erb10}. Also, the associated increase in \lhei-column density amplifies the effects of fluorescence, causing \heiopta/\heioptb\ to drop and \heioptc/\heioptb\ to rise \citep[Fig.~\ref{fig:baton3u_opt}d; see ][]{izotov98}.

\item[] \textit{Carbon-to-oxygen abundance ratio}. Increasing the \CO\ ratio from 0.17 to $\COsol=0.44$ (blue segments in Figs~\ref{fig:baton3z_uv}--\ref{fig:baton3u_opt}) is achieved in our model by raising the carbon abundance and lowering the abundances of all other metallic elements -- including oxygen -- at fixed total metallicity $Z$ (see section~2.3.1 of \citealt{gutkin2016}). This makes the \ciii\ and \civ\ lines stronger (Figs~\ref{fig:baton3z_uv} and \ref{fig:baton3u_uv}) and the \loiiopt, \loiiiopt\ and \lniiopt\ lines slightly weaker (Figs~\ref{fig:baton3z_opt} and \ref{fig:baton3u_opt}), while the H and He lines are negligibly affected (Figs~\ref{fig:baton3z_opt}d--f and \ref{fig:baton3u_opt}d--f ).

\item[] \textit{Dust-to-metal mass ratio, \xid}. Lowering the dust-to-mass ratio from $\xid=0.3$ to 0.1 (dark-green segments in Figs~\ref{fig:baton3z_uv}--\ref{fig:baton3u_opt}) increases the abundance of coolants in the gas phase. This causes line luminosities from the most abundant refractory coolants (such as O and C) to rise, while at the same time, the drop in electronic temperature reduces cooling through ultraviolet and optical transitions, an effect which becomes dominant at high metallicity. Hence, EW(\lciii) (Fig.~\ref{fig:baton3z_uv}g), EW(\lciv) (Fig.~\ref{fig:baton3z_uv}c), \lciii/\heii\ (Fig.~\ref{fig:baton3z_uv}b) and \loiiiopt/\ha\ (Fig.~\ref{fig:baton3z_opt}a) rise at low $Z$ but drop at high $Z$ when \xid\ declines from 0.3 to 0.1. In Fig.~\ref{fig:baton3z_uv}f, \lciii/\loiii\ increases because carbon is more depleted than oxygen (see, e.g., table~1 of \citealt{gutkin2016}). Lowering \xid\ also makes the dust optical depth smaller, causing \heiiopt/\hb\ to drop and EW(\hb) to rise in Figs~\ref{fig:baton3z_opt}e and \ref{fig:baton3u_opt}e (see discussion of \Uav\ above). We note that the dust optical depth is not simply proportional to the product $\xid{Z}$, as absorption of ionizing photons by dust when $Z$ increases also reduces the \hii-region radius, and hence \Nh\ (equation~\ref{eq:taudust} and Fig.~\ref{fig:NH0}).

\item[] \textit{Hydrogen gas density, \nh}. Increasing \nh\ from $10^2$ to $10^3\,$cm$^{-3}$ (yellow segments in Figs~\ref{fig:baton3z_uv}--\ref{fig:baton3u_opt}) enhances collisional excitation, but also favours collisional over radiative de-excitation of excited species. The cooling through infrared transitions is reduced and that through ultraviolet and optical transitions enhanced, because the critical density for collisional de-excitation is lower for infrared fine-structure than for ultraviolet and optical transitions. The effect is most visible at high metallicity, where infrared transitions tend to dominate the cooling \citep[e.g.,][]{oey93}. Thus, EW(\lciii) (Fig.~\ref{fig:baton3z_uv}g), \lciii/\heii\ (Fig.~\ref{fig:baton3z_uv}b), EW(\lciv) and \lciv/\heii\ (Fig.~\ref{fig:baton3z_uv}c), \loiii/\heii\ (inverse abscissa of Fig.~\ref{fig:baton3z_uv}h) and \lniiopt/\ha\ and \loiiiopt/\hb\ (Fig.~\ref{fig:baton3z_opt}a) rise together with \nh. Since increasing \nh\ at fixed other parameters in our model implies reducing the gas-filling factor as $\epsilon\propto1/\sqrt{\nh}$ (equation~\ref{eq:Udef}), this causes the dust optical depth to rise as $\sqrt{\nh}$ (equation~\ref{eq:taudust}), the effect of which is nonetheless subtle in Figs~\ref{fig:baton3z_uv}--\ref{fig:baton3u_opt}. In contrast, the \lhei\ lines (Figs~\ref{fig:baton3z_opt}d and \ref{fig:baton3u_opt}d) are quite sensitive to changes in \nh. This is because the $\lambda$7065 transition is much more responsive to collisional enhancement than the $\lambda6678$ transition, itself more so than the $\lambda$3889 transition \citep{izotov17oct}, causing \heiopta/\heioptb\ to drop and \heioptc/\heioptb\ to rise significantly as \nh\ increases.

\item[] \textit{Interstellar-line absorption}. The light-green segments in Figs~\ref{fig:baton3z_uv}--\ref{fig:baton3u_opt} show the effect of accounting for interstellar-line absorption in the \hii\ interiors and outer \hi\ envelopes of stellar birth clouds, following the prescription of \citet[][see Section~\ref{sec:ionb} above]{vidal17}. As expected, the effect is most striking for the \civ\ and \nv\ resonance lines, whose net emission can be drastically reduced and even entirely canceled -- as pictured by light-green aureolas around some benchmark models -- in Figs~\ref{fig:baton3z_uv} and \ref{fig:baton3u_uv}.

\item[] \textit{Stellar population age, $t$}. At constant star formation rate, the age of the stellar population sets the age of the oldest \hii\ regions contributing to the nebular emission from a galaxy in our model (equation~\ref{eq:flux_gal}). In an individual \hii\ region, the rate of ionizing photons, and hence, the ionization parameter (equation~\ref{eq:Udef}), drop sharply at ages after about 3\,Myr (e.g., Fig.~\ref{fig:fesc}d). For $\psi(t)=\mathrm{constant}$, therefore, the global effective ionization parameter of the population of \hii\ regions declines until a stationary population of ionizing stars is reached, which happens around $t=10\,$Myr in the C\&B models (but see also Section~\ref{sec:lycprod} below). Thus, increasing $t$ from 3 to 10\,Myr (brown segments in Figs~\ref{fig:baton3z_uv}--\ref{fig:baton3u_opt}) tends to have an effect similar on emission lines to that of decreasing \Uav\ (see above), such as making \lniiopt/\ha\ (Figs~\ref{fig:baton3z_opt}a and \ref{fig:baton3u_opt}a) and \loiiopt/\loiiiopt\ (Figs~\ref{fig:baton3z_opt}b and \ref{fig:baton3u_opt}b) larger and \heiiopt/\hb\ (Figs~\ref{fig:baton3z_opt}e and \ref{fig:baton3u_opt}e) smaller. The equivalent widths of \heii\ (Figs~\ref{fig:baton3z_uv}b and \ref{fig:baton3u_uv}b), \lciv\ (Figs~\ref{fig:baton3z_uv}c and \ref{fig:baton3u_uv}c) and \lciii\ (Fig.~\ref{fig:baton3z_uv}g and \ref{fig:baton3u_uv}g) drop because of the build-up of continuum flux from older stellar populations.

\item[] \textit{Upper mass cut-off of the IMF, \mup}. At fixed \Uav\ and other parameters, increasing \mup\ from 100, to 300, to 600\,\Msun\ (dark-purple segments in Figs~\ref{fig:baton3z_uv}--\ref{fig:baton3u_opt}) hardens the ionizing spectrum, because massive stars evolve at higher temperatures than lower-mass stars. The effect is much stronger from 100 to 300\,\Msun\ than from 300 to 600\,\Msun, because of the upturn of the upper main sequence in the Hertzsprung-Russell diagram. The hardening of the spectrum primarily implies larger ratios of \lheii-to-other lines, such as \heii/\loiii\ (Figs~\ref{fig:baton3z_uv}h and \ref{fig:baton3u_uv}h) and \heiiopt/\hb\ (Figs~\ref{fig:baton3z_opt}e and \ref{fig:baton3u_opt}e), and in turn, smaller \lciii/\heii\ (Figs~\ref{fig:baton3z_uv}b and \ref{fig:baton3u_uv}b) and \lciv/\heii\ (Figs~\ref{fig:baton3z_uv}c and \ref{fig:baton3u_uv}c). We note that, for $\mup=100\,\Msun$, \heiiopt/\hb\ increases steadily from $Z=0.008$, to 0.002, to 0.0005 (Fig.~\ref{fig:baton3z_opt}e), because of the higher effective temperatures of metal-poor relative to metal-rich stars (see above). Raising \mup\ also strengthens the equivalent width of \heii\ (Figs~\ref{fig:baton3z_uv}a and \ref{fig:baton3u_uv}a), and to a lesser extent, those of \lciv\ (Figs~\ref{fig:baton3z_uv}c and \ref{fig:baton3u_uv}c) and \lciii\ (Figs~\ref{fig:baton3z_uv}g and \ref{fig:baton3u_uv}g). Finally, we note that, for $\mup=600\,\Msun$, a stellar \heii\ wind feature can arise even at the metallicity $Z=0.0005$ in the models of Figs~\ref{fig:baton3z_uv}--\ref{fig:baton3u_opt} (with predicted equivalent width $\sim1.7\,$\AA\ and full width at half-maximum $\sim1800\,$km\,s$^{-1}$).

\item[] \textit{Stellar population synthesis model}. The light-purple and magenta segments in Figs~\ref{fig:baton3z_uv}--\ref{fig:baton3u_opt} show the effect of using the \bpass\ single- and binary-star models, respectively, in place of the \CB\ model, to compute $S_{\lambda}(\tprime)$ in equation~\eqref{eq:flux_gal}. At the considered age of  3\,Myr, both versions of the \bpass\ model tend to produce slightly softer ionizing radiation than the \CB\ model, which incorporates recent evolutionary tracks and model atmospheres for massive stars (Section~\ref{sec:ionb}; see also Section~\ref{sec:lycprod} below). As a result, in all panels of Figs~\ref{fig:baton3z_uv}--\ref{fig:baton3u_opt}, changing from the \CB\ to \bpass\ models has an effect on line ratios and equivalent widths similar to that of lowering \mup\ (see above).\footnote{The photoionization modelling in the comparison of \CB\ and \bpass\,v2.2.1 models presented in Figs~\ref{fig:baton3z_uv}--\ref{fig:baton3u_opt} is fully self-consistent and includes dust physics. We note that \citet{Xiao18} compare the \citet{gutkin2016} models, which include dust physics, with dust-free photoionization models computed using \bpass\,v2.1 (\citealt{Xiao18} also inadvertently plotted \lciv/\heii\ in place of \lciii/\heii\ from the \citealt{gutkin2016} models in their fig.~B1; E. Stanway, private communication).} As expected, the \bpass\ binary-star model produces harder radiation than the \bpass\ single-star model \citep[e.g.,][]{stanway19}, the effect increasing toward later ages (not shown). It is also worth noting that, since a majority of massive stars are expected to undergo binary interactions \citep[e.g.,][]{Sana12}, the \lheii-line strengths predicted by the single-star \CB\ models should be considered as lower limits.

\item[] \textit{Fraction of escaping LyC photons, \fesc}. The light-blue segments in Figs~\ref{fig:baton3z_uv}--\ref{fig:baton3u_opt} show the effect of decreasing \tauref, the zero-age optical depth of \hii\ regions to LyC photons with wavelength $\lambda=570\,$\AA\ (Section~\ref{sec:fesc}), from +1.0 to $-1.0$. This is equivalent to increasing \fesc\ from zero to nearly unity (Fig.~\ref{fig:fesc}c). As seen in Section~\ref{sec:fesc} (Figs~\ref{fig:struction_carbon} and \ref{fig:struction_oxygen}), increasing \fesc\ progressively removes the outer low-ionization zones of \hii\ regions, causing \lciv/\lciii\ (Figs~\ref{fig:baton3z_uv}c and \ref{fig:baton3u_uv}c) and \loiiiopt/\loiiopt\ (inverse abscissa of Figs~\ref{fig:baton3z_opt}b and \ref{fig:baton3u_opt}b) to rise, while \lciii/\heii\ (Figs~\ref{fig:baton3z_uv}b and \ref{fig:baton3u_uv}b) and the equivalent widths of \lciii\ (Figs~\ref{fig:baton3z_uv}g and \ref{fig:baton3u_uv}g) and \hb\ (Figs~\ref{fig:baton3z_opt}e and \ref{fig:baton3u_opt}e) drop sharply (at ages $t>3\,$Myr for large \fesc, the effect on \loiiiopt/\loiiopt\ would be inverted; see Fig.~ \ref{fig:convol}f). Also, as noted by \citet{izotov17oct}, increasing \fesc\ makes \heiopta/\heioptb\ larger because of the high sensitivity of the $\lambda$3889 transition to fluorescence (which increases the line luminosity as the optical depth decreases), while the effect on \heioptc/\heioptb\ is weaker (Figs~\ref{fig:baton3z_opt}d and \ref{fig:baton3u_opt}d). This led \citet{izotov17oct} to argue that \lhei\ lines could be a promising alternative to \loiiiopt/\loiiopt\ to constrain \fesc\ in star-forming galaxies. 

\item[] \textit{AGN component}. The orange segments in Figs~\ref{fig:baton3z_uv}--\ref{fig:baton3u_opt} show the effect of adding an AGN component contributing from 0 to 99 per cent of the total \heii\ emission, using the prescription of Section~\ref{sec:agn} (this corresponds roughly to a contribution from 0 to 40--80 per cent of the total \hb\ emission, depending on the model). In these Seyfert 2-galaxy models, the AGN narrow-line region contributes to the nebular (line and recombination-continuum) emission, but not the underlying ultraviolet and optical emission. Thus, the equivalent widths of \heii\ (Figs~\ref{fig:baton3z_uv}a and \ref{fig:baton3u_uv}a), \lciv\ (Figs~\ref{fig:baton3z_uv}c and \ref{fig:baton3u_uv}c), \lciii\ (Figs~\ref{fig:baton3z_uv}h and \ref{fig:baton3u_uv}h) and \hb\ (Figs~\ref{fig:baton3z_opt}e and \ref{fig:baton3u_opt}e) all rise. The much harder spectra of AGN relative to stars at high energies \citep[e.g., fig.~1 of][]{Feltre2016} imply larger ratios of \lheii-to-other lines and larger \lciv/\lciii\ (Figs~\ref{fig:baton3z_uv}e and \ref{fig:baton3u_uv}e) and \loiiiopt/\loiiopt\ (inverse abscissa of Figs~\ref{fig:baton3z_opt}b and \ref{fig:baton3u_opt}b), the AGN component accounting for nearly all the \heii\ emission in the most extreme models. We note that \lnv/\heii\ drops in Figs~\ref{fig:baton3z_uv}d and \ref{fig:baton3u_uv}d, because of the conversion of N$^{4+}$ into N$^{5+}$. In these figures, the \citet{dors14} and \citet{Mignoli19} observations of powerful AGN hosted by massive galaxies can be reproduced by models with metallicity $Z\approx0.008$ and high ionization parameters, $-2\lesssim \log \Uav \lesssim -1$ \citep[see also section~4.2 of][]{Mignoli19}. Finally, the larger \nh\ of the AGN models ($10^3\,{\rm cm}^{-3}$) relative to \hii-region models ($10^2\,{\rm cm}^{-3}$) makes \heioptc/\heioptb\ rise because of collisional enhancement when the AGN contribution rises, while the inclusion of microturbulence in the AGN models (Section~\ref{sec:agn}) reduces the $\lambda$3889-line optical depth and hence the effects of fluorescence \citep{Benjamin02}, causing \heiopta/\heioptb\ to also rise (Figs~\ref{fig:baton3z_opt}d and \ref{fig:baton3u_opt}d).

\item[] \textit{Shock component}. The series of red symbols of different darkness in Figs~\ref{fig:baton3z_uv}--\ref{fig:baton3u_opt} show the effect of adding a radiative-shock component contributing 90 per cent of the total \heii\ emission, using the prescription of Section~\ref{sec:shocks}. The symbol shape corresponds to the metallicity of the associated benchmark model (upside-down triangle: $Z=0.0005$; circle: $Z=0.002$;  square: $Z=0.008$) and the darkness to the shock velocity (from $10^2\,$km\,s$^{-1}$: light; to $10^3\,$km\,s$^{-1}$: dark). The signatures of a shock component are very similar to those identified above for an AGN component, in particular very strong ratios of \lheii-to-other lines, the fraction of, for example, total \hb\ luminosity shocks account for being typically less than 15 per cent in Figs~\ref{fig:baton3z_uv}--\ref{fig:baton3u_opt}. This is because collisional ionization in the high-temperature ($\Te>10^6\,$K) radiative zone of a shock produces He$^{2+}$ along with other highly-ionized species (e.g., C$^{4+}$ to C$^{6+}$, N$^{4+}$ to N$^{7+}$, O$^{4+}$ to O$^{8+}$), whose recombination generates strong \lheii\ emission and extreme ultraviolet and soft X-ray emission capable of producing lower-ionization species upstream and downstream of the shock \citep[see figs.~8--11 of][]{allen08}. Hence, the main effect of adding a contribution by shock-ionized gas is to raise the \heii\ equivalent width (Figs~\ref{fig:baton3z_uv}b and \ref{fig:baton3u_uv}b) and all ratios of \lheii-to-other lines, such as \heii/\loiii\ (Figs~\ref{fig:baton3z_uv}h and \ref{fig:baton3u_uv}h) and \heiiopt/\hb\ (Figs~\ref{fig:baton3z_opt}e and \ref{fig:baton3u_opt}e), and the inverse of \lciii/\heii\ (Figs~\ref{fig:baton3z_uv}b and \ref{fig:baton3u_uv}b) and \lciv/\heii\ (Figs~\ref{fig:baton3z_uv}c and \ref{fig:baton3u_uv}c). The intensity and hardness of the ionizing radiation increases with shock velocity. We have checked that adopting different pre-shock densities (in the range $1\leq \nh\leq 10^4$\,cm$^{-2}$; see Section~\ref{sec:shocks}) and transverse magnetic fields ($10^{-4}\leq B\leq 10\,\mu$G) has a negligible effect on the results of Figs~\ref{fig:baton3z_uv}--\ref{fig:baton3u_opt}, except for the low-ionization \lniiopt\ and \loiiopt\ lines (Figs~\ref{fig:baton3z_opt}a,b and \ref{fig:baton3u_opt}a,b), whose fluxes tend to decrease when \nh\ rises and $B$ drops.

\end{itemize}


\section{Constraints on the production and escape of ionizing radiation}\label{sec:constraints}

In the previous section, we described the ultraviolet and optical emission-line signatures of a wide range of ISM, stellar-population, AGN and radiative-shock parameters in metal-poor star-forming galaxies. We now interpret these results to investigate emission-line diagnostics of the production and escape of ionizing radiation in such galaxies. Specifically, we wish to assess the hints provided by the reference observational sample of Section~\ref{sec:obs} on the sources dominating the production of ionizing photons (Section~\ref{sec:lycprod}) and on LyC-photon leakage (Section~\ref{sec:lycfesc}) in these galaxies. We mention along the way how our findings compare to those of previous studies relying on investigations of often fewer emission lines with different models. 

\subsection{Diagnostics of ionizing sources}\label{sec:lycprod}

\begin{figure}
\begin{center}
\resizebox{\hsize}{!}{\includegraphics{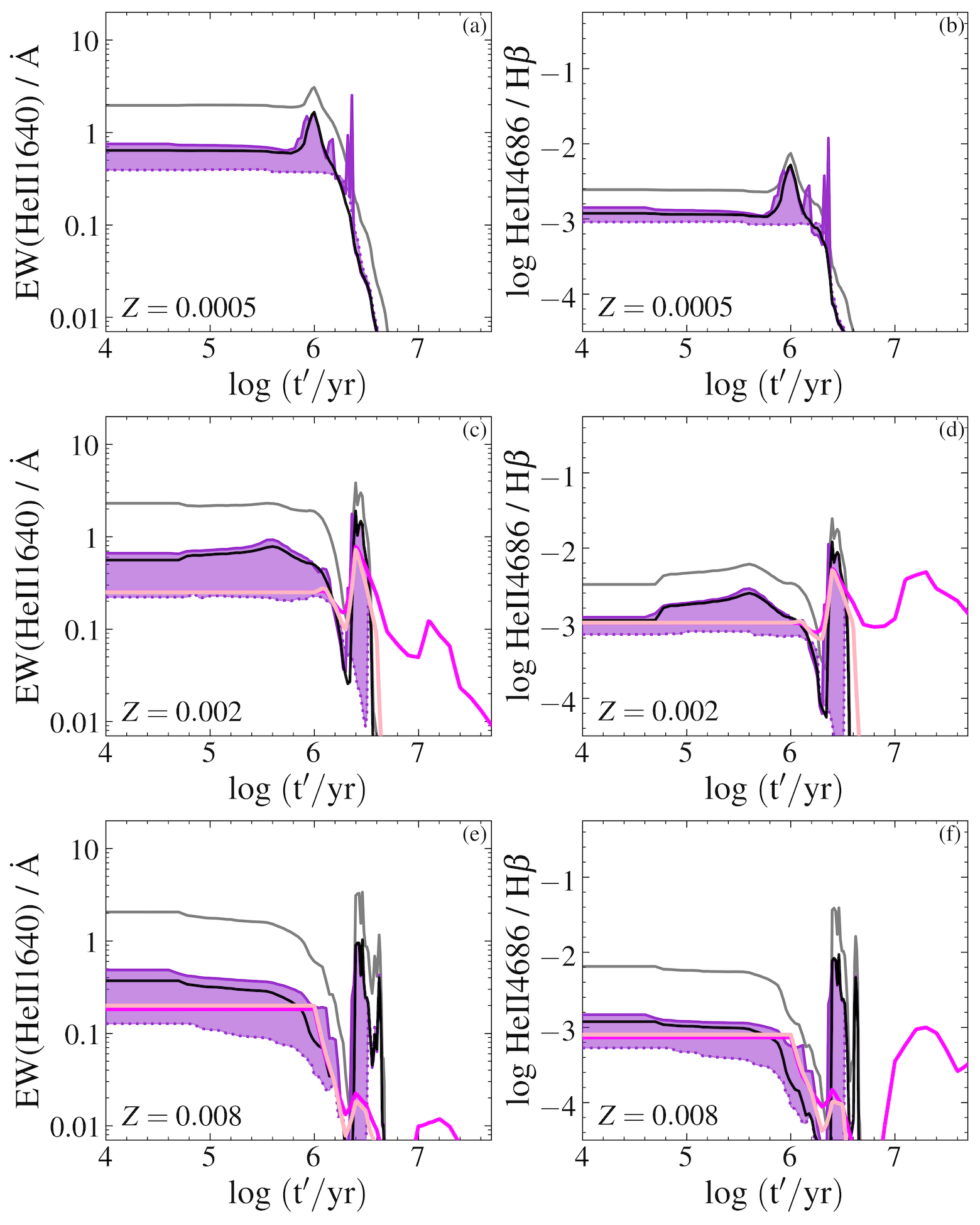}}
\end{center}
\caption{Evolution of the \heii\ equivalent width (left) and \heiiopt/\hb\ ratio (right) in models of ionization-bounded \hii\ regions powered by different types of SSPs. The black curves show the \CB-based benchmark models with $\log\Uav=-2$ of Section~\ref{sec:params}, for $Z=0.0005$ (top), 0.002 (middle) and 0.008 (bottom). The light-purple and magenta curves show the corresponding models powered by \bpass\ single- and binary-star SSPs, respectively (\bpass\ models are not available for $Z=0.0005$). The dotted and solid dark-purple lines show SSP models with the same parameters as black curves, but for IMF upper mass cut-offs $\mup=100$ and 600\,\Msun, respectively (the area between these two models has been shaded in purple, for clarity).The grey curves show the same models as the black curves, but for $\log\Uav=-1$.}
\label{fig:HeII_ssp_age}
\end{figure}

\begin{figure}
\begin{center}
\resizebox{\hsize}{!}{\includegraphics{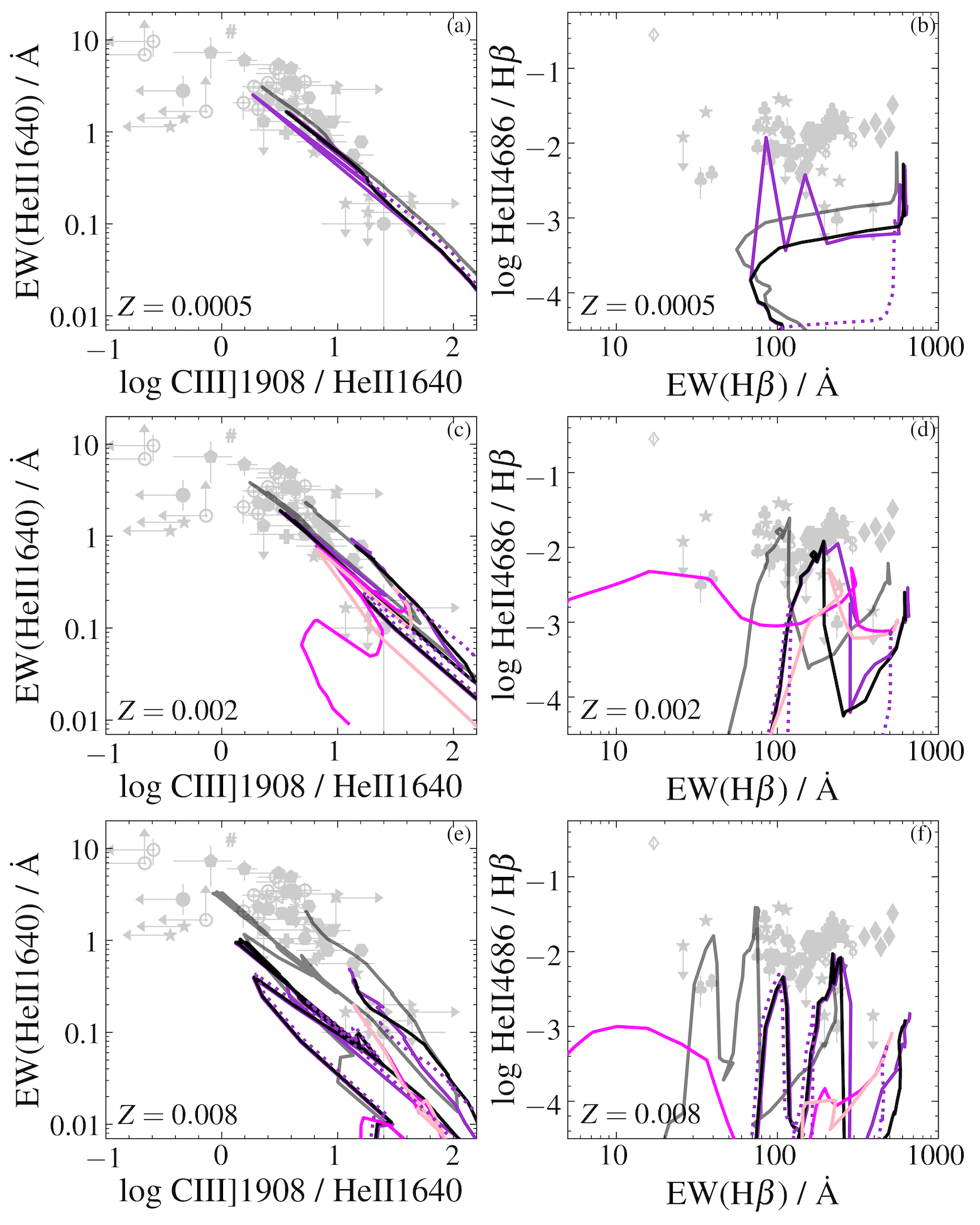}}
\end{center}
\caption{EW(\heii) plotted against \lciii/\heii\ (left; as in Fig.~\ref{fig:obs_uv}b) and \heiiopt/\hb\ plotted against EW(\hb) (right; as in Fig.~\ref{fig:obs_opt}e). The observations (greyed for clarity) are the same as in Figs.~\ref{fig:obs_uv}b and \ref{fig:obs_opt}e, while the models are the same as in Fig.~\ref{fig:HeII_ssp_age} (without the purple shading between models for $\mup=100$ and 600\,\Msun).}
\label{fig:HeII_ssp_obs}
\end{figure}

\begin{figure}
\begin{center}
\resizebox{\hsize}{!}{\includegraphics{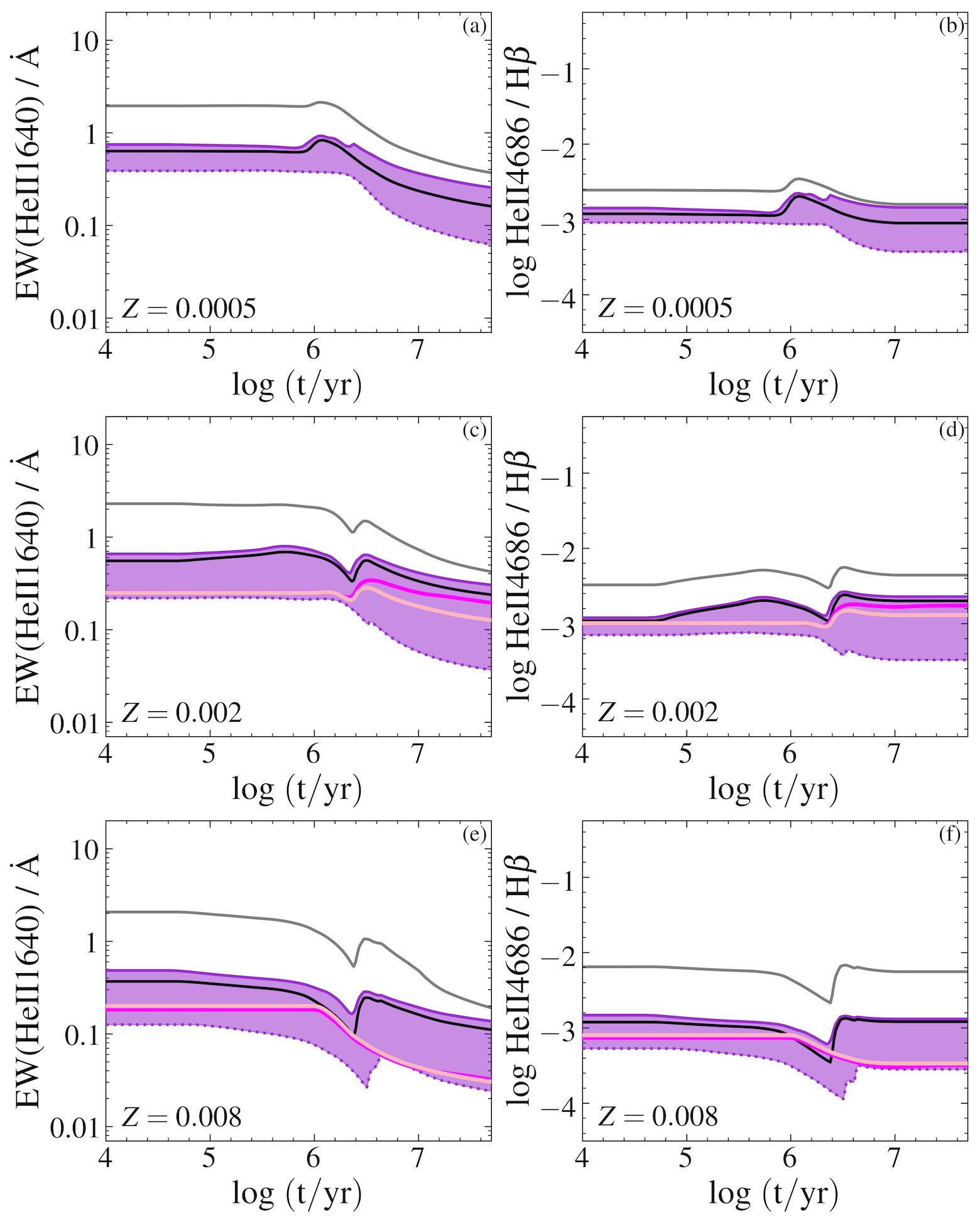}}
\end{center}
\caption{Same as Fig.~\ref{fig:HeII_ssp_age}, but for models with constant star formation rate.}
\label{fig:HeII_cst_age}
\end{figure}

Among the most challenging lines to reproduce in the spectra of metal-poor star-forming galaxies are the \lheii\ recombination lines, whose strength can be much stronger than predicted by standard models (Section~\ref{sec:intro}). The signatures of the wide collection of models considered in Section~\ref{sec:params} in spectral diagnostic diagrams involving \lheii\ lines therefore provide potentially useful hints on the sources powering the ionizing radiation in such galaxies. For example, we saw that, in ionization-bounded models, the equivalent widths of \heii\ and \hb\ and the \heiiopt/\hb\ ratio depend only moderately on ISM parameters other than \Uav, making these observables selectively sensitive to the source of ionizing radiation. Instead, the ratios of \lheii-to-metallic lines are also strongly affected by metallicity, the C/O ratio (in the case of \lciii/\heii\ and \lciv/\heii) and interstellar-line absorption (in the case of \lciv/\heii). With this in mind, we investigate below the extent to which stellar populations, AGN and radiative shocks can account for the emission-line signatures of metal-poor, star-forming galaxies. We also discuss X-ray binaries as potential sources of ionizing radiation in these galaxies.

\subsubsection{Stellar populations}\label{stelpops}

If stars are the main source of ionizing radiation in the galaxies of the reference sample of Section~\ref{sec:obs}, the equivalent widths of \heii\ and \hb\ and the \heiiopt/\hb\ ratio will depend sensitively on the upper mass cut-off of the IMF, \mup, the stellar population age, $t$, the stellar population model itself and \Uav\ (Figs~\ref{fig:baton3z_uv}a, \ref{fig:baton3z_opt}e, \ref{fig:baton3u_uv}a and \ref{fig:baton3u_opt}e). To further characterise this dependence, we show in Fig.~\ref{fig:HeII_ssp_age} the evolution of EW(\heii) and \heiiopt/\hb\ for ionization-bounded \hii\ regions powered by different types of SSPs, while Fig.~\ref{fig:HeII_ssp_obs} shows these models in the same panels as in Figs~\ref{fig:obs_uv}b and \ref{fig:obs_opt}e defined by EW(\heii), \ciii/\heii, \heiiopt/\hb\ and EW(\hb). The black curves in Figs~\ref{fig:HeII_ssp_age} and \ref{fig:HeII_ssp_obs} show SSP models with the same parameters as the benchmark models with $\log\Uav=-2$ of Section~\ref{sec:params}, for $Z=0.0005$ (top panels), 0.002 (middle panels) and 0.008 (bottom panels). In Fig.~\ref{fig:HeII_ssp_age}, these models show how, as $Z$ increases, the drop in effective temperature of massive stars on the early main sequence (for $\tprime\ll1\,$Myr), the rise in mass-loss rate (a $300\,\Msun$ star leaving the main sequence, around $\tprime\approx2\,$Myr, has lost 10, 25 and 70 per cent of its mass for $Z=0.0005$, 0.002 and 0.008, respectively) and the development of the WR phase (around $\tprime\approx3\,$Myr) shape the evolution of EW(\heii) and \heiiopt/\hb\ (see also Section~\ref{sec:params}). 

In Figs~\ref{fig:HeII_ssp_obs}a, \ref{fig:HeII_ssp_obs}c and \ref{fig:HeII_ssp_obs}e, the above models reach a region populated by galaxies with more extreme EW(\heii) and \lciii/\heii\ than could be attained by models with constant star formation rate in Fig.~\ref{fig:baton3z_uv}b. This is because, as Fig.~\ref{fig:HeII_cst_age} shows, continuous star formation smoothes out the evolution of the spectral features in Fig.~\ref{fig:HeII_ssp_age}. However, while the WR phase of the model with $Z=0.002$ hardly reaches the high observed $\heiiopt/\hb\approx0.01$ around $\mathrm{EW(\hb)}\approx200\,$\AA\ (Fig.~\ref{fig:HeII_ssp_obs}d), none of the reference SSP models can account for the extreme $\heiiopt/\hb\approx0.02$ of galaxies with $\mathrm{EW(\hb)}\approx500\,$\AA\ in the \citet{izotov17oct} sample (Figs~\ref{fig:HeII_ssp_obs}b, \ref{fig:HeII_ssp_obs}d and \ref{fig:HeII_ssp_obs}f). Increasing the ionization parameter from $\log\Uav=-2$ to $-1$ (grey curves) significantly boosts EW(\heii) and \heiiopt/\hb\ in Fig.~\ref{fig:HeII_ssp_age} (see Section~\ref{sec:params}), but this improves only moderately the discrepancy between observed and predicted \heiiopt/\hb\ for galaxies with $\mathrm{EW(\hb)}\approx500\,$\AA\ in Fig.~\ref{fig:HeII_ssp_obs}.

The light-purple and magenta curves in Figs~\ref{fig:HeII_ssp_age}--\ref{fig:HeII_cst_age} show the predictions of the \bpass\ single- and binary-star models, respectively, for SSPs with the same parameters as the \CB\ reference models, for $Z=0.002$ and 0.008 (\bpass\ models are not available for $Z=0.0005$). These models start at $\tprime=1\,$Myr, hence the flat evolution of EW(\heii) and \heiiopt/\hb\ at younger ages in Fig.~\ref{fig:HeII_ssp_obs}. From 1 to 3\,Myr, both \bpass\ models show qualitatively the same evolution as the \CB\ models, albeit with a weaker WR phase implying smaller EW(\heii)  and  \heiiopt/\hb. Then, as the \lheii\ emission dies off in the single-star \bpass\ and \CB\ models, the production of hard ionizing radiation is maintained through binary mass transfer in the binary-star \bpass\ model. The effect is particularly striking in the generation of strong \heiiopt/\hb\ at small EW(\hb) in Figs~\ref{fig:HeII_ssp_obs}d and \ref{fig:HeII_ssp_obs}f. However, this has no effect on the discrepancy between models and observations, which appears to be even more severe at larger EW(\hb) for the \bpass\ than for the \CB\ models. We note in passing that \citet{BPASSv21} used observations of \loiiiopt/\hb\ versus EW(\hb) (Fig.~\ref{fig:obs_opt}c above) to highlight the better performance of binary- versus single-stellar population models. As the brown (age) and magenta (\bpass) segments in Figs~\ref{fig:baton3z_opt}c and \ref{fig:baton3u_opt}c suggest, star-forming galaxies with low EW(\hb) and high \loiiiopt/\hb\ can be reached by both types of models for continuous star formation at ages $t\gg10\,$Myr.\footnote{In the version of Fig.~\ref{fig:obs_opt}c published by \citet[][fig.~38 in their paper]{BPASSv21}, only SSP models were presented, and the \hb\ equivalent widths from \citet{schenker13} were inadvertently corrected twice for redshift (J.~J. Eldridge, private communication).} In fact, we have checked for example that the model with $\log\Uav=-2$ and $Z=0.002$ in these figures reaches $\mathrm{EW(\hb)}\approx50$\,\AA\ (40\,\AA) after 1\,Gyr (2\,Gyr) of constant star formation, at nearly constant \loiiiopt/\hb.

The other parameter strongly affecting the ionizing radiation from young stellar populations is the upper mass cut-off of the IMF. The dotted and solid dark-purple lines in Figs~\ref{fig:HeII_ssp_age}--\ref{fig:HeII_cst_age} show SSP models with the same parameters as the \CB\ reference models, but for $\mup=100$ and 600\,\Msun, respectively. In Figs~\ref{fig:HeII_ssp_age} and \ref{fig:HeII_cst_age}, the area between these two models has been shaded in purple, for clarity. While raising \mup\ hardens the ionizing radiation, the effect is modest from $\mup=300$ and 600\,\Msun\ (Section~\ref{sec:params}). Fine-tuning the upper IMF therefore does not look promising to improve significantly the agreement between models and observations of EW(\heii), \lciii/\heii, \heiiopt/\hb\ and EW(\hb) in Fig.~\ref{fig:HeII_ssp_obs}. This is consistent with the conclusions reached by \citet{stanway19} based simply on the ratio of \lheii-to-\hi\ ionizing photons.

We note that models with LyC-photon leakage ($\fesc>0$) can reach larger ratios of \lheii-to-low ionization lines, such as \heii/\lciii\ (inverse abscissa of Figs~\ref{fig:baton3z_uv}d and \ref{fig:baton3u_uv}d), \heii/(\lciv+\lciii) (inverse abscissa of Figs~\ref{fig:baton3z_uv}e and \ref{fig:baton3u_uv}e), \heii/\loiii\ (Figs~\ref{fig:baton3z_uv}h and \ref{fig:baton3u_uv}h) and \heiiopt/\hb\ (Figs~\ref{fig:baton3z_opt}e and \ref{fig:baton3u_opt}e). However, such models fail to account simultaneously for the large equivalent widths of low-ionization lines observed in many galaxies (see, e.g., Figs~\ref{fig:baton3z_uv}h and \ref{fig:baton3u_uv}h for \lciii, and Figs~\ref{fig:baton3z_opt}e and \ref{fig:baton3u_opt}e for \hb). 

\subsubsection{AGN and radiative shocks}\label{agnshocks}

Figs~\ref{fig:baton3z_uv}--\ref{fig:baton3u_opt} show how introducing either an AGN or radiative-shock component allows models to reproduce observations of galaxies with high \lheii\ emission in nearly all ultraviolet and optical line-ratio diagrams. This is not surprising, given the strong \lheii\ emission produced by AGN and radiative shocks (Section~\ref{sec:params}), which has long made them good candidate sources of hard ionizing radiation in metal-poor star-forming galaxies (Section~\ref{sec:intro}). The novelty of Figs~\ref{fig:baton3z_uv}--\ref{fig:baton3u_opt} is to illustrate at once, and with a self-consistent modelling of interstellar abundances, the influence of these components on a wide range of ultraviolet and optical emission lines. Also, having assembled a substantial observational reference sample (Section~\ref{sec:obs}) allows us to highlight general trends and derive more robust conclusions than based on individual objects.

It is not obvious from Figs~\ref{fig:baton3z_uv}--\ref{fig:baton3u_opt} which of an AGN or radiative-shock component can best account for the properties of metal-poor star-forming galaxies with strong \lheii\ emission. \citet{izotov12} find that the production of high-ionization \nevopt\ emission ($E_\mathrm{ion}>97.2$\,eV) in the spectra of 8 blue compact dwarf galaxies with $\logoh=7.3$--7.7 and strong \lheii\ emission ($\heiiopt/\hb\gtrsim0.01$) requires a contribution of about 10 per cent of the total ionizing radiation by AGN or radiative shocks. While these authors favour supernova-driven radiative shocks with velocities around 300--500\,km\,s$^{-1}$ as the source of this emission, they cannot rule out an AGN origin. \citet{stasinska15} also note that shocks can naturally account for the high \oiopt/\loiiiopt\ ratios observed in the spectra of blue compact dwarf galaxies with high \loiiiopt/\loiiopt\ ratios, as density-bounded models producing high \loiiiopt/\loiiopt\ would imply low \loiopt/\loiiiopt\ (see also Section~\ref{sec:fesc}).

\begin{figure}
\begin{center}
\resizebox{\hsize}{!}{\includegraphics{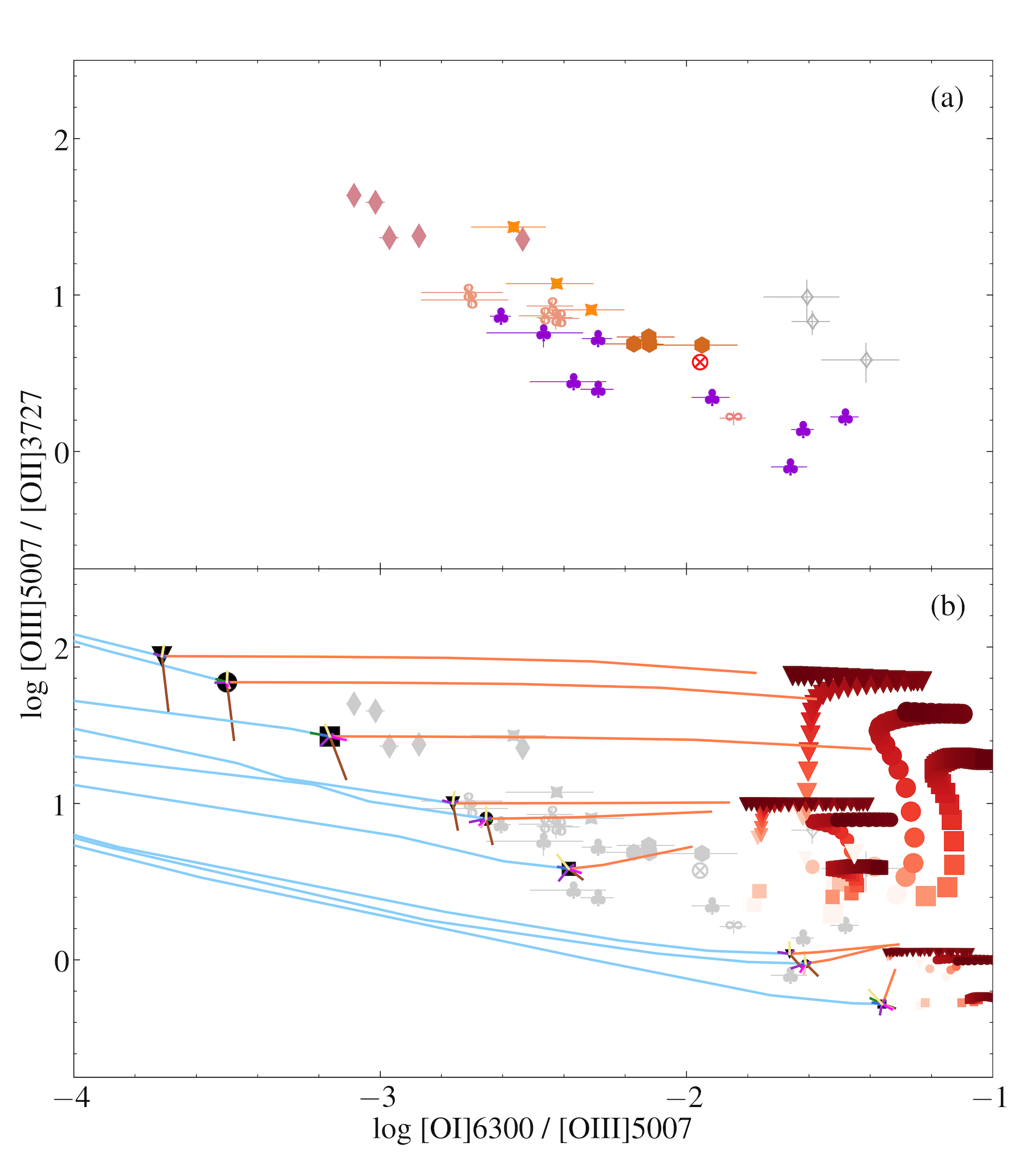}}
\end{center}
\caption{\loiiiopt/\loiiopt\ plotted against \loiopt/\loiiiopt\ for: (a) the galaxies in the reference sample of Section~\ref{sec:obs} (available in practice only for the subsample of LyC leakers and two AGN); and (b) the complete set of models from Figs~\ref{fig:baton3z_uv}--\ref{fig:baton3u_opt}, along with observations from panel (a) greyed for clarity.}
\label{fig:OI}
\end{figure}

To further investigate this issue, in Fig.~\ref{fig:OI}, we plot \loiiiopt/\loiiopt\ versus \loiopt/\loiiiopt\ for the complete set of models from Figs~\ref{fig:baton3z_uv}--\ref{fig:baton3u_opt}, along with observations from the sample of Section~\ref{sec:obs} -- available in practice only for the subsample of LyC leakers \citep{leitet11, jaskot13, izotov16oct, izotov16jan, izotov17oct, izotov18mar, izotov18aug, chisholm17}, a few Wolf-Rayet galaxies \citep{lopezsanchez08} and three AGN \citep{dors14}. As in other line-ratio diagrams, contributions by AGN and radiative shocks to the ionizing radiation of model galaxies have roughly similar signatures in Fig.~\ref{fig:OI}, increasing \loiopt/\loiiiopt\ typically far more than \loiiiopt/\loiiopt\ (except for very low ionization parameter). Hence, these line ratios cannot either help discriminate at first glance between AGN and shock ionization in a galaxy. A more striking feature of Fig.~\ref{fig:OI} is that nearly all observations of (confirmed and candidate) LyC leakers exhibit higher \loiopt/\loiiiopt\ than the benchmark ionization-bounded models in the full explored ranges of $-3\leq\log\Uav\leq-1$ and $0.0005\leq Z\leq0.008$ at fixed \loiiiopt/\loiiopt. Tuning the stellar population parameters, including the star formation history, can bring the models only slightly closer to the data. In density-bounded models, the ratio of low- to high-ionization lines further drops (Section~\ref{sec:fesc}), worsening the agreement between models and observations (light-blue segments in Fig.~\ref{fig:OI}). The only way to account for the observed properties of LyC leakers in Fig.~\ref{fig:OI} is to invoke a significant contribution by an AGN or radiative shocks (or X-ray binaries, but see Section~\ref{sec:xrb} below) to the ionizing radiation. This is because the hard penetrating X-ray and extreme-ultraviolet radiation from such sources produces higher electronic temperatures than stellar radiation in the outskirts of \hii\ regions, thereby enhancing \loi\ collisional excitation (we note that, in young shocks which have not yet developed a cool tail, \loiopt/\loiiiopt\ can be significantly reduced and \lheii/\hb\ slighly enhanced relative to the models shown in Fig.~\ref{fig:OI}; see \citealt{3MdBs19}). This can arise in the context of both density-bounded and ionization-bounded (in the picket-fence leakage scenario; see Section~\ref{sec:fesc}) models. This conclusion is consistent with that drawn by \citet{stasinska15} from the analysis of a sample of blue compact dwarf galaxies with very high excitation. 

\begin{figure}
\begin{center}
\resizebox{\hsize}{!}{\includegraphics{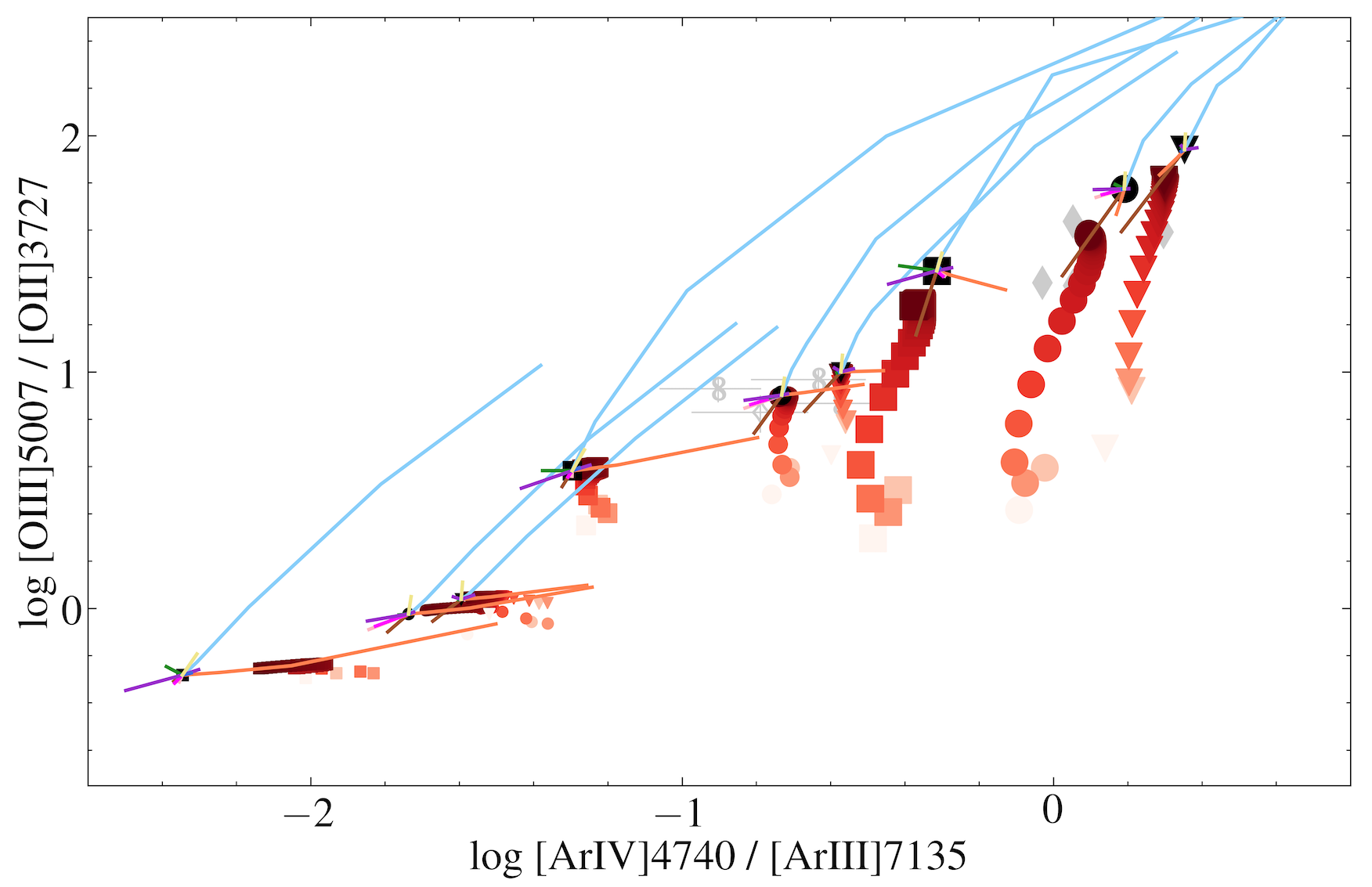}}
\end{center}
\caption{Same as Fig.~\ref{fig:OI}b, but for \loiiiopt/\loiiopt\ plotted against \larivopt/\lariiiopt.}
\label{fig:Ar}
\end{figure}

\citet{stasinska15} also pointed out the interest of the \ariiiopt\ and \arivopt\ (hereafter simply \lariiiopt\ and \larivopt) lines to probe ionizing-photon energies greater than the ionization potential of Ar$^{2+}$ (40.7\,eV), which lies between the ionization potentials of O$^+$ (35.1\,eV) and He$^+$ (54.4\,eV). In Fig.~\ref{fig:Ar}, we plot \loiiiopt/\loiiopt\ against \larivopt/\lariiiopt\ for the same models and observations as in Fig.~\ref{fig:OI}; in practice, data are available only for a few candidate LyC leakers \citep{jaskot13,izotov17oct} and an AGN \citep{dors14}. The benchmark ionization-bounded models appear to overlap with the data in this diagram, as do density-bounded models, eventually combined with an AGN or radiative-shock component. Along with this smaller dispersion of models relative to Fig.~\ref{fig:OI}, is worth noting that, in Fig.~\ref{fig:Ar}, only models with a radiative-shock component can reach $\larivopt/\lariiiopt\sim1$ around $\loiiiopt/\loiiopt\sim10$, where some extreme-excitation galaxies can be found in the \citet{stasinska15} sample (see their fig.~6).

Radiative shocks from expanding \hii\ regions and supernova blast waves are an appealing natural hypothesis for the origin of hard ionizing radiation in actively star-forming, metal-poor galaxies \citep[e.g.,][]{Thuan05,stasinska15}. Fig.~\ref{fig:OI} supports the idea that shocks may be intimately related to the leakage of ionizing photons from such galaxies. Interestingly, the presence of shocks will increase primarily the luminosities of \lheii\ and very-high-ionization lines, such as \nevopt\ (but not so much \nv, as N$^{4+}$ is converted into N$^{5+}$; see Section~\ref{sec:params}), while the luminosities of lower-ionization lines (including \hb) remain largely controlled by stellar radiation. In fact, \citet{izotov12} find no significant correlation between \lnev\ and \hb\ emission in the 8 galaxies of their sample. In this context, the absence of correlation between \lheii\ and \hb\ emission in the sample of extremely metal-poor galaxies studied by \citet[][see also \citealt{senchyna19}]{SenchynaStark19} could be consistent with a radiative-shock origin of the \lheii\ emission.

AGN and radiative-shock components are sometimes discarded as sources of hard ionizing radiation on the basis of spectral fits. For example, \citet{berg18} conclude that an AGN or radiative-shock component is unlikely to account for the strong \heii\ emission in the extreme star-forming galaxy SL2SJ021737-051329, as this would make \lciii/\loiii\ too small and \lciv/\lciii\ too large, based on AGN models by \citet{groves04} and shock models by \citet[][see Table~\ref{tab:analogs}]{allen08}. While the \hii-region, AGN and shock models used by \citet{berg18} were computed using different ISM prescriptions, in the framework of our models, as can be guessed from Figs~\ref{fig:baton3z_uv}--\ref{fig:baton3u_opt}, a combination of $\log\Uav\lesssim-2$, $\CO\gtrsim0.17$, $\xid\approx0.1$ and an AGN (or radiative-shock) contribution of $\sim$8 per cent of the total \hb\ emission turns out to accommodate the observed ultraviolet and optical nebular spectrum of this galaxy (see Figs~\ref{fig:obs_uv} and \ref{fig:obs_opt} to locate the galaxy in all panels, the oxygen abundance corresponding to a metallicity between $Z=0.0005$ and 0.002).\footnote{A pure SSP of age $\tprime\approx2.5\,$Myr with the same \Uav, \CO\ and $\xid$ as this composite model can also approximate closely all observations of SL2SJ021737-051329 in Figs~\ref{fig:baton3z_uv}--\ref{fig:baton3u_opt}, except for the \hb\ equivalent width [$\mathrm{EW(\hb)}\approx200\,$\AA\ instead of the observed 517\,\AA]. We consider this model less likely because of the very specific age required.} Hence, in some cases, the assessment of the potential presence of an AGN or radiative-shock component in a galaxy may depend on the adopted model prescription, highlighting once more the importance of a physically-consistent modelling of nebular emission from different sources (Section~\ref{sec:models}). We also recall that the AGN models presented in this paper were computed for a typical ionizing-spectrum slope $\alpha=-1.7$ (Section~\ref{sec:agn}), and that $\alpha$ variations could imply significant dispersion in predicted ultraviolet and optical line ratios \citep[e.g.,][]{Feltre2016}.

It is worth noting that while a radiative-shock or AGN component can readily accommodate the emission-line properties of many observed galaxies with strong \heii\ emission in Figs~\ref{fig:baton3z_uv}--\ref{fig:baton3u_opt}, including those with weak \lciv/\lciii\ (see above; there is also the possibility for \lciv\ emission to be reduced via interstellar absorption; Section~\ref{sec:params}), some outlier galaxies exhibit properties not sampled by the limited set of models presented here. We have checked that some models can account for the properties of such galaxies. For example, we find that galaxies with $\mathrm{EW(\heii)}\gtrsim2\,$\AA\ and $\lciii/\heii\gtrsim4$ (Figs~\ref{fig:baton3z_uv}b and \ref{fig:baton3u_uv}b) can be reached by models with $\CO\gtrsim0.17$ and a radiative-shock or AGN component. In Figs~\ref{fig:baton3z_uv}c and \ref{fig:baton3u_uv}c, the observed $\mathrm{EW(\lciv)}\gtrsim20\,$\AA\ and $\lciv/\heii\approx10$ of the lensed double-super star cluster ID14 \citep[][whose properties approach those of the \hii\ galaxy of \citealt{fosbury03}]{vanzella17} can be accommodated by young ($t\sim1\,$Myr), high-ionization ($\log\Uav\sim-1\,$) models with $\CO\gtrsim0.17$ and $\xid\lesssim0.3$, also compatible with the other emission-line properties of this object. Very young models with $\CO\gtrsim0.17$ can also reach galaxies with high EW(\lciii) at small \lciv/\lciii\ in Figs~\ref{fig:baton3z_uv}g and \ref{fig:baton3u_uv}g, while the \citet{laporte17} galaxy, with low \lciii/\heii\ and high \lnv/\heii, could well be a LyC-photon leaker powered by radiative shocks or an AGN (Figs~\ref{fig:baton3z_uv}g and \ref{fig:baton3u_uv}g). The above rough exploration of the parameter space will need to be refined by more robust spectral fits of each galaxy in the sample, using tools such as \beagle\ \citep{Chevallard2016}, extended to incorporate AGN and radiative-shock prescriptions.

\subsubsection{X-ray binaries}\label{sec:xrb}

X-ray binaries, in which a compact object (neutron star or stellar-mass black hole) accretes material from a massive O/B companion, have been proposed as natural sources of hard ionizing photons in metal-poor star-forming galaxies \citep[e.g.,][]{Garnett91}. An argument supporting this hypothesis is the observed increase in hard X-ray luminosity with decreasing oxygen abundance (at a fixed star formation rate) in nearby metal-poor star-forming galaxies \citep[][and references therein]{Brorby16}, which goes in the same sense as the increase in EW(\heii) (Fig.~\ref{fig:obs_uv}a) and \heiiopt/\hb\ (Fig.~\ref{fig:obs_opt}f). Also, the non-correlation of the equivalent width of \heiiopt\ with that of \hb\ and other emission lines in the sample of extremely metal-poor galaxies studied by \citet{SenchynaStark19} suggests that He$^+$-ionizing photons are produced by sources with timescales greater than massive O/B stars, such as stripped stars produced by close binary evolution and X-ray binaries. The accretion physics of X-ray binaries presents similarities to that of AGN \citep[see the review by][]{Gilfanov14}, which are in fact often considered as scaled-up versions of X-ray binaries \citep[e.g.,][]{McHardy06}. Hence,  X-ray binaries are expected to produce ionizing spectra similar to those of AGN \citep[see also fig.~C5 of][]{stasinska15}, implying effects on emission-line ratios and equivalent widths similar to those found for an AGN component in Figs~\ref{fig:baton3z_uv}--\ref{fig:baton3u_opt}.

 Recently, \citet{schaerer19}  computed the time evolution of \heiiopt/\hb\ and EW(\hb) for SSPs including X-ray binaries at different metallicities, by combining the \citet[][see also \citealt{Madau17}]{Fragos13} stellar population synthesis models of X-ray binaries with \bpass~v2.1, and adopting an approximate conversion between X-ray luminosity and rate of He$^+$-ionizing photons. This model reproduces roughly the trend of increasing \heiiopt/\hb\ with decreasing \logoh\ in nearby metal-poor star-forming galaxies \citep[fig.~1 of][]{schaerer19}. However, it fails to account for the high \heiiopt/\hb\ ratios of galaxies with large EW(\hb), just as the other stellar population synthesis models considered in Fig.~\ref{fig:HeII_ssp_obs} above \citep[see fig.~3 of][]{schaerer19}. This is consistent with the finding that X-ray binaries have spectra too soft to account for the very hard ionizing radiation of some metal-poor star-forming galaxies \citep[e.g.,][]{Thuan05,izotov12,stasinska15}, that they appear on too-long timescales to account for the emission-line properties of Green-Pea galaxies \citep{jaskot13} and with the stringent observational upper limit from {\it Chandra} on the presence of X-ray binaries in the most extreme \lheii-emitter observed by \citet{senchyna17}. We conclude that, while X-ray binaries may provide a natural source of hard ionizing photons in metal-poor star-forming galaxies, they cannot account for the entire emission observed in the most extreme, highest-ionization cases. 

\subsection{Diagnostics of LyC-photon leakage}\label{sec:lycfesc}

We now focus on the models of density-bounded \hii\ regions in Figs~\ref{fig:baton3z_uv}--\ref{fig:baton3u_opt}, to assess whether the emission-line properties of metal-poor star-forming galaxies can provide useful constraints on the fraction of escaping LyC photons, \fesc. As seen in Section~\ref{sec:fesc} (and references therein), increasing \fesc\ removes the outer low-ionization zones of \hii\ regions, making ratios of high- to low-ionization lines (e.g. \loiiiopt/\loiiopt) rise and the equivalent widths of lines with low ionization potential (e.g. \lciii) drop. The interpretation of these signatures in galaxy spectra is unfortunately complicated by the competing effects of other galaxy physical parameters, in particular the nature of the ionizing source, the ionization parameter, \Uav, metallicity, $Z$, and to a lesser extent the gas density, \nh, dust-to-metal mass ratio, \xid, and \CO\ ratio \citep[Figs~\ref{fig:baton3z_uv}--\ref{fig:baton3u_opt}; see also][]{jaskot13, nakajimaetouchi14, stasinska15, jaskot16, izotov17oct}. These degeneracies between the spectral signatures of \fesc\ and other parameters are the reason why LyC leakers appear to overlap with the rest of the population of actively star-forming galaxies in Figs~\ref{fig:obs_uv} and \ref{fig:obs_opt} (see Section~\ref{sec:obsprop}).

Several diagnostics must therefore be combined to potentially discriminate the effects of \fesc\ from those of other parameters on emission-line ratios. That  \loiiiopt/\loiiopt\ alone is not a sufficient condition for LyC leakage is also illustrated by the fact that, as seen in  Section~\ref{sec:fesc} (Fig.~\ref{fig:convol}f), this ratio for a density-bounded galaxy with constant star formation can actually be smaller than that of an ionization-bounded one for large \fesc\ and \Uav\ (see age effect on model with $\log\Uav=-1$ in Fig.~\ref{fig:baton3u_opt}b). \citet{jaskot16} suggest that, for example, high \loiiiopt/\loiiopt\ ($\gtrsim10$) and low EW(\lciii) ($\lesssim4\,$\AA) will tend to select density-bounded galaxies, although they do acknowledge that the scaling of \fesc\ with EW(\lciii) will depend sensitively on metal abundances and stellar population age, as Figs~\ref{fig:baton3z_uv}g and \ref{fig:baton3z_opt}b show. We note in this context that the \lhei-based \fesc\ diagnostic proposed by \citet[][see Figs~\ref{fig:baton3z_opt}d and \ref{fig:baton3u_opt}d]{izotov17oct} requires independent constraints on \nh, \Uav\ and $Z$. In practice, Figs~\ref{fig:baton3z_uv}--\ref{fig:baton3u_opt} reveal that few observations fall in regions of diagrams populated purely by density-bounded models (in Figs~\ref{fig:HeII_ssp_obs}d and \ref{fig:HeII_ssp_obs}f, galaxies with low \hb\ equivalent width and high \loiiiopt/\hb\ can be accounted for by ionization-bounded models with ages greater than 10\,Myr; see Section~\ref{stelpops}).

It is also interesting to note that, for the low-mass star-forming galaxy BX418 with low $\mathrm{EW(\hb)}\approx44\,$\AA\ and high $\loiiiopt/\loiiopt>26$ (using the 1$\sigma$ limit on the \loiiopt\ flux), \citet{erb10} constrain an age less than 100\,Myr from ultraviolet and \ha\ observations as well as dynamical arguments. This young age, despite the location of BX418 at low EW(\hb) and high \loiiiopt/\hb\ in Figs~\ref{fig:baton3z_opt}c and \ref{fig:baton3u_opt}c, is suggestive of the fact that the galaxy might be leaking LyC photons, which would be compatible with the other properties of the galaxy in Figs~\ref{fig:baton3z_uv}--\ref{fig:baton3u_opt} .\footnote{An ionization-bounded model with $\log\Uav=-1$ can reach $\mathrm{EW(\hb)}\approx44\,$\AA\ after about 100\,Myr of constant star formation, although the corresponding \heii/\loiii\ is too small relative to the observed one in Fig.~\ref{fig:baton3u_uv}h (which pertains to the 25-per-cent nebular contribution to the total \lheii\ emission of this object; see \citealt{erb10}).)}  In comparison, the confirmed, per-cent level LyC leakers Haro~11, Tol-0440-381 and Tol-1247-232 \citep[][see Table~\ref{tab:leakers}]{leitet11,chisholm17} also exhibit somewhat low $\mathrm{EW(\hb)}\sim40$--100\,\AA\ and high \loiiiopt/\hb\ in these figures, but with more modest \loiiiopt/\loiiopt\ around 1.5--4.0 (Figs~\ref{fig:baton3z_opt}b and \ref{fig:baton3u_opt}b), consistent with a picket-fence leakage scenario \citep[Section~\ref{sec:fesc}; see also][]{leitet11}, in addition to density-bounded \hii\ regions.

Hence, assessing whether a galaxy is leaking LyC photons based on the emission-line diagrams in Figs~\ref{fig:baton3z_uv}--\ref{fig:baton3u_opt} is not  straightforward at first glance. Several diagnostics must be examined simultaneously to discriminate the signatures of \fesc\ from those of other physical parameters, which can be best achieved with a full spectral analysis tool incorporating density-bounded models. 

\section{Conclusions}\label{sec:conclu}

We have explored the constraints on the production and escape of ionizing photons in young galaxies by investigating the ultraviolet and optical emission-line properties of a broad collection of models relative to the observations of a reference sample of metal-poor star-forming galaxies and LyC leakers at various redshifts. A main feature of our study is the adoption of models of \hii\ regions, AGN narrow-line regions and radiative shocks computed all using the same physically-consistent description of element abundances and depletion on to dust grains down to metallicities of a few per cent of solar \citep[from][]{gutkin2016}. We computed ionizing spectra of single- and binary-star populations using the most recent versions of the \citet{Bruzual2003} and \citet{BPASSv21} stellar population synthesis codes and explored models of ionization-bounded as well as density-bounded (i.e., optically thin to LyC photons) \hii\ regions. To compute emission-line spectra of AGN narrow-line regions, we appealed to an updated version of the \citet{Feltre2016} models, while for radiative shocks we adopted the recent computations  of \citet{3MdBs19}. 

The observational sample assembled to constrain these models incorporates data from 13 subsamples of metal-poor star-forming galaxies (Table~\ref{tab:analogs}), 9 subsamples of confirmed and candidate LyC leakers (Table~ \ref{tab:leakers}), as well as a few more quiescent star-forming galaxies and AGN at redshifts out to $z=7.1$. The combined sample of closest known analogues to reionization-era galaxies in Tables~\ref{tab:analogs} and \ref{tab:leakers} allows the simultaneous exploration of diagnostic diagrams involving the \nv, \civd, \heii, \oiiid, \ciiid, \oiid, \heiopta, \heiiopt, \hb, \oiiiopt, \ha, \niiopt, \heioptb\ and \heioptc\ emission lines, of which only a few are typically available at once for individual subsamples. This sample shows that, overall, metal-poor star-forming galaxies in wide ranges of redshift populate similar regions of the diagrams \citep[but see][]{senchyna19}, while LyC leakers tend to overlap with the most extreme star-forming galaxies (Figs~\ref{fig:obs_uv} and \ref{fig:obs_opt}).

In agreement with many previous studies, we find that current single- and binary-star population synthesis models do not produce hard-enough ionizing radiation to account for the strong \lheii\ emission observed in the most metal-poor star-forming galaxies, even when tuning the IMF. Interestingly, the updated \CB\ version of the \citet{Bruzual2003} single-star models used here, which differs from that described by \citet[][see also \citealt{vidal17}]{gutkin2016} in the inclusion of updated spectra for hot massive stars, produces altogether more \lheii-ionizing radiation than the binary-star \bpass\,v2.2.1 models, providing slightly better agreement with the observations (Section~\ref{stelpops} and Figs~\ref{fig:HeII_ssp_age}--\ref{fig:HeII_cst_age}). Since a majority of massive stars are expected to undergo binary interactions \citep[e.g.,][]{Sana12}, we consider the \lheii\ luminosity predicted by the single-star \CB\ models as a lower limit, which binary-star models (currently under development) will likely exceed. Also, for completeness, since the \oivfir\ line is often discussed in the same context as the \lheii\ line \citep[e.g.][]{SchaerStas99}, we checked that the \oivfir/\oiiiopt\ ratio in our models behaves similarly to the \heiiopt/\hb\ ratio (this is even more true for the \oivfir/\oiiifir\ ratio, which is less sensitive to electronic temperature).

Introducing hard ionizing radiation from either an AGN or radiative-shock component allows models to overlap with observations of galaxies with high \lheii\ emission in nearly all the ultraviolet and optical line-ratio diagrams we investigated. On an object-by-object basis, we find that the conclusion drawn about the potential presence of such ionizing sources using our models can differ from those derived previously using libraries of AGN and radiative-shock models computed with inconsistent descriptions of element abundances. Both AGN and radiative-shock components have very similar signatures in all diagrams, which prevents a simple discrimination between the two at first glance. Similarly, no diagram provides a simple discrimination between LyC-leaking and ionization-bounded galaxies, because of degeneracies in the signatures of \fesc\ and other galaxy physical parameters. This is the case also in the \oiiiopt/\oiiopt\ versus \oiopt/\oiiiopt\ diagram, in which all observations of (confirmed and candidate) LyC leakers exhibit higher \loiopt/\loiiiopt\ than benchmark ionization-bounded models. This is surprising, because density-bounded models produce lower \loiopt/\loiiiopt\ than ionization-bounded ones at fixed \loiiiopt/\loiiopt\ (Section~\ref{sec:fesc} and Fig.~\ref{fig:OI}; see also \citealt{stasinska15}). The only way to account for the observed properties of LyC leakers in this diagram is to invoke a systematic significant contribution by a source of hard ionizing radiation. 

Another potential source of hard ionizing radiation is X-ray binaries, the predicted growing importance of these systems toward low metallicity being supported by the observed increase in hard X-ray luminosity with decreasing oxygen abundance (at fixed star formation rate) in nearby metal-poor star-forming galaxies \citep{Fragos13,Brorby16}. Adopting an approximate conversion of X-ray luminosity into rate of He$^+$-ionizing photons allows one to reproduce roughly the observed rise in \heiiopt/\hb\ ratio with decreasing oxygen abundance in such galaxies \citep{schaerer19}. However, like other stellar population synthesis models, this fails to account for the high observed \heiiopt/\hb\ ratios of galaxies with large EW(\hb). A source of harder ionizing radiation must be invoked in these extreme objects, such as an AGN or radiative-shock component. 

So far, no predictive model has been proposed to link shocks to other galaxy properties and account for, notably, the increase in \lheii-emission strength with decreasing metallicity. Potential avenues to be explored might be an IMF bias toward massive stars at low metallicities \citep[e.g.,][]{Marks12} or the higher specific star formation rates of metal-poor dwarf galaxies relative to their more metal-rich, massive counterparts \citep[e.g.,][]{Kauffmann06,Yates12}. Both effects would tend to enhance the incidence of radiative shocks from massive stars and supernova blast waves in metal-poor relative to metal-rich galaxies. We also note that gas compression associated with radiative shocks will generate high densities \citep[e.g.,][]{allen08}. In this context, the high gas densities ($\nh\gtrsim10^4$\,cm$^{-3}$) measured from the \ciiid\ doublet in some distant, low-metallicity, actively star-forming galaxies \citep[e.g.,][]{Maseda17,James18} could be suggestive of the presence of radiative shocks. The possibility that fast radiative shocks provide the hard radiation necessary to power strong \lheii\ emission in metal-poor star-forming galaxies may be tested using high-quality observations of nearby galaxies. In a related paper (Chevallard et al., in preparation), we appeal to spatially-resolved observations of the extremely metal-poor compact dwarf galaxy SBS0335-052E to quantify the relative contributions from supernova-driven radiative shocks and massive stars to the total \lheii-ionizing emission from this galaxy.

While the ultraviolet and optical emission-line diagrams of Figs~\ref{fig:obs_uv} and \ref{fig:obs_opt} do not allow simple by-eye diagnostics of the nature of ionizing sources and the escape of LyC photons in metal-poor star-forming galaxies, differences exist in the spectral signatures of these parameters, which should enable more stringent constraints from simultaneous fits of several lines. This can be best achieved in a Bayesian framework using versatile spectral analysis tools incorporating a physically-consistent description of the sources and transfer of radiation in a galaxy, such as the \beagle\ tool \citep{Chevallard2016}. Although this tool was already shown to reproduce remarkably well the fluxes of 20 ultraviolet and optical (not including \lheii) emission lines in 10 extreme nearby star-forming regions \citep{Chevallard18}, the current version of the code does not incorporate models for density-bounded \hii-regions, narrow-line regions of AGN and radiative shocks. The implementation of these components, in progress, should enable valuable constraints on the production and escape of ionizing radiation from the emission-line spectra of metal-poor star-forming galaxies, and soon of reionization-era galaxies observed by \JWST.


\section*{Acknowledgements}

We are grateful to D.~Erb, M.~Hirschmann, P.~Petitjean, P.~Senchyna, D.~Stark and A.~Wofford for helpful discussions. We also thank M.~Mignoli for providing us with line-flux measurement in the average spectrum of type-2 AGN from \citet{Mignoli19}.  AF, SC, GB, AF and AVG acknowledge financial support from the European Research Council (ERC) via an Advanced Grant under grant agreement no. 321323--NEOGAL. AF acknowledges support from the ERC via an Advanced Grant under grant agreement no. 339659-MUSICOS.  GB acknowledges financial support from DGAPA-UNAM through PAPIIT project IG100319. CM acknowledges financial support through grant CONACyT-CB2015-254132.



\bibliographystyle{mnras}
\bibliography{biblio} 


\appendix

\begin{table*}
\begin{threeparttable}
\caption{PoWR models used in the present paper.\tnote{\it a}}
\centering
\begin{tabular}{llccccl}
\hline
Metallicity  		&    Type   & $\log (\Tstar/\mathrm{K})$\tnote{\it b} & $\log(\Rt/\Rsun)$\tnote{\it c}    & $\vinf/\mathrm{km\,s}^{-1}$    &    $D$       &     PARSEC-track metallicity       \\
\hline
Galaxy     			& WNL-H20\tnote{\it d}  \ \ \    & [4.40\,,\,5.00]  & [0.0\,,\,1.7] & 1000 &  4  &               \\
($Z\approx0.014$) 		& WNE                              & [4.50\,,\,5.30]  & [0.0\,,\,1.7] & 1600 &  4 & $Z \ge 0.014$ \\
                       		& WC                                &\ \ \  [4.60\,,\,5.30]  \ \ \ & [$-0.5$\,,\,+1.6]   & 2000 & 10 &               \\
                       		& WO                                &   5.30              &   0.3          & 1000 &  4  &               \\
\hline
LMC   			& WNL-H20\tnote{\it d}  & [4.40\,,\,5.05] & [0.0\,,\,1.9] & 1000 & 10 &                         \\
($Z\approx0.006$)      	& WNE                           & [4.60\,,\,5.25] & [$-0.2$\,,\,+1.9]   & 1600 & 10 & $0.006 \le Z \le 0.010$ \\
                         		& WC                             & [4.65\,,\,5.30] &  [$-0.8$\,,\,+1.6]    & 2000 & 10 &                         \\
\hline
SMC  			& WNL-H20\tnote{\it d}   & [4.60\,,\,5.05] &  [0.6\,,\,1.7] & 1600 & 10 &                         \\
($Z\approx0.003$)    & WNE                          & [4.60\,,\,5.15] &  [0.2\,,\,1.7] & 1600 & 10 & $0.002 \le Z \le 0.004$ \\
\hline
SubSMC 			& WNL-H20\tnote{\it d}  & [4.50\,,\,5.05] &  [0.6\,,\,1.9] & 1600 & 10 &                         \\
($Z\approx0.001$)   & WNE                           & [4.50\,,\,5.10] &  [0.0\,,\,1.9] & 1600 & 10 & $ Z \le 0.001$          \\
\hline
\end{tabular}
\label{tab:wrmodels}
\begin{tablenotes}
\item [{\it a}]  Available from \url{http://www.astro.physik.uni-potsdam.de/~PoWR}.
\item [{\it b}]  Models available in steps of $\Delta(\log \Tstar)=0.05$ over the quoted range. The effective temperature \Tstar\ is defined at the stellar radius $R_\star$ (corresponding to a radial Rosseland continuum optical depth of 20).
\item [{\it c}]  Models available in steps of $\Delta(\log \Rt)=0.10$ over the quoted range. 
\item [{\it d}]  Adopted for all WNL stars in the models (conservatively identified as stars with $T_\star>25,000\,$K and H-surface abundance by mass $X<0.3$ at all metallicities in the PARSEC tracks).
\end{tablenotes}
\end{threeparttable}
\end{table*}
\section{Library of WR-star spectra}\label{app:wrmodels}

The \CB\ stellar population model used in this paper incorporates stellar evolutionary tracks computed with the PARSEC code of \citet{Bressan2012} for stars with initial masses up to 600\,\Msun, which include the evolution through the WR phase \citep[][see also \citealt{gutkin2016}]{Chen2015}. To compute the emission from WR stars, we appeal to the library of high-resolution synthetic spectra from the Potsdam Wolf-Rayet (PoWR) group \citep{Graefener2002, Hamann2003, Hamann2004, Sander2012, Hainich2014, Hainich2015, Sander2015, Todt2015}. These line-blanketed, non-LTE, spherically expanding models are available for WNE-, WNL-, WC- and WO-type stars in various ranges of stellar effective temperatures, \Tstar, and `transformed radii', \Rt, at 4 metallicities, $Z\approx0.001$, 0.003, 0.006 and 0.014 (see Table~\ref{tab:wrmodels}). The transformed radius is a convenient, luminosity-independent parametrization of the stellar wind in terms of the stellar radius, $R_\star$, terminal wind velocity, \vinf, mass-loss rate, $\dot{M}$, and `density contrast', $D$ (defined as the factor by which the density in the clumps of an inhomogeneous wind is enhanced relative to that of a homogeneous wind of same $\dot{M}$), through the formula  
\begin{equation}
\Rt= R_\star\left(\frac{\vinf}{2500\,\mathrm{km\,s}^{-1}}\Bigg/\frac{\dot{M}\sqrt{D}}{10^{-4}\Msun\,\mathrm{yr}^{-1}}\right)^{2/3}\,.
\label{eq:Rt}
\end{equation}
At fixed \Rt, models with same chemical composition and effective temperature \Tstar\ have similar emission-line equivalent widths \citep[see, e.g.,][]{Schmutz1989,Hamann2004}.

In the \CB\ model, the quantities $R_\star$, \Tstar\ and $\dot{M}$ for a WR star of metallicity $Z$ are taken from the PARSEC tracks, and the spectrum assigned to that star is taken to be the PoWR model with closest  \Tstar, $\dot{M}$ and $Z$, as indicated in Table~\ref{tab:wrmodels} (rightmost column). Each PoWR model consists of a high-resolution spectrum (resolving power of $\sim10,000$) over the wavelength range from 200\,\AA\ to 8$\mu$m and a low-resolution spectrum (resolving power $\lesssim300$) covering from 5 to 200\,\AA. The rates of \hi-, \lhei- and \lheii-ionizing photons (with wavelengths  $\lambda \le911.33$, 504.3 and 227.85\,\AA, respectively) integrated from the combined spectra do not always match the reference values (computed in an independent way) tabulated for each model on the PoWR website, because of numerical errors in the spectra. In particular, for \lheii-ionizing photons, whose rate is typically several orders of magnitude lower than those of \lhei- and \hi-ionizing photons, the tabulated reference value can often be zero, while spectral integration gives non-zero values.  Setting the \lheii-ionizing flux to zero in the corresponding model spectra, the agreement between integrated and tabulated values for all other spectra used in the \CB\ models is always better than 10 per cent. 


\label{lastpage}
\end{document}